\documentclass[10pt]{extarticle}

\usepackage[english]{babel}
\usepackage{graphicx}
\usepackage{framed}
\usepackage[normalem]{ulem}
\usepackage{indentfirst,ragged2e}
\usepackage{amsmath,amsthm,amssymb,amsfonts,wasysym,verbatim,bbm}
\usepackage[italicdiff]{physics}
\usepackage[T1]{fontenc}
\usepackage{lmodern,mathrsfs}
\usepackage{mathdots} 
\usepackage{wrapfig}
\usepackage[inline,shortlabels]{enumitem}
\setlist{topsep=2pt,itemsep=2pt,parsep=0pt,partopsep=0pt}
\usepackage[dvipsnames]{xcolor}
\usepackage[utf8]{inputenc}
\usepackage[a4paper,top=0.7in,bottom=0.7in,left=0.7in,right=0.7in]{geometry}
\usepackage{multicol}
\usepackage[most]{tcolorbox}
\usepackage{nicematrix} 
\usepackage{tikz,tikz-3dplot,tikz-cd,tkz-tab,tkz-euclide,tikzsymbols,pgf,pgfplots}
\pgfplotsset{compat=newest}
\usepackage{subfiles}
\graphicspath{{images/}{../images/}} 
\usepackage[backend=bibtex,style=numeric]{biblatex}
\usepackage[draft]{minted} 
\usepackage{xurl} 
\usepackage[colorlinks,linkcolor=teal,citecolor=blue,urlcolor=violet]{hyperref}
\usepackage[nameinlink]{cleveref}

\makeatletter
\def\PYGdefault@reset{\let\PYGdefault@it=\relax \let\PYGdefault@bf=\relax%
    \let\PYGdefault@ul=\relax \let\PYGdefault@tc=\relax%
    \let\PYGdefault@bc=\relax \let\PYGdefault@ff=\relax}
\def\PYGdefault@tok#1{\csname PYGdefault@tok@#1\endcsname}
\def\PYGdefault@toks#1+{\ifx\relax#1\empty\else%
    \PYGdefault@tok{#1}\expandafter\PYGdefault@toks\fi}
\def\PYGdefault@do#1{\PYGdefault@bc{\PYGdefault@tc{\PYGdefault@ul{%
    \PYGdefault@it{\PYGdefault@bf{\PYGdefault@ff{#1}}}}}}}
\def\PYGdefault#1#2{\PYGdefault@reset\PYGdefault@toks#1+\relax+\PYGdefault@do{#2}}

\expandafter\def\csname PYGdefault@tok@gd\endcsname{\def\PYGdefault@tc##1{\textcolor[rgb]{0.63,0.00,0.00}{##1}}}
\expandafter\def\csname PYGdefault@tok@gu\endcsname{\let\PYGdefault@bf=\textbf\def\PYGdefault@tc##1{\textcolor[rgb]{0.50,0.00,0.50}{##1}}}
\expandafter\def\csname PYGdefault@tok@gt\endcsname{\def\PYGdefault@tc##1{\textcolor[rgb]{0.00,0.27,0.87}{##1}}}
\expandafter\def\csname PYGdefault@tok@gs\endcsname{\let\PYGdefault@bf=\textbf}
\expandafter\def\csname PYGdefault@tok@gr\endcsname{\def\PYGdefault@tc##1{\textcolor[rgb]{1.00,0.00,0.00}{##1}}}
\expandafter\def\csname PYGdefault@tok@cm\endcsname{\let\PYGdefault@it=\textit\def\PYGdefault@tc##1{\textcolor[rgb]{0.25,0.50,0.50}{##1}}}
\expandafter\def\csname PYGdefault@tok@vg\endcsname{\def\PYGdefault@tc##1{\textcolor[rgb]{0.10,0.09,0.49}{##1}}}
\expandafter\def\csname PYGdefault@tok@m\endcsname{\def\PYGdefault@tc##1{\textcolor[rgb]{0.40,0.40,0.40}{##1}}}
\expandafter\def\csname PYGdefault@tok@mh\endcsname{\def\PYGdefault@tc##1{\textcolor[rgb]{0.40,0.40,0.40}{##1}}}
\expandafter\def\csname PYGdefault@tok@go\endcsname{\def\PYGdefault@tc##1{\textcolor[rgb]{0.53,0.53,0.53}{##1}}}
\expandafter\def\csname PYGdefault@tok@ge\endcsname{\let\PYGdefault@it=\textit}
\expandafter\def\csname PYGdefault@tok@vc\endcsname{\def\PYGdefault@tc##1{\textcolor[rgb]{0.10,0.09,0.49}{##1}}}
\expandafter\def\csname PYGdefault@tok@il\endcsname{\def\PYGdefault@tc##1{\textcolor[rgb]{0.40,0.40,0.40}{##1}}}
\expandafter\def\csname PYGdefault@tok@cs\endcsname{\let\PYGdefault@it=\textit\def\PYGdefault@tc##1{\textcolor[rgb]{0.25,0.50,0.50}{##1}}}
\expandafter\def\csname PYGdefault@tok@cp\endcsname{\def\PYGdefault@tc##1{\textcolor[rgb]{0.74,0.48,0.00}{##1}}}
\expandafter\def\csname PYGdefault@tok@gi\endcsname{\def\PYGdefault@tc##1{\textcolor[rgb]{0.00,0.63,0.00}{##1}}}
\expandafter\def\csname PYGdefault@tok@gh\endcsname{\let\PYGdefault@bf=\textbf\def\PYGdefault@tc##1{\textcolor[rgb]{0.00,0.00,0.50}{##1}}}
\expandafter\def\csname PYGdefault@tok@ni\endcsname{\let\PYGdefault@bf=\textbf\def\PYGdefault@tc##1{\textcolor[rgb]{0.60,0.60,0.60}{##1}}}
\expandafter\def\csname PYGdefault@tok@nl\endcsname{\def\PYGdefault@tc##1{\textcolor[rgb]{0.63,0.63,0.00}{##1}}}
\expandafter\def\csname PYGdefault@tok@nn\endcsname{\let\PYGdefault@bf=\textbf\def\PYGdefault@tc##1{\textcolor[rgb]{0.00,0.00,1.00}{##1}}}
\expandafter\def\csname PYGdefault@tok@no\endcsname{\def\PYGdefault@tc##1{\textcolor[rgb]{0.53,0.00,0.00}{##1}}}
\expandafter\def\csname PYGdefault@tok@na\endcsname{\def\PYGdefault@tc##1{\textcolor[rgb]{0.49,0.56,0.16}{##1}}}
\expandafter\def\csname PYGdefault@tok@nb\endcsname{\def\PYGdefault@tc##1{\textcolor[rgb]{0.00,0.50,0.00}{##1}}}
\expandafter\def\csname PYGdefault@tok@nc\endcsname{\let\PYGdefault@bf=\textbf\def\PYGdefault@tc##1{\textcolor[rgb]{0.00,0.00,1.00}{##1}}}
\expandafter\def\csname PYGdefault@tok@nd\endcsname{\def\PYGdefault@tc##1{\textcolor[rgb]{0.67,0.13,1.00}{##1}}}
\expandafter\def\csname PYGdefault@tok@ne\endcsname{\let\PYGdefault@bf=\textbf\def\PYGdefault@tc##1{\textcolor[rgb]{0.82,0.25,0.23}{##1}}}
\expandafter\def\csname PYGdefault@tok@nf\endcsname{\def\PYGdefault@tc##1{\textcolor[rgb]{0.00,0.00,1.00}{##1}}}
\expandafter\def\csname PYGdefault@tok@si\endcsname{\let\PYGdefault@bf=\textbf\def\PYGdefault@tc##1{\textcolor[rgb]{0.73,0.40,0.53}{##1}}}
\expandafter\def\csname PYGdefault@tok@s2\endcsname{\def\PYGdefault@tc##1{\textcolor[rgb]{0.73,0.13,0.13}{##1}}}
\expandafter\def\csname PYGdefault@tok@vi\endcsname{\def\PYGdefault@tc##1{\textcolor[rgb]{0.10,0.09,0.49}{##1}}}
\expandafter\def\csname PYGdefault@tok@nt\endcsname{\let\PYGdefault@bf=\textbf\def\PYGdefault@tc##1{\textcolor[rgb]{0.00,0.50,0.00}{##1}}}
\expandafter\def\csname PYGdefault@tok@nv\endcsname{\def\PYGdefault@tc##1{\textcolor[rgb]{0.10,0.09,0.49}{##1}}}
\expandafter\def\csname PYGdefault@tok@s1\endcsname{\def\PYGdefault@tc##1{\textcolor[rgb]{0.73,0.13,0.13}{##1}}}
\expandafter\def\csname PYGdefault@tok@sh\endcsname{\def\PYGdefault@tc##1{\textcolor[rgb]{0.73,0.13,0.13}{##1}}}
\expandafter\def\csname PYGdefault@tok@sc\endcsname{\def\PYGdefault@tc##1{\textcolor[rgb]{0.73,0.13,0.13}{##1}}}
\expandafter\def\csname PYGdefault@tok@sx\endcsname{\def\PYGdefault@tc##1{\textcolor[rgb]{0.00,0.50,0.00}{##1}}}
\expandafter\def\csname PYGdefault@tok@bp\endcsname{\def\PYGdefault@tc##1{\textcolor[rgb]{0.00,0.50,0.00}{##1}}}
\expandafter\def\csname PYGdefault@tok@c1\endcsname{\let\PYGdefault@it=\textit\def\PYGdefault@tc##1{\textcolor[rgb]{0.25,0.50,0.50}{##1}}}
\expandafter\def\csname PYGdefault@tok@kc\endcsname{\let\PYGdefault@bf=\textbf\def\PYGdefault@tc##1{\textcolor[rgb]{0.00,0.50,0.00}{##1}}}
\expandafter\def\csname PYGdefault@tok@c\endcsname{\let\PYGdefault@it=\textit\def\PYGdefault@tc##1{\textcolor[rgb]{0.25,0.50,0.50}{##1}}}
\expandafter\def\csname PYGdefault@tok@mf\endcsname{\def\PYGdefault@tc##1{\textcolor[rgb]{0.40,0.40,0.40}{##1}}}
\expandafter\def\csname PYGdefault@tok@err\endcsname{\def\PYGdefault@bc##1{\setlength{\fboxsep}{0pt}\fcolorbox[rgb]{1.00,0.00,0.00}{1,1,1}{\strut ##1}}}
\expandafter\def\csname PYGdefault@tok@kd\endcsname{\let\PYGdefault@bf=\textbf\def\PYGdefault@tc##1{\textcolor[rgb]{0.00,0.50,0.00}{##1}}}
\expandafter\def\csname PYGdefault@tok@ss\endcsname{\def\PYGdefault@tc##1{\textcolor[rgb]{0.10,0.09,0.49}{##1}}}
\expandafter\def\csname PYGdefault@tok@sr\endcsname{\def\PYGdefault@tc##1{\textcolor[rgb]{0.73,0.40,0.53}{##1}}}
\expandafter\def\csname PYGdefault@tok@mo\endcsname{\def\PYGdefault@tc##1{\textcolor[rgb]{0.40,0.40,0.40}{##1}}}
\expandafter\def\csname PYGdefault@tok@kn\endcsname{\let\PYGdefault@bf=\textbf\def\PYGdefault@tc##1{\textcolor[rgb]{0.00,0.50,0.00}{##1}}}
\expandafter\def\csname PYGdefault@tok@mi\endcsname{\def\PYGdefault@tc##1{\textcolor[rgb]{0.40,0.40,0.40}{##1}}}
\expandafter\def\csname PYGdefault@tok@gp\endcsname{\let\PYGdefault@bf=\textbf\def\PYGdefault@tc##1{\textcolor[rgb]{0.00,0.00,0.50}{##1}}}
\expandafter\def\csname PYGdefault@tok@o\endcsname{\def\PYGdefault@tc##1{\textcolor[rgb]{0.40,0.40,0.40}{##1}}}
\expandafter\def\csname PYGdefault@tok@kr\endcsname{\let\PYGdefault@bf=\textbf\def\PYGdefault@tc##1{\textcolor[rgb]{0.00,0.50,0.00}{##1}}}
\expandafter\def\csname PYGdefault@tok@s\endcsname{\def\PYGdefault@tc##1{\textcolor[rgb]{0.73,0.13,0.13}{##1}}}
\expandafter\def\csname PYGdefault@tok@kp\endcsname{\def\PYGdefault@tc##1{\textcolor[rgb]{0.00,0.50,0.00}{##1}}}
\expandafter\def\csname PYGdefault@tok@w\endcsname{\def\PYGdefault@tc##1{\textcolor[rgb]{0.73,0.73,0.73}{##1}}}
\expandafter\def\csname PYGdefault@tok@kt\endcsname{\def\PYGdefault@tc##1{\textcolor[rgb]{0.69,0.00,0.25}{##1}}}
\expandafter\def\csname PYGdefault@tok@ow\endcsname{\let\PYGdefault@bf=\textbf\def\PYGdefault@tc##1{\textcolor[rgb]{0.67,0.13,1.00}{##1}}}
\expandafter\def\csname PYGdefault@tok@sb\endcsname{\def\PYGdefault@tc##1{\textcolor[rgb]{0.73,0.13,0.13}{##1}}}
\expandafter\def\csname PYGdefault@tok@k\endcsname{\let\PYGdefault@bf=\textbf\def\PYGdefault@tc##1{\textcolor[rgb]{0.00,0.50,0.00}{##1}}}
\expandafter\def\csname PYGdefault@tok@se\endcsname{\let\PYGdefault@bf=\textbf\def\PYGdefault@tc##1{\textcolor[rgb]{0.73,0.40,0.13}{##1}}}
\expandafter\def\csname PYGdefault@tok@sd\endcsname{\let\PYGdefault@it=\textit\def\PYGdefault@tc##1{\textcolor[rgb]{0.73,0.13,0.13}{##1}}}

\makeatother

\makeatletter
\def\PYG@reset{\let\PYG@it=\relax \let\PYG@bf=\relax%
    \let\PYG@ul=\relax \let\PYG@tc=\relax%
    \let\PYG@bc=\relax \let\PYG@ff=\relax}
\def\PYG@tok#1{\csname PYG@tok@#1\endcsname}
\def\PYG@toks#1+{\ifx\relax#1\empty\else%
    \PYG@tok{#1}\expandafter\PYG@toks\fi}
\def\PYG@do#1{\PYG@bc{\PYG@tc{\PYG@ul{%
    \PYG@it{\PYG@bf{\PYG@ff{#1}}}}}}}
\def\PYG#1#2{\PYG@reset\PYG@toks#1+\relax+\PYG@do{#2}}

\expandafter\def\csname PYG@tok@gd\endcsname{\def\PYG@tc##1{\textcolor[rgb]{0.63,0.00,0.00}{##1}}}
\expandafter\def\csname PYG@tok@gu\endcsname{\let\PYG@bf=\textbf\def\PYG@tc##1{\textcolor[rgb]{0.50,0.00,0.50}{##1}}}
\expandafter\def\csname PYG@tok@gt\endcsname{\def\PYG@tc##1{\textcolor[rgb]{0.00,0.27,0.87}{##1}}}
\expandafter\def\csname PYG@tok@gs\endcsname{\let\PYG@bf=\textbf}
\expandafter\def\csname PYG@tok@gr\endcsname{\def\PYG@tc##1{\textcolor[rgb]{1.00,0.00,0.00}{##1}}}
\expandafter\def\csname PYG@tok@cm\endcsname{\let\PYG@it=\textit\def\PYG@tc##1{\textcolor[rgb]{0.25,0.50,0.50}{##1}}}
\expandafter\def\csname PYG@tok@vg\endcsname{\def\PYG@tc##1{\textcolor[rgb]{0.10,0.09,0.49}{##1}}}
\expandafter\def\csname PYG@tok@m\endcsname{\def\PYG@tc##1{\textcolor[rgb]{0.40,0.40,0.40}{##1}}}
\expandafter\def\csname PYG@tok@mh\endcsname{\def\PYG@tc##1{\textcolor[rgb]{0.40,0.40,0.40}{##1}}}
\expandafter\def\csname PYG@tok@go\endcsname{\def\PYG@tc##1{\textcolor[rgb]{0.53,0.53,0.53}{##1}}}
\expandafter\def\csname PYG@tok@ge\endcsname{\let\PYG@it=\textit}
\expandafter\def\csname PYG@tok@vc\endcsname{\def\PYG@tc##1{\textcolor[rgb]{0.10,0.09,0.49}{##1}}}
\expandafter\def\csname PYG@tok@il\endcsname{\def\PYG@tc##1{\textcolor[rgb]{0.40,0.40,0.40}{##1}}}
\expandafter\def\csname PYG@tok@cs\endcsname{\let\PYG@it=\textit\def\PYG@tc##1{\textcolor[rgb]{0.25,0.50,0.50}{##1}}}
\expandafter\def\csname PYG@tok@cp\endcsname{\def\PYG@tc##1{\textcolor[rgb]{0.74,0.48,0.00}{##1}}}
\expandafter\def\csname PYG@tok@gi\endcsname{\def\PYG@tc##1{\textcolor[rgb]{0.00,0.63,0.00}{##1}}}
\expandafter\def\csname PYG@tok@gh\endcsname{\let\PYG@bf=\textbf\def\PYG@tc##1{\textcolor[rgb]{0.00,0.00,0.50}{##1}}}
\expandafter\def\csname PYG@tok@ni\endcsname{\let\PYG@bf=\textbf\def\PYG@tc##1{\textcolor[rgb]{0.60,0.60,0.60}{##1}}}
\expandafter\def\csname PYG@tok@nl\endcsname{\def\PYG@tc##1{\textcolor[rgb]{0.63,0.63,0.00}{##1}}}
\expandafter\def\csname PYG@tok@nn\endcsname{\let\PYG@bf=\textbf\def\PYG@tc##1{\textcolor[rgb]{0.00,0.00,1.00}{##1}}}
\expandafter\def\csname PYG@tok@no\endcsname{\def\PYG@tc##1{\textcolor[rgb]{0.53,0.00,0.00}{##1}}}
\expandafter\def\csname PYG@tok@na\endcsname{\def\PYG@tc##1{\textcolor[rgb]{0.49,0.56,0.16}{##1}}}
\expandafter\def\csname PYG@tok@nb\endcsname{\def\PYG@tc##1{\textcolor[rgb]{0.00,0.50,0.00}{##1}}}
\expandafter\def\csname PYG@tok@nc\endcsname{\let\PYG@bf=\textbf\def\PYG@tc##1{\textcolor[rgb]{0.00,0.00,1.00}{##1}}}
\expandafter\def\csname PYG@tok@nd\endcsname{\def\PYG@tc##1{\textcolor[rgb]{0.67,0.13,1.00}{##1}}}
\expandafter\def\csname PYG@tok@ne\endcsname{\let\PYG@bf=\textbf\def\PYG@tc##1{\textcolor[rgb]{0.82,0.25,0.23}{##1}}}
\expandafter\def\csname PYG@tok@nf\endcsname{\def\PYG@tc##1{\textcolor[rgb]{0.00,0.00,1.00}{##1}}}
\expandafter\def\csname PYG@tok@si\endcsname{\let\PYG@bf=\textbf\def\PYG@tc##1{\textcolor[rgb]{0.73,0.40,0.53}{##1}}}
\expandafter\def\csname PYG@tok@s2\endcsname{\def\PYG@tc##1{\textcolor[rgb]{0.73,0.13,0.13}{##1}}}
\expandafter\def\csname PYG@tok@vi\endcsname{\def\PYG@tc##1{\textcolor[rgb]{0.10,0.09,0.49}{##1}}}
\expandafter\def\csname PYG@tok@nt\endcsname{\let\PYG@bf=\textbf\def\PYG@tc##1{\textcolor[rgb]{0.00,0.50,0.00}{##1}}}
\expandafter\def\csname PYG@tok@nv\endcsname{\def\PYG@tc##1{\textcolor[rgb]{0.10,0.09,0.49}{##1}}}
\expandafter\def\csname PYG@tok@s1\endcsname{\def\PYG@tc##1{\textcolor[rgb]{0.73,0.13,0.13}{##1}}}
\expandafter\def\csname PYG@tok@sh\endcsname{\def\PYG@tc##1{\textcolor[rgb]{0.73,0.13,0.13}{##1}}}
\expandafter\def\csname PYG@tok@sc\endcsname{\def\PYG@tc##1{\textcolor[rgb]{0.73,0.13,0.13}{##1}}}
\expandafter\def\csname PYG@tok@sx\endcsname{\def\PYG@tc##1{\textcolor[rgb]{0.00,0.50,0.00}{##1}}}
\expandafter\def\csname PYG@tok@bp\endcsname{\def\PYG@tc##1{\textcolor[rgb]{0.00,0.50,0.00}{##1}}}
\expandafter\def\csname PYG@tok@c1\endcsname{\let\PYG@it=\textit\def\PYG@tc##1{\textcolor[rgb]{0.25,0.50,0.50}{##1}}}
\expandafter\def\csname PYG@tok@kc\endcsname{\let\PYG@bf=\textbf\def\PYG@tc##1{\textcolor[rgb]{0.00,0.50,0.00}{##1}}}
\expandafter\def\csname PYG@tok@c\endcsname{\let\PYG@it=\textit\def\PYG@tc##1{\textcolor[rgb]{0.25,0.50,0.50}{##1}}}
\expandafter\def\csname PYG@tok@mf\endcsname{\def\PYG@tc##1{\textcolor[rgb]{0.40,0.40,0.40}{##1}}}
\expandafter\def\csname PYG@tok@err\endcsname{\def\PYG@bc##1{\setlength{\fboxsep}{0pt}\fcolorbox[rgb]{1.00,0.00,0.00}{1,1,1}{\strut ##1}}}
\expandafter\def\csname PYG@tok@kd\endcsname{\let\PYG@bf=\textbf\def\PYG@tc##1{\textcolor[rgb]{0.00,0.50,0.00}{##1}}}
\expandafter\def\csname PYG@tok@ss\endcsname{\def\PYG@tc##1{\textcolor[rgb]{0.10,0.09,0.49}{##1}}}
\expandafter\def\csname PYG@tok@sr\endcsname{\def\PYG@tc##1{\textcolor[rgb]{0.73,0.40,0.53}{##1}}}
\expandafter\def\csname PYG@tok@mo\endcsname{\def\PYG@tc##1{\textcolor[rgb]{0.40,0.40,0.40}{##1}}}
\expandafter\def\csname PYG@tok@kn\endcsname{\let\PYG@bf=\textbf\def\PYG@tc##1{\textcolor[rgb]{0.00,0.50,0.00}{##1}}}
\expandafter\def\csname PYG@tok@mi\endcsname{\def\PYG@tc##1{\textcolor[rgb]{0.40,0.40,0.40}{##1}}}
\expandafter\def\csname PYG@tok@gp\endcsname{\let\PYG@bf=\textbf\def\PYG@tc##1{\textcolor[rgb]{0.00,0.00,0.50}{##1}}}
\expandafter\def\csname PYG@tok@o\endcsname{\def\PYG@tc##1{\textcolor[rgb]{0.40,0.40,0.40}{##1}}}
\expandafter\def\csname PYG@tok@kr\endcsname{\let\PYG@bf=\textbf\def\PYG@tc##1{\textcolor[rgb]{0.00,0.50,0.00}{##1}}}
\expandafter\def\csname PYG@tok@s\endcsname{\def\PYG@tc##1{\textcolor[rgb]{0.73,0.13,0.13}{##1}}}
\expandafter\def\csname PYG@tok@kp\endcsname{\def\PYG@tc##1{\textcolor[rgb]{0.00,0.50,0.00}{##1}}}
\expandafter\def\csname PYG@tok@w\endcsname{\def\PYG@tc##1{\textcolor[rgb]{0.73,0.73,0.73}{##1}}}
\expandafter\def\csname PYG@tok@kt\endcsname{\def\PYG@tc##1{\textcolor[rgb]{0.69,0.00,0.25}{##1}}}
\expandafter\def\csname PYG@tok@ow\endcsname{\let\PYG@bf=\textbf\def\PYG@tc##1{\textcolor[rgb]{0.67,0.13,1.00}{##1}}}
\expandafter\def\csname PYG@tok@sb\endcsname{\def\PYG@tc##1{\textcolor[rgb]{0.73,0.13,0.13}{##1}}}
\expandafter\def\csname PYG@tok@k\endcsname{\let\PYG@bf=\textbf\def\PYG@tc##1{\textcolor[rgb]{0.00,0.50,0.00}{##1}}}
\expandafter\def\csname PYG@tok@se\endcsname{\let\PYG@bf=\textbf\def\PYG@tc##1{\textcolor[rgb]{0.73,0.40,0.13}{##1}}}
\expandafter\def\csname PYG@tok@sd\endcsname{\let\PYG@it=\textit\def\PYG@tc##1{\textcolor[rgb]{0.73,0.13,0.13}{##1}}}

\makeatother 

\newcommand{\hide}[1]{} 

\definecolor{contcol}{HTML}{8757E6}
\definecolor{conttxtcol}{HTML}{052A78}
\definecolor{convcol}{HTML}{3259C2}
\definecolor{abscol}{HTML}{212694}

\makeatletter
\def\my@vector #1,#2\@eolst{
    \ifx\relax#2\relax
        #1
    \else
        #1\my@delim
        \my@vector #2\@eolst
    \fi}
\newcommand\vcstring[2][\\]{
    \global\def\my@delim{#1}
        \my@vector #2,\relax\noexpand\@eolst}
\newcommand\cvc[2][p]{
    \global\def\my@delim{\\}
        \ensuremath{\begin{#1matrix} 
            \my@vector #2,\relax\noexpand\@eolst
        \end{#1matrix}}}
\newcommand\rvc[2][p]{
    \global\def\my@delim{&}
        \ensuremath{\begin{#1matrix} 
            \my@vector #2,\relax\noexpand\@eolst
        \end{#1matrix}}}
\newcommand{\mat}[2][p]{
    \def\myenv{#1matrix} 
    \def\my@delim{&}
        \begin{\myenv} 
            \my@vector #2,\relax\noexpand\@eolst
            \@ifnextchar\bgroup{\passtonextarg}{\end{\myenv}}} 
    \newcommand{\passtonextarg}[1]{\\ \my@vector #1,\relax\noexpand\@eolst
        \@ifnextchar\bgroup{\passtonextarg}{\end{\myenv}}} 
\makeatother

\usepackage{fancyhdr}
\fancypagestyle{plain}{
\fancyhf{}

\fancyfoot[C]{\sffamily\thepage}
}

\newcommand{\ep}{\varepsilon}
\newcommand{\vp}{\varphi}
\newcommand{\lam}{\lambda}
\newcommand{\Lam}{\Lambda}
\renewcommand{\ip}[1]{\ensuremath{\left\langle#1\right\rangle}}

\newcommand{\C}{\mathbb{C}}

\newcommand{\N}{\mathbb{N}}

\newcommand{\R}{\mathbb{R}}

\newcommand{\As}{\mathcal{A}}
\newcommand{\Bs}{\mathcal{B}}

\newcommand{\Ds}{\mathcal{D}}

\newcommand{\Gs}{\mathcal{G}}
\newcommand{\Hs}{\mathcal{H}}

\newcommand{\Ps}{\mathcal{P}}

\newcommand{\xb}{\textbf{x}}

\newcommand{\limn}{\lim_{n\to\infty}}

\newcommand{\sumn}[1][1]{\sum_{n=#1}^\infty}

\newcommand{\emp}{\varnothing}

\newcommand{\sub}{\subseteq}
\newcommand{\sups}{\supseteq}

\newcommand{\cuppn}[1][1]{\bigcup_{n=#1}^\infty}

\newcommand{\dx}{\,dx}
\newcommand{\dy}{\,dy}

\let\Re\relax

\DeclareMathOperator{\Re}{\text{Re}}

\newcommand{\dg}{^\dagger}

\newtheoremstyle{mystyle}{}{}{}{}{\sffamily\bfseries}{.}{ }{}
\newtheoremstyle{cstyle}{}{}{}{}{\sffamily\bfseries}{.}{ }{\thmnote{#3}}
\makeatletter
\renewenvironment{proof}[1][\proofname] {\par\pushQED{\qed}{\normalfont\sffamily\bfseries\topsep6\p@\@plus6\p@\relax #1\@addpunct{.} }}{\popQED\endtrivlist\@endpefalse}
\makeatother
\theoremstyle{mystyle}{\newtheorem{definition}{Definition}[section]}
\theoremstyle{mystyle}{\newtheorem{proposition}[definition]{Proposition}}
\theoremstyle{mystyle}{\newtheorem{theorem}[definition]{Theorem}}
\theoremstyle{mystyle}{\newtheorem{lemma}[definition]{Lemma}}
\theoremstyle{mystyle}{}
\theoremstyle{mystyle}{\newtheorem*{remark}{Remark}}
\theoremstyle{mystyle}{}
\theoremstyle{mystyle}{\newtheorem*{example}{Example}}
\theoremstyle{mystyle}{}
\theoremstyle{definition}{}
\theoremstyle{cstyle}{\newtheorem*{cthm}{}}
\newtheoremstyle{warn}{}{}{}{}{\normalfont}{}{ }{}
\theoremstyle{warn}
\newtheorem*{warning}{\warningsign{0.2}\relax} 

\newcommand{\warningsign}[1]{\tikz[scale=#1,every node/.style={transform shape}]{
\draw[-,line width={#1*0.8mm},red,fill=yellow,rounded corners={#1*2.5mm}] (0,0)--(1,{-sqrt(3)})--(-1,{-sqrt(3)})--cycle;
\node at (0,-1) {\fontsize{48}{60}\selectfont\bfseries!};}}

\newtcolorbox{constpotentialbox}[1]{colback=white,colframe=gray!50!black,title=#1,fonttitle=\sffamily\bfseries,coltitle=white,top=1mm,bottom=1mm,left=1mm,right=1mm,sharp corners,before skip=10pt,after skip=10pt}

\newtcbox{\bubble}[2][20]{colback=#2!10,colframe=#2,hbox,center,halign=center,valign=center,height=#1pt,bean arc,top=2pt,bottom=2pt,left=2pt,right=2pt,boxrule=1pt,before skip=10pt,after skip=10pt}

\newtcolorbox{algotext}[1][]{title={#1},fonttitle=\sffamily\bfseries,coltitle=white,colbacktitle=gray!20!black,toptitle=4pt,bottomtitle=4pt,boxrule=1pt,boxsep=0pt,colback=gray!10,colframe=gray!20!black,enhanced jigsaw, sharp corners,before skip=10pt,after skip=10pt,breakable}

\newtcbox{\personbox}[2]{
title={#2},
flip title={interior hidden}, 
fonttitle=\sffamily\selectfont,center title,
colframe=#1,coltitle=black,
hbox,boxsep=0mm, 
top=0mm,bottom=0mm,left=0mm,right=0mm,toptitle=1mm,bottomtitle=1mm, 
sharp corners, 
clip upper, 
enhanced}

\makeatletter
\def\ifemptyarg#1{%
  \if\relax\detokenize{#1}\relax 
    \expandafter\@firstoftwo
  \else
    \expandafter\@secondoftwo
  \fi}
\makeatother

\tcolorboxenvironment{definition}{boxrule=0pt,boxsep=0pt,colback={red!10},left=8pt,right=8pt,enhanced jigsaw, borderline west={2pt}{0pt}{red},sharp corners,before skip=10pt,after skip=10pt,breakable}
\tcolorboxenvironment{proposition}{boxrule=0pt,boxsep=0pt,colback={Orange!10},left=8pt,right=8pt,enhanced jigsaw, borderline west={2pt}{0pt}{Orange},sharp corners,before skip=10pt,after skip=10pt,breakable}
\tcolorboxenvironment{theorem}{boxrule=0pt,boxsep=0pt,colback={blue!10},left=8pt,right=8pt,enhanced jigsaw, borderline west={2pt}{0pt}{blue},sharp corners,before skip=10pt,after skip=10pt,breakable}
\tcolorboxenvironment{lemma}{boxrule=0pt,boxsep=0pt,colback={Cyan!10},left=8pt,right=8pt,enhanced jigsaw, borderline west={2pt}{0pt}{Cyan},sharp corners,before skip=10pt,after skip=10pt,breakable}
\tcolorboxenvironment{corollary}{boxrule=0pt,boxsep=0pt,colback={violet!10},left=8pt,right=8pt,enhanced jigsaw, borderline west={2pt}{0pt}{violet},sharp corners,before skip=10pt,after skip=10pt,breakable}
\tcolorboxenvironment{proof}{boxrule=0pt,boxsep=0pt,blanker,borderline west={2pt}{0pt}{CadetBlue!80!white},left=8pt,right=8pt,sharp corners,before skip=10pt,after skip=10pt,breakable}
\tcolorboxenvironment{remark}{boxrule=0pt,boxsep=0pt,blanker,borderline west={2pt}{0pt}{Green},left=8pt,right=8pt,before skip=10pt,after skip=10pt,breakable}
\tcolorboxenvironment{remarks}{boxrule=0pt,boxsep=0pt,blanker,borderline west={2pt}{0pt}{Green},left=8pt,right=8pt,before skip=10pt,after skip=10pt,breakable}
\tcolorboxenvironment{example}{boxrule=0pt,boxsep=0pt,blanker,borderline west={2pt}{0pt}{Black},left=8pt,right=8pt,sharp corners,before skip=10pt,after skip=10pt,breakable}
\tcolorboxenvironment{examples}{boxrule=0pt,boxsep=0pt,blanker,borderline west={2pt}{0pt}{Black},left=8pt,right=8pt,sharp corners,before skip=10pt,after skip=10pt,breakable}
\tcolorboxenvironment{cthm}{boxrule=0pt,boxsep=0pt,colback={gray!10},left=8pt,right=8pt,enhanced jigsaw, borderline west={2pt}{0pt}{gray},sharp corners,before skip=10pt,after skip=10pt,breakable}

\newenvironment{talign*}{\csname align*\endcsname}{\endalign}

\usepackage[explicit]{titlesec}
\titleformat{\section}{\fontsize{24}{30}\sffamily\bfseries}{\thesection}{16pt}{#1}
\titleformat{\subsection}{\fontsize{16}{20}\sffamily\Large\bfseries}{\thesubsection}{12pt}{#1}
\titleformat{\subsubsection}{\sffamily\large\bfseries}{\thesubsection}{8pt}{#1}

\titlespacing*{\section}{0pt}{5pt}{5pt}
\titlespacing*{\subsection}{0pt}{5pt}{5pt}
\titlespacing*{\subsubsection}{0pt}{5pt}{5pt}

\DeclareMathAlphabet\mathbfcal{OMS}{cmsy}{b}{n}
\makeatletter
\g@addto@macro\normalsize{
\setlength\abovedisplayskip{3pt}
\setlength\belowdisplayskip{3pt}
\setlength\abovedisplayshortskip{0pt}
\setlength\belowdisplayshortskip{0pt}}
\makeatother

\makeatletter
\def\supervisor#1{\gdef\@supervisor{#1}}
\def\@supervisor{\@latex@warning@no@line{No \noexpand\supervisor given}}
\def\module#1{\gdef\@module{#1}}
\def\@module{\@latex@warning@no@line{No \noexpand\module given}}
\renewcommand\maketitle{
\begin{center}
{\fontsize{24}{28}\sffamily\bfseries\selectfont\@title}\\
\vspace{8mm}
{\fontsize{24}{28}\sffamily\selectfont\@author}\\
\vspace{12mm}
{\fontsize{20}{24}\sffamily\selectfont Project Supervisor: \@supervisor}\\
\vspace{6mm}
{\fontsize{18}{22}\sffamily\selectfont Module: \@module}\\
\vspace{6mm}
{\fontsize{18}{22}\sffamily\selectfont\@date} 
\end{center}}
\makeatother

\renewenvironment{abstract}{ 
\begin{tcolorbox}[
    enhanced,width=6in,center,title=Abstract,
    frame hidden,colback=white,colbacktitle=white,
    fonttitle=\sffamily\bfseries,coltitle=black,
        borderline north={1.5pt}{0pt}{abscol},
        attach boxed title to top center={yshift=-0.25mm-\tcboxedtitleheight/2,yshifttext=2mm-\tcboxedtitleheight/2},
        boxed title style={boxrule=0.5mm,
        frame code={\path[tcb fill frame,abscol] ([xshift=-4mm]frame.west)--(frame.north west)--(frame.north east)--([xshift=4mm]frame.east)--(frame.south east)--(frame.south west)--cycle;},
        interior code={\path[tcb fill interior] ([xshift=-2mm]interior.west)--(interior.north west)--(interior.north east)--([xshift=2mm]interior.east)--(interior.south east)--(interior.south west)--cycle;}},
    top=\tcboxedtitleheight/2,bottom=0pt,left=0pt,right=0pt]}
{\end{tcolorbox}}

\title{Supersymmetric Quantum Mechanics:\\Light at the End of the (Quantum) Tunnel}
\author{Senan Sekhon}
\supervisor{Christopher Herzog}
\module{7CCM461A}
\date{2021/22}

\begin{document}

\begin{titlepage}
\vspace*{20mm}

\begin{center}

\begin{tikzpicture}[complexnode/.pic={
\tdplotsetmaincoords{52}{118} 
\begin{scope}[tdplot_main_coords]
    \foreach \n in {0,1,...,9}
    {\pgfmathsetmacro{\perc}{90-6*\n} 
    \fill[colb!\perc] (0,{cos(15*\n)},{sin(15*\n)})--(-1.5,{cos(15*\n)},{sin(15*\n)})--(-1.5,{cos(15*(\n+1))},{sin(15*(\n+1))})--(0,{cos(15*(\n+1))},{sin(15*(\n+1))})--cycle;} 
    \foreach \n in {0,1,...,10}
    \draw (0,{cos(15*\n)},{sin(15*\n)})--(-1.5,{cos(15*\n)},{sin(15*\n)}); 
    \begin{scope}[canvas is yz plane at x=0]
        \fill[left color=gray!60!black,right color=black] (1,0) arc(0:180:1) --cycle; 
    \end{scope}
    \begin{scope}[canvas is yz plane at x=-0.25]
        \foreach \t in {15,45,...,135}
        \draw (\t:1) arc(\t:{\t+15}:1); 
    \end{scope}
    \begin{scope}[canvas is yz plane at x=-0.5]
        \foreach \t in {0,30,...,120}
        \draw (\t:1) arc(\t:{\t+15}:1); 
    \end{scope}
    \begin{scope}[canvas is yz plane at x=-0.75]
        \foreach \t in {15,45,...,135}
        \draw (\t:1) arc(\t:{\t+15}:1); 
    \end{scope}
    \begin{scope}[canvas is yz plane at x=-1]
        \foreach \t in {0,30,...,120}
        \draw (\t:1) arc(\t:{\t+15}:1); 
    \end{scope}
    \begin{scope}[canvas is yz plane at x=-1.25]
        \foreach \t in {15,45,...,135}
        \draw (\t:1) arc(\t:{\t+15}:1); 
    \end{scope}
    \begin{scope}[canvas is yz plane at x=-1.5]
        \draw (1,0) arc(0:150:1); 
    \end{scope}
\end{scope}
}]

\pgfmathdeclarefunction{psi1}{2}{\pgfmathparse{0.18*sin((#2+1)*180*(#1+0.5))}}; 
\pgfmathdeclarefunction{psi2}{2}{\pgfmathparse{0.18*sqrt(1/((#2+1)*(#2+3)))*((#2+2)*cos((#2+2)*180*(#1+0.5))+tan(180*#1)*sin((#2+2)*180*(#1+0.5))}}; 

\definecolor{col0}{HTML}{851E0C} 
\definecolor{col1}{HTML}{C4B408} 
\definecolor{col2}{HTML}{097D2A} 
\definecolor{col3}{HTML}{112CB8} 
\definecolor{col4}{HTML}{541163} 
\definecolor{colb}{HTML}{A12803} 


\fill[gray!50] (-1,-0.5)--(7.58755,7.90416)--(8.29733,7.66756)--(0,-1)--cycle;
\fill[gray!50] (4.05,0.05)--(9.97807,7.10732)--(10.67163,6.87613)--(5.05,-0.45)--cycle;
\draw[very thick,colb!80!black] (-1,-0.5)--(0,-1); 
\draw[very thick,colb!80!black] (4.05,0.05)--(5.05,-0.45); 

\draw (9.1,7.4) pic[scale=2] {complexnode}; 

\foreach \x/\y/\n in {0/0/0,2.5/2.5/1,4.5/4.5/2,6.1/6.1/3,7.38/7.38/4}
{\begin{scope}[shift={(\x,\y)},scale={(0.9)^(\n)},even odd rule]
\fill[white] (0,0) circle (0.5); 
\shade[ball color=red,fill opacity=0.1] (0,0) circle (0.5); 
\draw[very thick,smooth,col\n,domain=-0.5:0.5,samples=200] plot(\x,{psi1(\x,\n)});
\shade[ball color=red] (0,0) circle (0.55) (0,0) circle (0.5); 
\end{scope}}

\foreach \x/\y/\n in {5/0.5/1,7/3/2,8.6/5/3,9.88/6.6/4} 
{\begin{scope}[shift={(\x,\y)},scale={(0.9)^(\n)},even odd rule]
\fill[white] (0,0) circle (0.5); 
\shade[ball color=blue,fill opacity=0.1] (0,0) circle (0.5); 
\draw[very thick,smooth,col\n,domain=-0.499:0.499,samples=200] plot(\x,{psi2(\x,\n)});
\shade[ball color=blue] (0,0) circle (0.55) (0,0) circle (0.5); 
\end{scope}}

\begin{scope}[shift={(2.5,2.5)},scale={sqrt((5-2.5)^2+(0.5-2.5)^2)},rotate={atan((0.5-2.5)/(5-2.5))}]
    \draw[-latex,very thick,gray!50!black] (0.26,0.1) to[bend left=20] (0.76,0.1);
    \draw[-latex,very thick,gray!50!black] (0.76,-0.1) to[bend left=20] (0.26,-0.1);
\end{scope}

\begin{scope}[shift={(4.5,4.5)},scale={sqrt((7-4.5)^2+(3-4.5)^2)},rotate={atan((3-4.5)/(7-4.5))}]
    \draw[-latex,very thick,gray!50!black] (0.28,0.1) to[bend left=20] (0.77,0.1);
    \draw[-latex,very thick,gray!50!black] (0.77,-0.1) to[bend left=20] (0.26,-0.1);
\end{scope}

\begin{scope}[shift={(6.1,6.1)},scale={sqrt((8.6-6.1)^2+(5-6.1)^2)},rotate={atan((5-6.1)/(8.6-6.1))}]
    \draw[-latex,very thick,gray!50!black] (0.31,0.1) to[bend left=20] (0.77,0.1);
    \draw[-latex,very thick,gray!50!black] (0.76,-0.1) to[bend left=20] (0.27,-0.1);
\end{scope}

\begin{scope}[shift={(7.38,7.38)},scale={sqrt((9.88-7.38)^2+(6.6-7.38)^2)},rotate={atan((6.6-7.38)/(9.88-7.38))}]
    \draw[-latex,very thick,gray!50!black] (0.33,0.1) to[bend left=20] (0.79,0.1);
    \draw[-latex,very thick,gray!50!black] (0.76,-0.1) to[bend left=20] (0.28,-0.1);
\end{scope}
\end{tikzpicture} 
\end{center}

\vspace*{15mm}
\maketitle

\end{titlepage}
\thispagestyle{empty} 
\addtocounter{page}{-1} 

\begin{tcolorbox}[colback=contcol!10,colframe=contcol!40,title=Contents, fonttitle=\huge\sffamily\bfseries\selectfont,coltitle=black,top=2mm,bottom=2mm,left=2mm,right=2mm,drop fuzzy shadow,enhanced,breakable]
\hypersetup{linkcolor=conttxtcol}
\makeatletter
\sffamily\selectfont\@starttoc{toc}
\makeatother
\end{tcolorbox}

\vspace*{10mm}

\begin{tcolorbox}[colback=convcol!10,colframe=convcol!40,title=Conventions,fonttitle=\large\sffamily\bfseries\selectfont,coltitle=black,top=2mm,bottom=2mm,left=2mm,right=2mm,drop fuzzy shadow,enhanced]
\begin{itemize}[leftmargin=0.15in]
    \item $\N$ denotes the set $\{1,2,3,...\}$ of natural numbers (excluding $0$).
    \item Inner products are assumed to be linear in the first argument and conjugate linear in the second.
    \item $\Hs$ denotes a (complex) Hilbert space.
\end{itemize}
\end{tcolorbox}

\vspace*{20mm}

\begin{abstract}
    In this project, we will develop the foundations of quantum mechanics using the methods of supersymmetry. We will discuss the use of the superpotential to derive the supersymmetric partner of a potential in one dimension, and explore several key examples with an emphasis on shape invariant potentials. We will then discuss the modeling of supersymmetric quantum systems using matrices and operators, and how it relates to the eigenstate thermalization hypothesis.
\end{abstract}

\vfill

\begin{tcolorbox}[enhanced,center,frame hidden,colback=white,fontupper=\sffamily,
    borderline north={0.8pt}{0pt}{Green!40!black},borderline south={0.8pt}{0pt}{Green!40!black},
    top=0pt,bottom=0pt,left=0pt,right=0pt]
The cover image was created by the author using {\rmfamily Ti\textit{k}Z} and {\rmfamily\scshape PGF}. It depicts two sets of quantum states, on the left are the eigenstates of the particle in a box and on the right are those of its supersymmetric partner. The arrows depict how the supersymmetric operators realize the transition from one state to another.
\end{tcolorbox}

\section{Introduction}
Quantum mechanics (QM) is one of the cornerstones of modern physics. It was first developed in the early 20\textsuperscript{th} century by \href{https://mathshistory.st-andrews.ac.uk/Biographies/Schrodinger/}{Erwin Schrödinger} (1887--1961), \href{https://mathshistory.st-andrews.ac.uk/Biographies/Heisenberg/}{Werner Heisenberg} (1901--1976), \href{https://mathshistory.st-andrews.ac.uk/Biographies/Pauli/}{Wolfgang Pauli} (1900--1958), \href{https://mathshistory.st-andrews.ac.uk/Biographies/Broglie/}{Louis de Broglie} (1892--1987) and many others. It was developed mainly to explain the results of various experiments of the late 19\textsuperscript{th} and early 20\textsuperscript{th} centuries, such as blackbody radiation in 1859 by \href{https://mathshistory.st-andrews.ac.uk/Biographies/Kirchhoff/}{Gustav Robert Kirchhoff} (1824--1887), the photoelectric effect in 1887 by \href{https://mathshistory.st-andrews.ac.uk/Biographies/Hertz_Heinrich/}{Heinrich Hertz} (1857--1894), and the gold foil experiment in 1908 by \href{https://nzhistory.govt.nz/people/ernest-rutherford}{Ernest Rutherford}\footnote{Rutherford was a student of \href{https://www.sciencehistory.org/historical-profile/joseph-john-j-j-thomson}{J. J. Thomson} (1856--1940), who discovered the electron in 1897. Subsequently, Rutherford discovered the proton in 1919, and his student, \href{https://www.atomicheritage.org/profile/james-chadwick}{James Chadwick} (1891--1974) discovered the neutron in 1932.} (1871--1937).\\

From these experiments through to the present day, quantum mechanics has been observed repeatedly and consistently. While it has many interpretations that physicists (as well as mathematicians and philosophers) disagree on, it is usually introduced in a very standard way, through the \emph{Schrödinger equation}:
\begin{equation*}
    i\hbar\pdv{\Psi(\xb,t)}{t}=\widehat{H}\Psi(\xb,t)
\end{equation*}
This approach has since been dubbed `standard QM', to distinguish it from the dozens of other formulations of QM that have since emerged.\\

Supersymmetry (SUSY) is a quantum field theory concerning the relations between elementary particles of modern physics. It was first formulated in 1971 by \href{https://inspirehep.net/authors/992334}{Pierre Ramond} (1943--), \href{https://inspirehep.net/authors/989584}{John Henry Schwarz} (1941--) and \href{https://inspirehep.net/authors/995808}{André Neveu} (1946--) \cite{neveu,ramond}, building on ideas by \href{https://inspirehep.net/authors/997187}{Hironari Miyazawa} (1927--), \href{https://inspirehep.net/authors/1008543}{Jean-Loup Gervais} (1936--), \href{https://inspirehep.net/authors/990608}{Bunji Sakita} (1930--2002), \href{https://www.worldscientific.com/worldscibooks/10.1142/9281}{Yuri A. Golfand}\footnote{Not to be confused with \href{https://mathshistory.st-andrews.ac.uk/Biographies/Gelfand/}{Israel M. Gelfand} (1913--2009), known for the Gelfand representation and the Gelfand-Naimark theorem, or \href{https://mathshistory.st-andrews.ac.uk/Biographies/Gelfond/}{Alexander O. Gelfond} (1906--1968), known for the Gelfond-Schneider theorem.} (1922--1994) and \href{https://inspirehep.net/authors/1045725}{Evgeny P. Likhtman} (1946--) \cite{gervais,golfand,miyazawa}. It was (and largely still is) favored over its predecessor, string theory, mainly for its mathematical elegance and far-reaching predictions, such as dark matter \cite{jungman} and the mass of the Higgs boson \cite{higgsboson}.\\

As nice as it is, supersymmetry has never been observed in nature. Several experiments, such as the \href{https://atlas.cern/}{ATLAS} experiment at \href{https://home.cern/}{CERN}, have tried to find evidence of supersymmetry, with no success\footnote{As of 2022, several models of supersymmetry have been disproved, but the general theory remains open.}. Some physicists see this as a sign that SUSY may be a stepping stone to a grand unified theory of particle physics. Others simply make the excuse that we just need to build larger particle colliders.\\

Since the 1970s, supersymmetry has been developed as a theoretical concept, invoking many areas of mathematics, such as Lie algebras and topological manifolds. Despite the lack of experimental evidence for supersymmetry, it is still promising and respected by theoretical and experimental physicists alike \cite{failedprediction}.\\

In the standard model of particle physics, upon which most of modern physics is now based, there are two main types of elementary particles: \textit{fermions} and \textit{bosons}.

\vspace{5mm}

\begin{tcolorbox}[enhanced,frame hidden,sidebyside,colback=white,colframe=red!40!black,
fontupper=\sffamily,fontlower=\sffamily,
top=0mm,bottom=-3mm,left=0mm,right=0mm, 
]
\begin{center}
{\fontsize{16}{20}\sffamily\bfseries\selectfont Fermions}
\begin{itemize}[leftmargin=1.1in]
    \item Comprise \emph{matter}
    \item Have \emph{half-integer} spin
    \item Obey \emph{Fermi-Dirac} statistics
\end{itemize}
\begin{multicols}{2}
\personbox{white}{
\href{https://mathshistory.st-andrews.ac.uk/Biographies/Fermi/}
    {Enrico Fermi}\\
    (1901--1954)}
    {\includegraphics[width=1.3in]{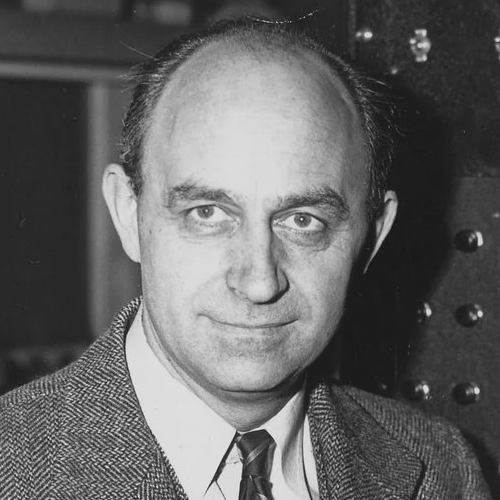}}\label{photo:fermi} 

\personbox{white}{
\href{https://mathshistory.st-andrews.ac.uk/Biographies/Dirac/}
    {Paul Dirac}\\
    (1902--1984)}
    {\includegraphics[width=1.3in]{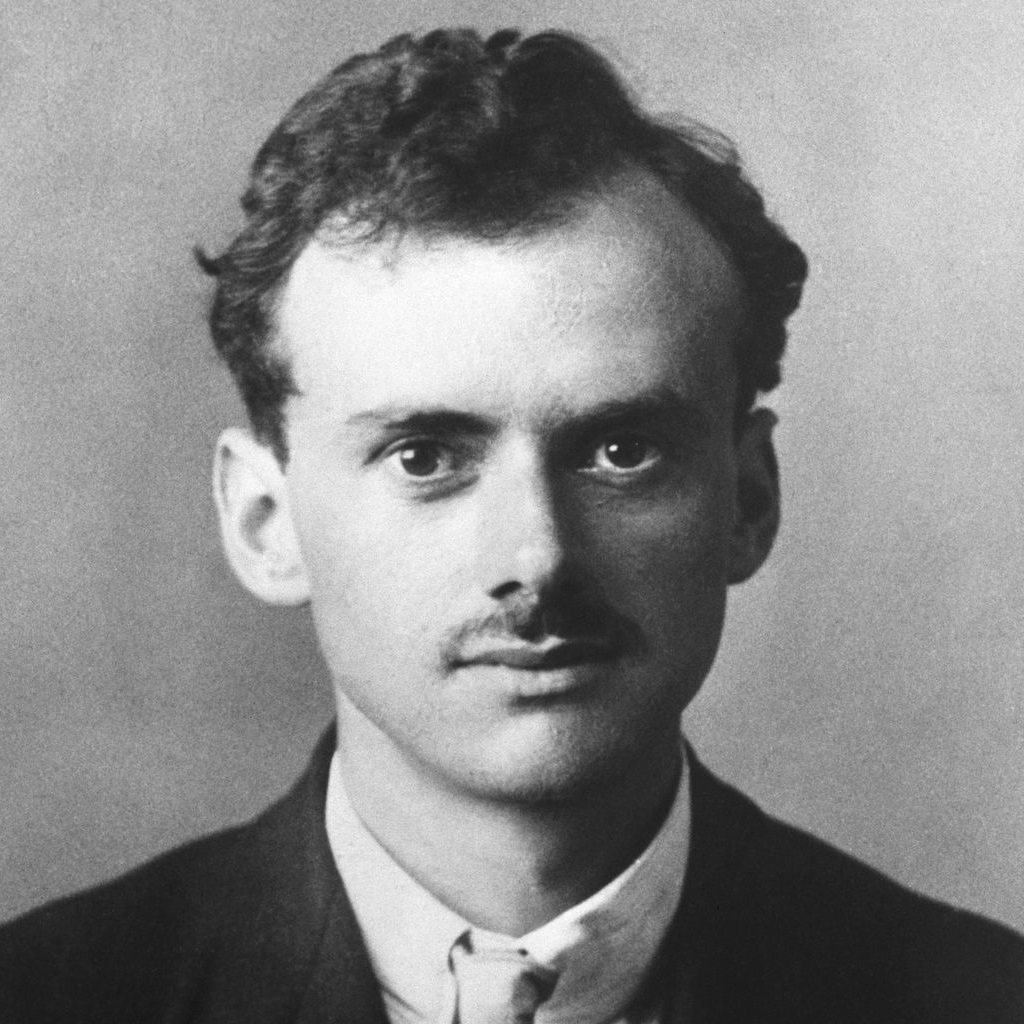}}\label{photo:dirac} 
\end{multicols}
\end{center}
\tcblower 
\begin{center}
{\fontsize{16}{20}\sffamily\bfseries\selectfont Bosons}
\begin{itemize}[leftmargin=1.2in]
    \item Comprise \emph{force}
    \item Have \emph{integer} spin
    \item Obey \emph{Bose-Einstein} statistics
\end{itemize}
\begin{multicols}{2}
\personbox{white}{
\href{https://mathshistory.st-andrews.ac.uk/Biographies/Bose/}
    {Satyendra Nath Bose}\\
    (1894--1974)}
    {\includegraphics[width=1.3in]{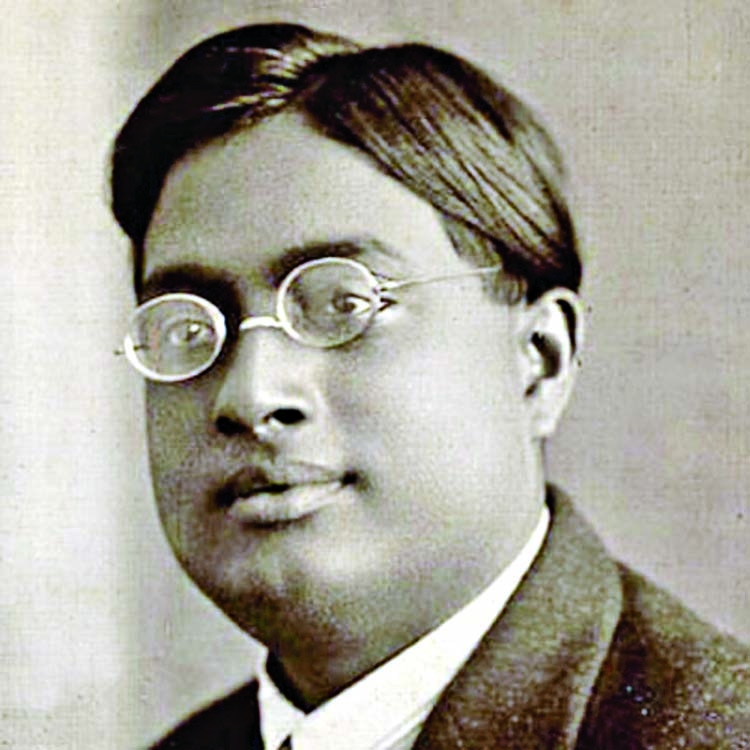}}\label{photo:bose} 

\personbox{white}{
\href{https://mathshistory.st-andrews.ac.uk/Biographies/Einstein/}
    {Albert Einstein}\\
    (1879--1955)}
    {\includegraphics[width=1.3in]{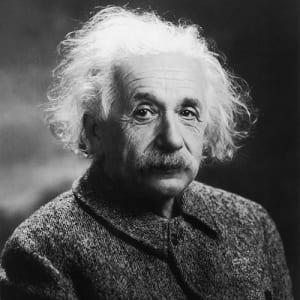}}\label{photo:einstein} 
\end{multicols}
\end{center}
\end{tcolorbox}

\newpage

Supersymmetry posits that every fermion has a partner boson, and every boson has a partner fermion. Of course, this is not possible within the standard model, as there is an imbalance of particles on either side. What supersymmetry suggests instead is that there is \emph{another copy} of the standard model containing the partners of the particles currently within the standard model. These, taken together, are known as the \textbf{Minimal Supersymmetric Standard Model} (MSSM) \cite{baer,terning}.

\begin{center}
\begin{tikzpicture}[font=\large\sffamily\selectfont]
\definecolor{quark}{HTML}{F28FDB}
\definecolor{lepton}{HTML}{72E0CA}
\definecolor{gaugeboson}{HTML}{E88B51}
\definecolor{higgs}{HTML}{94E058}
\definecolor{squark}{HTML}{D7DB56}
\definecolor{slepton}{HTML}{E66AB6}
\definecolor{gaugino}{HTML}{37D9E2}
\definecolor{higgsino}{HTML}{E66372}

\begin{scope}[shift={(0,0)},every node/.style={transform shape}] 
\shade[ball color=quark] (0,2.4) circle (0.4) node{u}; 
\shade[ball color=quark] (0.8,2.4) circle (0.4) node{c}; 
\shade[ball color=quark] (1.6,2.4) circle (0.4) node{t}; 
\shade[ball color=quark] (0,1.6) circle (0.4) node[yshift=1pt]{d}; 
\shade[ball color=quark] (0.8,1.6) circle (0.4) node{s}; 
\shade[ball color=quark] (1.6,1.6) circle (0.4) node[yshift=1pt]{b}; 
\shade[ball color=lepton] (0,0.8) circle (0.4) node{e}; 
\shade[ball color=lepton] (0.8,0.8) circle (0.4) node[yshift=-1pt]{$\mu$}; 
\shade[ball color=lepton] (1.6,0.8) circle (0.4) node{$\tau$}; 
\shade[ball color=lepton] (0,0) circle (0.4) node{$\nu_\text{e}$}; 
\shade[ball color=lepton] (0.8,0) circle (0.4) node[yshift=-1pt]{$\nu_\mu$}; 
\shade[ball color=lepton] (1.6,0) circle (0.4) node{$\nu_\tau$}; 
\shade[ball color=gaugeboson] (2.8,2.4) circle (0.4) node{g}; 
\shade[ball color=gaugeboson] (2.8,1.6) circle (0.4) node{$\gamma$}; 
\shade[ball color=gaugeboson] (2.8,0.8) circle (0.4) node{Z}; 
\shade[ball color=gaugeboson] (2.8,0) circle (0.4) node{W}; 
\shade[ball color=higgs] (3.6,2.4) circle (0.4) node{H}; 

\draw[dashed] (2.2,-0.4)--(2.2,3.3)
    node[left,yshift=-2mm,font=\small\sffamily\itshape] {Fermions}
    node[right,yshift=-2mm,font=\small\sffamily\itshape] {Bosons}; 
\draw[decorate,decoration={brace,mirror,amplitude=6pt}] (-0.4,2.8)--(-0.4,1.2)
    node[midway,left=5pt,font=\small\sffamily] {Quarks}; 
\draw[decorate,decoration={brace,mirror,amplitude=6pt}] (-0.4,1.2)--(-0.4,-0.4)
    node[midway,left=5pt,font=\small\sffamily] {Leptons}; 
\draw[decorate,decoration={brace,mirror,amplitude=6pt}] (2.4,-0.4)--(3.2,-0.4)
    node[midway,below=5pt,font=\small\sffamily] {Gauge bosons}; 
\node[align=center,font=\small\sffamily] at (4.2,1.7) {Higgs\\boson}; 
\end{scope}

\begin{scope}[shift={(8,0)},every node/.style={transform shape}] 
\shade[ball color=squark] (0,2.4) circle (0.4) node[yshift=1pt]{$\widetilde{\text{u}}$}; 
\shade[ball color=squark] (0.8,2.4) circle (0.4) node[yshift=1pt]{$\widetilde{\text{c}}$}; 
\shade[ball color=squark] (1.6,2.4) circle (0.4) node[yshift=1pt]{$\widetilde{\text{t}}$}; 
\shade[ball color=squark] (0,1.6) circle (0.4) node[yshift=2pt]{$\widetilde{\text{d}}$}; 
\shade[ball color=squark] (0.8,1.6) circle (0.4) node[yshift=1pt]{$\widetilde{\text{s}}$}; 
\shade[ball color=squark] (1.6,1.6) circle (0.4) node[yshift=2pt]{$\widetilde{\text{b}}$}; 
\shade[ball color=slepton] (0,0.8) circle (0.4) node[yshift=1pt]{$\widetilde{\text{e}}$}; 
\shade[ball color=slepton] (0.8,0.8) circle (0.4) node{$\widetilde{\mu}$}; 
\shade[ball color=slepton] (1.6,0.8) circle (0.4) node[yshift=1pt]{$\widetilde{\tau}$}; 
\shade[ball color=slepton] (0,0) circle (0.4) node[yshift=1pt]{$\widetilde{\nu_\text{e}}$}; 
\shade[ball color=slepton] (0.8,0) circle (0.4) node{$\widetilde{\nu_\mu}$}; 
\shade[ball color=slepton] (1.6,0) circle (0.4) node[yshift=1pt]{$\widetilde{\nu_\tau}$}; 
\shade[ball color=gaugino] (2.8,2.4) circle (0.4) node[yshift=1pt]{$\widetilde{\text{g}}$}; 
\shade[ball color=gaugino] (2.8,1.6) circle (0.4) node[yshift=1pt]{$\widetilde{\gamma}$}; 
\shade[ball color=gaugino] (2.8,0.8) circle (0.4) node[yshift=1pt]{$\widetilde{\text{Z}}$}; 
\shade[ball color=gaugino] (2.8,0) circle (0.4) node[yshift=1pt]{$\widetilde{\text{W}}$}; 
\shade[ball color=higgsino] (3.6,2.4) circle (0.4) node[yshift=1pt]{$\widetilde{\text{H}}$}; 

\draw[dashed] (2.2,-0.4)--(2.2,3.3)
    node[left,yshift=-2mm,font=\small\sffamily\itshape] {Sfermions}
    node[right,yshift=-2mm,font=\small\sffamily\itshape] {Bosinos}; 
\draw[decorate,decoration={brace,mirror,amplitude=6pt}] (-0.4,2.8)--(-0.4,1.2)
    node[midway,left=5pt,font=\small\sffamily] {Squarks}; 
\draw[decorate,decoration={brace,mirror,amplitude=6pt}] (-0.4,1.2)--(-0.4,-0.4)
    node[midway,left=5pt,font=\small\sffamily] {Sleptons}; 
\draw[decorate,decoration={brace,mirror,amplitude=6pt}] (2.4,-0.4)--(3.2,-0.4)
    node[midway,below=5pt,font=\small\sffamily] {Gauginos}; 
\node[font=\small\sffamily] at (4.05,1.75) {Higgsino}; 
\end{scope}

\draw[thick,rounded corners=2mm] (-2,-1.1) rectangle (5,3.4); 
\node at (3.2,-1.4) {\bfseries Standard Model}; 
\draw[thick,rounded corners=2mm] (-2.2,-1.7) rectangle (13,3.6); 
\node at (12.2,-1.4) {\bfseries MSSM}; 
\end{tikzpicture}
\end{center}

Note that sfermions are \emph{bosons} (as the partners of ordinary fermions), while bosinos are \emph{fermions} (as the partners of ordinary bosons). Yes, physicists are not very creative when it comes to naming things\footnote{Take Dirac, for example, who simply split up the word `bracket' to get `bra' and `ket'.}.\\

So what is \textit{\textsf{supersymmetric quantum mechanics}}? Supersymmetric quantum mechanics (SUSY QM) is an alternative formulation of quantum mechanics using the ideas of supersymmetry. Motivated by the notion of supersymmetric particles, we surmise that every quantum mechanical particle is related to another particle via its Hamiltonian. More specifically, given a Hamiltonian that describes one particle, we aim to find a ``partner Hamiltonian'' that describes another particle and is related to the original Hamiltonian in a supersymmetric manner. In some cases, the Schrödinger equation for the new Hamiltonian may be easier to solve than for the old one. In other cases, there may be symmetries between the two Hamiltonians that we can exploit to solve both of them simultaneously. All in all, SUSY QM will allow us to solve some quantum systems that we could not solve with standard QM, or even if we could, we will be able to do so much more cleanly and efficiently with SUSY QM.\\

In \Cref{sec:systems}, we will use these methods to solve some important examples of one-dimensional quantum systems, such as the particle in a box, the harmonic oscillator and the hydrogen atom. In \Cref{sec:matricesop}, we will take the ideas of SUSY QM we have developed and carry them over to the setting of matrices and operators. In a sense, this chapter will be less physical and more mathematical. We will apply these techniques to random matrices and show how they lead to the foundations of the eigenstate thermalization hypothesis. We will also take a peek into the theory of self-adjoint operators on a Hilbert space, and discuss the use of boundary conditions and their effect on the validity of the supersymmetric treatment of quantum systems.\\

So now that we have laid out this all-encompassing powerful tool that is SUSY QM, let's find the light at the end of the (quantum) tunnel...


\vfill

\subsection*{Acknowledgments}
I would like to thank my project supervisor, Prof. Christopher Herzog, for his invaluable feedback and comments on my project. I would also like to thank Prof. Benjamin Doyon, Prof. Gerard Watts and Dr. Paul Cook for their input in office discussions, some of which have inspired various parts of this project. Finally, I would like to thank my peers for their support throughout my years at King's College London.

\vspace{6mm}

\begin{flushright} 
Senan Sekhon\\
March 24, 2022
\end{flushright}

\vspace{8mm}

\section{The Schrödinger Equation}\label[chapter]{sec:schrodinger}
The starting point of supersymmetric quantum mechanics is the same as that of standard quantum mechanics: The Schrödinger equation. First formulated in 1925 by \href{https://mathshistory.st-andrews.ac.uk/Biographies/Schrodinger/}{Erwin Schrödinger} (1887--1961), it has become the poster child of quantum mechanics, appearing not just in physicists' blackboards, but even on T-shirts\footnote{Or keychains, or coffee mugs, or any consumer item you wish.}.\\

\noindent {\sffamily\bfseries Warm-up example.} Consider the following second-order ODE:
\begin{equation}\label{warmupode}
    \dv[2]{y}{x}-\alpha^2 y=0
\end{equation}
Where $\alpha>0$ is a constant. In a first course on differential equations, one would learn to solve this by first solving the \emph{characteristic equation} (or \emph{auxiliary equation}) $\lam^2-\alpha^2=0$. Using its solutions $\lam=\pm\alpha$, one would then write the solution to \eqref{warmupode} as $y=c_1e^{\alpha x}+c_2e^{-\alpha x}$.\\

\noindent Another way to solve this ODE would be to factor the operator acting on $y$:
\begin{align*}
    \dv[2]{y}{x}-\alpha^2 y=0 \qquad\longrightarrow\qquad \left(\dv[2]{x}-\alpha^2\right)y=0 \qquad\longrightarrow\qquad \left(\dv{x}+\alpha\right)\left(\dv{x}-\alpha\right)y=0
\end{align*}
We can then reduce our problem to solving the first-order ODEs $\left(\dv{x}+\alpha\right)y=0$ and $\left(\dv{x}-\alpha\right)y=0$ individually, then combining the solutions of each to get the general solution.

\vspace*{12mm}

We will now attempt to apply this technique to solve the Schrödinger equation. We start with the (time-independent) Schrödinger equation:
\begin{equation*}
    H\psi=E\psi
\end{equation*}
For a particle moving in one space dimension under a potential $V(x)$, the Hamiltonian is given by:
\begin{equation*}
    H=-\frac{\hbar^2}{2m}\dv[2]{x}+V(x)
\end{equation*}
With this, the Schrödinger equation can be rewritten as:
\begin{equation}\label{sch:1}
    H\psi(x)=-\frac{\hbar^2}{2m}\psi''(x)+V(x)\psi(x)=E\psi(x)
\end{equation}
In a first course on quantum mechanics, one learns to solve \eqref{sch:1} under certain sets of boundary conditions to find the allowed values of $E$, known as the \emph{energy eigenvalues} (the eigenvalues of $H$) and the corresponding functions $\psi$, known as the \emph{energy eigenstates} (the eigenfunctions of $H$). Since the Hamiltonian is always self-adjoint, all of its eigenvalues are real. There is also a smallest allowed value of $E$, namely the energy $E_0$ of the ground state $\psi_0$.\\

We now define a shifted Hamiltonian $H^{(1)}$ and a shifted potential $V^{(1)}$:
\begin{align*}
    H^{(1)}=H-E_0 && V^{(1)}=V-E_0
\end{align*}
This shift sets the ground state energy to zero. We now focus on solving the `shifted' Schrödinger equation:
\begin{equation}\label{sch:2}
    H^{(1)}=-\frac{\hbar^2}{2m}\dv[2]{x}+V^{(1)}(x)
\end{equation}
At first glance, this does not seem useful at all. Why do we want the ground state energy to be zero? One reason is that we will want to factor it into $H^{(1)}=A\dg A$, where $A$ is an operator to be determined. This factorization is only possible for \emph{positive} self-adjoint operators (simply because $\ip{A\dg A\psi,\psi}=\ip{A\psi,A\psi}=\norm{A\psi}^2\ge0$ is always non-negative). If the ground state energy $E_0$ were negative, we would not be able to perform this factorization without raising the energy levels first. Why we want the ground state energy to be \emph{zero} (rather than a positive number) will be addressed in \Cref{subsec:susyhamiltonian}.\\

By the form of \eqref{sch:1}, we make the following ansatz\footnote{Technically, we need to check that $A\dg$ really is the adjoint of $A$. However, this is apparent after rewriting them as $A=W(x)+\frac{i}{\sqrt{2m}}\hat{p}$ and $A\dg=W(x)-\frac{i}{\sqrt{2m}}\hat{p}$, where $\hat{p}=-i\hbar\dv{x}$ is the momentum operator (which is self-adjoint under appropriate boundary conditions).}:
\begin{align}\label{superpotansatz}
    A=\frac{\hbar}{\sqrt{2m}}\dv{x}+W(x) &&
    A\dg=-\frac{\hbar}{\sqrt{2m}}\dv{x}+W(x)
\end{align}
Where $W(x)$ is a function known as the \emph{superpotential}. To solve for $W(x)$, we compute $A\dg A$:
\begin{align*}
    A\dg A\psi(x)
    &=\left(-\frac{\hbar}{\sqrt{2m}}\dv{x}+W(x)\right)\left(\frac{\hbar}{\sqrt{2m}}\dv{x}+W(x)\right)\psi(x) \\
    &=\left(-\frac{\hbar}{\sqrt{2m}}\dv{x}+W(x)\right)\left(\frac{\hbar}{\sqrt{2m}}\psi'(x)+W(x)\psi(x)\right) \\
    &=-\frac{\hbar}{\sqrt{2m}}\dv{x}\left(\frac{\hbar}{\sqrt{2m}}\psi'(x)+W(x)\psi(x)\right)+W(x)\left(\frac{\hbar}{\sqrt{2m}}\psi'(x)+W(x)\psi(x)\right) \\
    &=-\frac{\hbar^2}{2m}\psi''(x)-\frac{\hbar}{\sqrt{2m}}W'(x)\psi(x)-\frac{\hbar}{\sqrt{2m}}W(x)\psi'(x)+\frac{\hbar}{\sqrt{2m}}W(x)\psi'(x)+W(x)^2\psi(x) \\
    &=\left(-\frac{\hbar^2}{2m}\dv[2]{x}-\frac{\hbar}{\sqrt{2m}}W'(x)+W(x)^2\right)\psi(x)
\end{align*}
Comparing this with \eqref{sch:2}, we get the following ODE for $W(x)$:
\begin{equation}\label{ricsuperpot}
    -\frac{\hbar}{\sqrt{2m}}W'(x)+W(x)^2=V^{(1)}(x)
\end{equation}
This is known as the time-independent \textbf{quantum Hamilton-Jacobi equation}. This is because the superpotential $W(x)$ arises implicitly in the WKB approximation for solving the Schrödinger equation, which we will explore in \Cref{subsec:swkb}. See \cite{girard} for a discussion on how this relates to the classical Hamilton-Jacobi equation.\\

Equation \eqref{ricsuperpot} is an example of a \emph{Riccati equation}, a first-order ODE of the form:
\begin{equation*}
    g(x)y'=p(x)y^2+q(x)y+r(x)
\end{equation*}
In general, it does not have an exact (analytical) solution, though solutions of special cases can be found in \cite{polyanin}.\\

For the ground state $\psi_0$ to have zero energy, we must have $H^{(1)}\psi_0=A\dg A\psi_0=0$. In other words:
\begin{equation*}
    0=\ip{A\dg A\psi_0,\psi_0}=\ip{A\psi_0,A\psi_0}=\norm{A\psi_0}^2 \quad\therefore\quad A\psi_0=0
\end{equation*}
This yields a separable (and linear) first-order ODE for $\psi_0$:
\begin{equation*}
    A\psi_0(x)=-\frac{\hbar}{\sqrt{2m}}\psi_0'(x)+W(x)\psi_0(x)=0
\end{equation*}
We can rewrite this as follows:
\begin{equation}\label{wpsi0}
    \frac{\psi_0'(x)}{\psi_0(x)}=-\frac{\sqrt{2m}}{\hbar}W(x)
\end{equation}
Solving it, we get:
\begin{equation}\label{psi0w}
    \psi_0(x)=C\exp\left(-\frac{\sqrt{2m}}{\hbar}\int_a^x W(y)\dy\right)
\end{equation}
Where $C$ is a normalization constant and $a$ is any point in the domain (usually the left endpoint, which may be $-\infty$). This gives us a solution for the ground state $\psi_0$ in terms of the superpotential $W(x)$.\\

If we apply the operators $A$ and $A\dg$ (separately) to $\psi_0$, we get:
\begin{align}\label{apsi0}
    A\psi_0(x)
    &=\frac{\hbar}{\sqrt{2m}}\psi_0'(x)+W(x)\psi_0(x)
    =\frac{\hbar}{\sqrt{2m}}\left(-\frac{\sqrt{2m}}{\hbar}W(x)\right)\psi_0(x)+W(x)\psi_0(x)
    =0 \\
    \label{adgpsi0} 
    A\dg\psi_0(x)
    &=-\frac{\hbar}{\sqrt{2m}}\psi_0'(x)+W(x)\psi_0(x)
    =-\frac{\hbar}{\sqrt{2m}}\left(-\frac{\sqrt{2m}}{\hbar}W(x)\right)\psi_0(x)+W(x)\psi_0(x)
    =2W(x)\psi_0(x)
\end{align}
Importantly, the operator $A$ annihilates $\psi_0$, while its adjoint $A\dg$ multiplies it by $2W(x)$. This is reminiscent of the creation and annihilation operators of the harmonic oscillator. We will see later that, in the case of the harmonic oscillator, these coincide with the ladder operators (up to a constant).\\

If we compute $AA\dg$, we get:
\begin{equation}\label{AAdg}
    AA\dg=-\frac{\hbar^2}{2m}\dv[2]{x}+\frac{\hbar}{\sqrt{2m}}W'(x)+W(x)^2
\end{equation}
This allows us to define a new Hamiltonian $H^{(2)}=AA\dg$.

\begin{definition}\label{susypartner}
The \textbf{supersymmetric partner} of $H^{(1)}$ is given by:
\begin{align*}
    H^{(2)}=-\frac{\hbar^2}{2m}\dv[2]{x}+V^{(2)}(x) && V^{(2)}(x)=\frac{\hbar}{\sqrt{2m}}W'(x)+W(x)^2
\end{align*}
The potential $V^{(2)}$ is known as the \textbf{partner potential}.
\end{definition}

In some cases, the Schrödinger equation for $H^{(2)}$ may be easier to solve than the Schrödinger equation for $H^{(1)}$. Note that this depends on the choice of the superpotential $W$, and not just the potential $V^{(1)}$. In other words, different operators $A$ can give the same Hamiltonian $H^{(1)}$, while also giving different partner Hamiltonians $H^{(2)}$.

\bubble[24]{Orchid}{From now on, we will denote the constant \raisebox{1pt}{$\frac{\hbar}{\sqrt{2m}}$} by $\eta$. We will often also set $\eta=1$ for convenience.} 

\begin{lemma}\label{riccatigs}
Suppose $W_0(x)$ is a solution of \eqref{ricsuperpot}. Then the general solution of \eqref{ricsuperpot} is given by:
\begin{equation*}
    W(x)=W_0(x)-\eta e^{\frac{2}{\eta}\int W_0(x)\dx}\left(\int e^{\frac{2}{\eta}\int W_0(x)\dx}\dx+C\right)^{-1}
\end{equation*}
The particular solution $W_0$ occurs when $C=\infty$.
\end{lemma}
This is a special case of the method for Riccati equations given in \cite[\S1.4.2, Page 15]{polyanin}.\\

We will now look at some simple examples of the possible behavior arising from the non-uniqueness of $W$. Here, we are no longer assuming the ground state energy is zero.

\begin{constpotentialbox}{Zero potential}
Suppose $V^{(1)}(x)\equiv0$ ($V^{(1)}$ is identically zero). Then $W_0(x)\equiv0$ is clearly a solution of \eqref{ricsuperpot}. By the previous lemma, the general solution is given by:
\begin{equation*}
    W(x)=-\frac{\eta}{x+C}
\end{equation*}
And so the partner potential is given by:
\begin{equation*}
    V^{(2)}(x)=\frac{2\eta^2}{(x+C)^2}
\end{equation*}
This is different for different values of $C$. Also, note that $V^{(2)}$ has a singularity at $x=-C$, even though $V^{(1)}$ does not. This shows that the process of constructing a partner potential can create singularities. It also shows that the zero potential (on the whole real line) does not have a supersymmetric partner, so we would need to restrict its domain to get a partner potential with normalizable eigenstates.
\end{constpotentialbox}

\begin{constpotentialbox}{Positive constant potential}
Suppose $V^{(1)}(x)\equiv\eta^2$ (any positive constant will do). Then $W_0(x)\equiv\eta$ is a solution of \eqref{ricsuperpot}. By the previous lemma, the general solution is given by:
\begin{equation*}
    W(x)=-\eta\tanh(x+C)
\end{equation*}
And so the partner potential is given by:
\begin{equation*}
    V^{(2)}(x)=\eta^2\left(1-2\sech(x+C)^2\right)
\end{equation*}
This has normalizable eigenstates for all finite values of $C$, as we will see later. Note that $C=\infty$ and $C=-\infty$ yield two distinct solutions for $W(x)$ ($-\eta$ and $\eta$ respectively), but these recombine to give the same partner potential $V^{(2)}(x)\equiv\eta^2$.
\end{constpotentialbox}

\begin{constpotentialbox}{Negative constant potential}
Suppose $V^{(1)}(x)\equiv-\eta^2$ (any negative constant will do). Then $W_0(x)\equiv\eta\tan(x)$ is a solution of \eqref{ricsuperpot}. By the previous lemma, the general solution is given by:
\begin{equation*}
    W(x)=\eta\tan(x+C)
\end{equation*}
And so the partner potential is given by:
\begin{equation*}
    V^{(2)}(x)=\eta^2\left(2\sec(x+C)^2-1\right)
\end{equation*}
This also has normalizable eigenstates for all finite values of $C$. In this example, the process of constructing the partner potential created infinitely many singularities. This time, the values $C=\pm\infty$ no longer make sense ($V^{(2)}(x)$ does not have a limit as $C\to\pm\infty$).
\end{constpotentialbox}

As the above examples show, we need to be careful in choosing $W(x)$ in order to have normalizable eigenstates (note that this also depends on the boundary conditions). If we look for eigenstates that are normalizable on the whole real line, then by \eqref{psi0w}, we need the following function to be square-intgrable:
\begin{equation*}
    \psi_0(x)=C\exp\left(-\frac{1}{\eta}\int_{-\infty}^x W(y)\dy\right)
\end{equation*}
Thus $\psi_0$ must vanish\footnote{There are square-integrable functions on $\R$ that do not vanish as $x\to\pm\infty$, but these would not normally appear as wavefunctions in QM (they fail to be uniformly continuous, and they do not remain square-integrable once time evolution is included).} as $x\to\pm\infty$, and so the argument of the exponential must tend to $-\infty$ as $x\to\pm\infty$. This will happen, for instance, if $W(y)$ is an odd increasing function.\\

We can relate the eigenstates of $H^{(2)}$ to those of $H^{(1)}$ using the following result:

\begin{theorem}\label{eigendual}
The eigenstates of $H^{(1)}$ and $H^{(2)}$ are related as follows:
\begin{itemize}
    \item If $\psi$ is an eigenstate of $H^{(1)}$ with energy $E\ne0$, then $A\psi$ is an eigenstate of $H^{(2)}$ with energy $E$.
    \item If $\psi$ is an eigenstate of $H^{(2)}$ with energy $E\ne0$, then $A\dg\psi$ is an eigenstate of $H^{(1)}$ with energy $E$.
\end{itemize}
\end{theorem}
\begin{proof}
Suppose $\psi$ is an eigenstate of $H^{(1)}$ with energy $E\ne0$, i.e. $H^{(1)}\psi=E\psi$. Since $E\ne0$, we have $E\psi=A\dg A\psi\ne0$, and so $A\psi\ne0$. We also have:
\begin{equation*}
    H^{(2)}(A\psi)=AA\dg(A\psi)=A(A\dg A)\psi=AH^{(1)}\psi=A(E\psi)=E(A\psi)
\end{equation*}
Thus $A\psi$ is an eigenstate of $H^{(2)}$ with energy $E$. The second statement follows similarly.
\end{proof}

In other words, given any eigenstate of either $H^{(1)}$ or $H^{(2)}$ (with nonzero energy), we can use the operators $A$ and $A\dg$ to transition to the corresponding eigenstate of the other Hamiltonian, i.e. the state with the same energy eigenvalue. Unfortunately, we cannot do this with the ground state $\psi_0$ of $H^{(1)}$, as $A\psi_0=0$ and so does not provide any new information.\\

Of course, we cannot possibly expect to gain more information just from $\psi_0$, as that would allow us to create new states out of thin air. What we can hope to do, however, is relate the spectrum of eigenstates of $H^{(2)}$ to those of $H^{(1)}$ (provided we know the entire spectrum of one of them). This is done in the next theorem:

\begin{theorem}\label{susyladder}
Suppose for each $n\in\N$, $\psi_n^{(1)}$ is the $n$\textsuperscript{th} excited state of $H^{(1)}$ with energy $E_n^{(1)}$, and $\psi_n^{(2)}$ is the $n$\textsuperscript{th} excited state of $H^{(2)}$ with energy $E_n^{(2)}$. Then for all $n\in\N$, we have:
\begin{align*}
    E_0^{(1)}=0
    &&
    E_n^{(2)}=E_{n+1}^{(1)}
    &&
    \psi_n^{(2)}=\frac{1}{\sqrt{E_{n+1}^{(1)}}}A\psi_{n+1}^{(1)}
    &&
    \psi_{n+1}^{(1)}=\frac{1}{\sqrt{E_n^{(2)}}}A\dg\psi_n^{(2)}
\end{align*}
\end{theorem}
\begin{proof}
By the definitions of $\psi_n^{(1)}$ and $\psi_n^{(2)}$, we have:
\begin{align}
    H^{(1)}\psi_n^{(1)}=E_n^{(1)}\psi_n^{(1)}
    &&
    H^{(2)}\psi_n^{(2)}=E_n^{(2)}\psi_n^{(2)}
\end{align}
By the previous theorem, we also have:
\begin{align}\label{partnerstate}
    H^{(2)}(A\psi_n^{(1)})=E_n^{(1)}A\psi_n^{(1)}
    &&
    H^{(1)}(A\dg\psi_n^{(2)})=E_n^{(2)}A\dg\psi_n^{(2)}
\end{align}
By assumption, we have $E_0^{(1)}=0$. By \eqref{apsi0}, we have $A\psi_0^{(1)}=0$. Thus the ground state of $H^{(2)}$ cannot be $A\psi_0^{(1)}$, it must be $A\psi_1^{(1)}$. In other words, $\psi_0^{(2)}=c_0A\psi_1^{(1)}$, where $c_0>0$ is a normalization constant. To find it, we simply compute the norm of $A\psi_1^{(1)}$:
\begin{equation*}
    \ip{A\psi_1^{(1)},A\psi_1^{(1)}}
    =\ip{A\dg A\psi_1^{(1)},\psi_1^{(1)}}
    =\ip{H^{(1)}\psi_1^{(1)},\psi_1^{(1)}}
    =\ip{E_1^{(1)}\psi_1^{(1)},\psi_1^{(1)}}
    =E_1^{(1)}
\end{equation*}
Thus $c_0=\frac{1}{\sqrt{E_1^{(1)}}}$.
Substituting this into \eqref{partnerstate}, we get:
\begin{equation*}
    A\dg\psi_0^{(2)}
    =\frac{1}{\sqrt{E_1^{(1)}}}A\dg A\psi_1^{(1)}
    =\frac{1}{\sqrt{E_1^{(1)}}}H^{(1)}\psi_1^{(1)}
    =\frac{1}{\sqrt{E_1^{(1)}}}E_1^{(1)}\psi_1^{(1)}
    =\sqrt{E_1^{(1)}}\psi_1^{(1)}
\end{equation*}
Thus $\psi_1^{(1)}=\frac{1}{\sqrt{E_1^{(1)}}}A\dg\psi_0^{(2)}$. Repeating this process for each excited state yields the result.
\end{proof}

\newpage 

The last theorem tells us that the two Hamiltonians form a supersymmetric `ladder', where every excited state of $H^{(1)}$ is matched with a state of $H^{(2)}$ and vice versa. Only the ground state of $H^{(1)}$ is unmatched.

\begin{center}
\begin{tikzpicture}
\foreach \a/\b in {-1.8/0,-1/0,1/0.8,1.8/0.8}
{\draw[thick] (\a,\b)--(\a,3.2);
\draw[-stealth,thick,dashed] (\a,3.2)--(\a,4);}
\draw (-1.8,1) node[left]{$E_1^{(1)}$}--(-1,1) (1,1)--(1.8,1) node[right]{$E_0^{(2)}=E_1^{(1)}$}; 
\draw (-1.8,1.6) node[left]{$E_2^{(1)}$}--(-1,1.6) (1,1.6)--(1.8,1.6) node[right]{$E_1^{(2)}=E_2^{(1)}$}; 
\draw (-1.8,2.1)--(-1,2.1) (1,2.1)--(1.8,2.1); 
\draw (-1.8,2.5)--(-1,2.5) (1,2.5)--(1.8,2.5); 
\draw (-1.8,2.8)--(-1,2.8) (1,2.8)--(1.8,2.8); 
\draw (-1.8,3)--(-1,3) (1,3)--(1.8,3); 
\draw (-1.8,0.2) node[left=7pt]{$E_0$}--(-1,0.2); 
\node[left] at (-1.8,3.8) {$H^{(1)}$};
\node[right] at (1.8,3.8) {$H^{(2)}$};
\draw[-latex] (-0.7,2.9) to[bend left=15] node[midway,above]{$A$} (0.7,2.9);
\draw[-latex] (0.7,2.6) to[bend left=15] node[midway,below]{$A\dg$} (-0.7,2.6);
\end{tikzpicture}\\
\sffamily The supersymmetric ladder
\end{center}

There is still one question we have not yet addressed. How do we know we can factor $H^{(1)}$ into $A\dg A$? The machinery behind this lies in the spectral theorem for self-adjoint operators (\Cref{spectraltheorem}). In \Cref{sec:spectral}, we will give a brief outline of the result and the theory around it.

\subsection{The Supersymmetric Hamiltonian}\label{subsec:susyhamiltonian}
Now that we have seen how to construct the supersymmetric partner of a Hamiltonian, we will investigate how supersymmetry naturally arises from this construction.

\begin{definition}
The \textbf{supersymmetric Hamiltonian} is given by:
\begin{equation}
    H=\mat{H^{(1)},0}{0,H^{(2)}}
\end{equation}
\end{definition}
Since $H^{(1)}$ and $H^{(2)}$ are positive self-adjoint operators, so is $H$. Moreover, its ground state energy is zero. We can also express $H$ as $H=\{Q,Q\dg\}=QQ\dg+Q\dg Q$, where:
\begin{align}
    Q=\mat{0,0}{A,0} && Q\dg=\mat{0,A\dg}{0,0}
\end{align}
Note that:
\begin{align*}
    [Q,H]
    &=QH-HQ
    =\mat{0,0}{A,0}\mat{A\dg A,0}{0,AA\dg}-\mat{A\dg A,0}{0,AA\dg}\mat{0,0}{A,0}
    =\mat{0,0}{AA\dg A,0}-\mat{0,0}{AA\dg A,0}
    =0 \\
    [Q\dg,H]
    &=Q\dg H-HQ\dg=(HQ-QH)\dg=-[Q,H]\dg=0
\end{align*}
In other words, both $Q$ and $Q\dg$ commute with the supersymmetric Hamiltonian. This shows that $Q$, $Q\dg$ and $H$ generate a \textbf{supersymmetry algebra}, given by:
\begin{align*}
    [Q,H]=0 && [Q\dg,H]=0 && \{Q,Q\dg\}=H
\end{align*}

\vspace*{6mm}

\noindent We also have:
\begin{equation*}
    Q^2=(Q\dg)^2=0
\end{equation*}
In other words, applying either $Q$ or $Q\dg$ twice will never result in a new state. This is reminiscent of the \emph{Pauli exclusion principle}: No two fermions can occupy the same quantum state at the same time. For this reason, we call the operators $Q$ and $Q\dg$ \textbf{fermionic}.\\

We can view the eigenstates $\psi^{(1)}$ and $\psi^{(2)}$ of $H^{(1)}$ and $H^{(2)}$ respectively as column vectors:
\begin{align*}
    \psi^{(1)}\rightarrow\cvc{\psi^{(1)},0} && \psi^{(2)}\rightarrow\cvc{0,\psi^{(2)}}
\end{align*}
Applying $Q$ and $Q\dg$ respectively to these states and using \Cref{susyladder}, we get:
\begin{align}\label{Qpsin1}
    Q\cvc{\psi_n^{(1)},0}
    &=\mat{0,0}{A,0}\cvc{\psi_n^{(1)},0}
    =\cvc{0,A\psi_n^{(1)}}
    =\frac{1}{\sqrt{E_{n-1}^{(2)}}}\cvc{0,A\psi_{n-1}^{(2)}}
    \\
    \label{Qpsin2} 
    Q\dg\cvc{0,\psi_n^{(2)}}
    &=\mat{0,A\dg}{0,0}\cvc{0,\psi_n^{(2)}}
    =\cvc{A\dg\psi^{(2)},0}
    =\frac{1}{\sqrt{E_{n+1}^{(1)}}}\cvc{A\dg\psi_{n+1}^{(1)},0}
\end{align}
In other words, the operators $Q$ and $Q\dg$ relate the states of $H^{(1)}$ and $H^{(2)}$ with the same energy. This can be viewed as transitioning between bosonic and fermionic states, exactly as is done in particle physics.\\

The operators $Q$ and $Q\dg$ generate the supersymmetry algebra of $H$. In essence, they play the role of supersymmetric charge. Consequently, in order to have supersymmetric states, the ground state must be preserved by $Q$ and $Q\dg$. If it is not preserved, supersymmetry is broken. Looking at \eqref{Qpsin1} and \eqref{Qpsin2}, we can only have unbroken supersymmetry if the following conditions hold:
\begin{align*}
    A\psi_0^{(1)}=0 && A\dg\psi_0^{(2)}=0
\end{align*}
In other words, the ground state energy must be zero, or else supersymmetry is broken.

\subsection{Climbing the Supersymmetric Ladder}
We started with a Hamiltonian $H^{(1)}$ with zero ground state energy and obtained another Hamiltonian $H^{(2)}$ by expressing $H^{(1)}$ as $A\dg A$ and setting $H^{(2)}=AA\dg$. We can continue this process to obtain more Hamiltonians. The basic idea is the same: We would like to factor $H^{(2)}$ into $H^{(2)}=B\dg B$. However, if the ground state energy of $H^{(2)}$ is positive, as it typically will be, we cannot hope to get any new states as supersymmetry will be broken. More formally, $BB\dg$ and $B\dg B$ will have exactly the same spectrum (see \cite[Lemma 2.44, Page 17]{sekhon}). To get around this, we first need to `reset' $H^{(2)}$ to have zero ground state energy.\\

Suppose $E_0^{(2)}$ is the ground state energy of $H^{(2)}$. We first need to subtract this from $H^{(2)}$ before we can factor it:
\begin{equation*}
    H^{(2)}-E_0^{(2)}=B\dg B
\end{equation*}
We can then repeat the construction of a partner potential for $B\dg B$. Before we proceed, we will relabel $W_1=W$, $A_1=A$ and $A_2=B$. We now make a similar ansatz as in \eqref{superpotansatz}:
\begin{align*}
    A_2=\eta\dv{x}+W_2(x) && A_2\dg=-\eta\dv{x}+W_2(x)
\end{align*}
Similar calculations yield the second superpotential $W_2(x)$ in terms of the ground state $\psi_0^{(2)}$ of $H^{(2)}$, analogous to \eqref{wpsi0}:
\begin{equation*}
    W_2(x)=-\eta\frac{{\psi_0^{(2)}}'(x)}{\psi_0^{(2)}(x)}
\end{equation*}
This automatically leads to a third Hamiltonian, a supersymmetric partner of $H^{(2)}$:
\begin{equation*}
    H^{(3)}=A_2A_2\dg+E_0^{(2)}
\end{equation*}
We can also express this as $H^{(3)}=-\eta^2\dv[2]{x}+V^{(3)}(x)$, where:
\begin{equation*}
    V^{(3)}(x)=\eta W_2'(x)+W_2(x)^2+E_0^{(2)}
\end{equation*}
Analogously to \eqref{ricsuperpot}, we have:
\begin{equation*}
    V^{(2)}(x)=-\eta W_2'(x)+W_2(x)^2+E_0^{(2)}
\end{equation*}
This allows us to rewrite $V^{(3)}(x)$ as follows:
\begin{align*}
    V^{(3)}(x)
    &=V^{(2)}(x)+2\eta W_2'(x) \\
    &=V^{(1)}(x)+2\eta W_2'(x)+2\eta W_1'(x)
\end{align*}
The last equation suggests a pattern. We can continue defining more partner potentials (and thus more partner Hamiltonians) as follows:

\begin{definition}\label{susypartnerext}
Suppose $m\in\N$. The $(m-1)$\textsuperscript{st} \textbf{supersymmetric partner} of $H^{(1)}$ is given by:
\begin{align*}
    H^{(m)}=-\eta^2\dv[2]{x}+V^{(m)}(x) && V^{(m)}(x)=V^{(1)}(x)+2\eta\sum_{k=1}^{m-1} W_k'(x)
\end{align*}
\end{definition}

This also allows us to continue the supersymmetric ladder through the chain of partner Hamiltonians. All in all, we can now extend \Cref{susyladder} to the following:

\begin{theorem}
Suppose for each $m,n\in\N$, $\psi_n^{(m)}$ is the $n$\textsuperscript{th} excited state of $H^{(m)}$ with energy $E_n^{(m)}$, and $\psi_n^{(m+1)}$ is the $n$\textsuperscript{th} excited state of $H^{(m+1)}$ with energy $E_n^{(m+1)}$. Then for all $m,n\in\N$, we have:
\begin{align*}
    E_n^{(m+1)}=E_{n+1}^{(m)} && \psi_n^{(m+1)}=\frac{1}{\sqrt{E_{n+1}^{(m)}-E_n^{(m)}}}A_m\psi_{n+1}^{(m)} && \psi_{n+1}^{(m)}=\frac{1}{\sqrt{E_n^{(m+1)}-E_n^{(m)}}}A_m\dg\psi_n^{(m+1)}
\end{align*}
\end{theorem}

Now that we know $E_n^{(m+1)}=E_{n+1}^{(m)}=E_{n+m}^{(1)}$ for all $m,n\in\N$, we will simply denote this by $E_{n+m}$. In other words, $E_k$ is the energy of the $k$\textsuperscript{th} excited state of $H^{(1)}$ (as well as the $(k-m+1)$\textsuperscript{st} excited state of $H^{(m)}$). Iterating the formulas above, we get:
\begin{align}\label{iterpartnerstate1}
    \psi_n^{(m)}
    &=\frac{1}{\sqrt{\left(E_{n+m}-E_{n+m-1}\right)\left(E_{n+m-1}-E_{n+m-2}\right)\cdots\left(E_{n+1}-E_n\right)}} \, A_{m-1}A_{m-2}\cdots A_1 \psi_{n+m-1}^{(1)} \\
    \label{iterpartnerstate2}
    \psi_{n+m-1}^{(1)}
    &=\frac{1}{\sqrt{\left(E_{n+m}-E_{n+m-1}\right)\left(E_{n+m-1}-E_{n+m-2}\right)\cdots\left(E_{n+1}-E_n\right)}} \, A_1\dg A_2\dg\cdots A_{m-1}\dg \psi_n^{(m)}
\end{align}
Note that the operators at play here are $A_{m-1}A_{m-2}\cdots A_1$ and $A_1\dg A_2\dg\cdots A_{m-1}\dg=(A_{m-1}A_{m-2}\cdots A_1)\dg$. This allows us to `jump steps' and get the spectrum of, say $H^{(72)}$, using the spectrum of $H^{(1)}$ and the composite operator $A_{71}A_{70}\cdots A_1$ (even if $H^{(72)}$ itself is not a supersymmetric partner of $H^{(1)}$).

\begin{center}
\begin{tikzpicture}
\foreach \a/\b in {-4.6/0,-3.8/0,-1.8/0.8,-1/0.8,1/1.4,1.8/1.4}
{\draw[thick] (\a,\b)--(\a,3.2);
\foreach \p in {2.2,3}
\node at (3.8,\p) {$\iddots$}; 
\draw[-stealth,thick,dashed] (\a,3.2)--(\a,4);}
\draw (-4.6,1) node[left]{$E_1$}--(-3.8,1) (-1.8,1) node[left]{$E_1$}--(-1,1) (1,1); 
\draw (-4.6,1.6) node[left]{$E_2$}--(-3.8,1.6) (-1.8,1.6) node[left]{$E_2$}--(-1,1.6) (1,1.6) node[left]{$E_2$}--(1.8,1.6); 
\draw (-4.6,2.1) node[left]{$E_3$}--(-3.8,2.1) (-1.8,2.1) node[left]{$E_3$}--(-1,2.1) (1,2.1) node[left]{$E_3$}--(1.8,2.1); 
\draw (-4.6,2.5)--(-3.8,2.5) (-1.8,2.5)--(-1,2.5) (1,2.5)--(1.8,2.5); 
\draw (-4.6,2.8)--(-3.8,2.8) (-1.8,2.8)--(-1,2.8) (1,2.8)--(1.8,2.8); 
\draw (-4.6,3)--(-3.8,3) (-1.8,3)--(-1,3) (1,3)--(1.8,3); 
\draw (-4.6,0.2) node[left]{$E_0$}--(-3.8,0.2); 
\node at (-4.2,4.4) {$H^{(1)}$};
\node at (-1.4,4.4) {$H^{(2)}$};
\node at (1.4,4.4) {$H^{(3)}$};
\draw[-latex] (-3.5,3.2) to[bend left=15] node[midway,above,font=\small]{$A_1$} (-2.1,3.2);
\draw[-latex] (-2.1,2.9) to[bend left=15] node[midway,below,font=\small]{$A_1\dg$} (-3.5,2.9);
\draw[-latex] (-0.7,3.2) to[bend left=15] node[midway,above,font=\small]{$A_2$} (0.7,3.2);
\draw[-latex] (0.7,2.9) to[bend left=15] node[midway,below,font=\small]{$A_2\dg$} (-0.7,2.9);
\end{tikzpicture}\\
\sffamily The extended supersymmetric ladder
\end{center}

This picture emphasizes the power of SUSY QM: In standard QM, we could only climb up and down a single ladder, but now, we also have the ability to jump from one ladder to another. Every ladder has exactly the same steps (energy levels) as the previous one, excluding the ground state. And as we remarked above, we can compose the supersymmetric operators $A_1,A_2,A_3,...$ to get the required operator that jumps from, say, $H^{(1)}$ to $H^{(72)}$. Therefore, it is better to visualize these ladders as fireman's ladders, as they can slide up and down past one another.\\

As a final remark, the picture of the supersymmetric ladder above generally only applies to \emph{bound} states. It is possible that, in addition to a discrete collection of bound states described by the ladders, there is also a continuum of scattering states with higher energies. This is the case with the finite potential well and, as we will see in \Cref{subsec:sech}, the sech potential.

\subsection{Shape Invariant Potentials}\label{subsec:shapeinvariance}
In SUSY QM, there is a class of potentials known as \emph{shape invariant} potentials. As the name suggests, these are potentials that in some sense `do not change shape'. To be more precise, they are the potentials whose supersymmetric partners have the same shape. For example, as we will see in \Cref{subsec:harmonic}, the supersymmetric partner of the harmonic oscillator is another harmonic oscillator with a higher ground state energy (i.e. the same parabola, shifted up). We will show here that shape invariant potentials are extremely nice to work with, at least from a supersymmetric point of view. Indeed, many of the potentials one would solve in a first course on quantum mechanics are shape invariant.

\begin{definition}
A \textbf{shape invariant potential} is a potential $V^{(1)}(x,a)$ whose supersymmetric partner $V^{(2)}(x,a)$ is given by:
\begin{equation}\label{shapeinvariance}
    V^{(2)}(x,a)=V^{(1)}(x,f(a))+R(a)
\end{equation}
Where $x$ is the space variable, $a$ is a parameter and $f$ and $R$ are functions of $a$ only.
\end{definition}
\begin{remark}
The purpose of $f(a)$ above is to allow $V^{(2)}$ to have a different parameter from $V^{(1)}$, such as being a translated or scaled copy of $V^{(1)}$. Likewise, the purpose of $R(a)$ is to allow for a different ground state energy.
\end{remark}

We have seen that for a given potential $V^{(1)}$, the superpotential $W$ is not unique, and neither is the partner potential $V^{(2)}$. You might be wondering if the property of shape invariance \eqref{shapeinvariance} depends on the choice of $V^{(2)}$. Interestingly, it does not. If a potential is shape invariant with respect to \emph{one} of its supersymmetric partners, it is shape invariant with respect to \emph{all} of them. See \cite{sandhya} for details.\\

Applying \eqref{shapeinvariance} repeatedly, we get:
\begin{equation*}
    V^{(m)}(x,a)=V^{(1)}(x,f^{m-1}(a))+\sum_{k=0}^{m-2} R(f^k(a))
\end{equation*}
Thus the $m$\textsuperscript{th} partner Hamiltonian is given by:
\begin{equation*}
    H^{(m)}=-\eta^2\dv[2]{x}+V^{(1)}(x,f^{m-1}(a))+\sum_{k=0}^{m-2} R(f^k(a))
\end{equation*}

\begin{theorem}\label{sipenergy}
Suppose $\psi^{(1)}(x,a)$ is an eigenstate (not necessarily the ground state) of the first Hamiltonian with energy $E^{(1)}$ Then $\psi^{(1)}(x,f^{m-1}(a))$ (the state $\psi^{(1)}(x,a)$, but with the parameter $f^{m-1}(a)$ in place of $a$) is an eigenstate of $H^{(m)}$, with energy:
\begin{equation*}
    E^{(1)}+\sum_{k=0}^{m-2} R(f^k(a))
\end{equation*}
\end{theorem}
\begin{warning}
This does NOT mean that $H^{(1)},H^{(2)},H^{(3)},...$ all have the same eigenstates. Their eigenstates are related to one another by repeatedly applying $f$ to the parameter $a$. For example, if $\psi^{(1)}(x,a)$ is an eigenstate of $H^{(1)}$, then $\psi^{(1)}(x,f^{31}(a))$ is an eigenstate of $H^{(32)}$.
\end{warning}
\begin{proof}
Applying $H^{(m)}$ to the state $\psi^{(1)}(x,f^{m-1}(a))$, we get:
\begin{align*}
    H^{(m)}\psi^{(1)}(x,f^{m-1}(a))
    =\textcolor{red!80!black}{\left(-\eta^2\dv[2]{x}+V^{(1)}(x,f^{m-1}(a))\right)\psi^{(1)}(x,f^{m-1}(a))}+\left(\sum_{k=0}^{m-2} R(f^k(a))\right)\psi^{(1)}(x,f^{m-1}(a))
\end{align*}
The \textcolor{red!80!black}{first term} above reduces to $H^{(1)}\psi^{(1)}(x,f^{m-1}(a))$, which equals $E^{(1)}\psi^{(1)}(x,f^{m-1}(a))$ by definition. This yields:
\begin{equation}
    H^{(m)}\psi^{(1)}(x,f^{m-1}(a))=\left(E^{(1)}+\sum_{k=0}^{m-2} R(f^k(a))\right)\psi^{(1)}(x,f^{m-1}(a))
\end{equation}
Thus $\psi^{(1)}(x,f^{m-1}(a))$ is also an eigenstate of $H^{(m)}$, with energy $E^{(1)}+\sum_{k=0}^{m-2} R(f^k(a))$.
\end{proof}

To find examples of shape invariant potentials, we need to solve the equations defining $V^{(1)}$ and $V^{(2)}$ together. In other words, we need to solve the following system:
\begin{align*}
    -W'(x,a)+W(x,a)^2=V^{(1)}(x,a) &&
    W'(x,a)+W(x,a)^2=V^{(1)}(x,f(a))+R(a)
\end{align*}
This is a system of \emph{functional differential equations}, which cannot be solved exactly except in a few special cases. See \cite{azbelev} for the general theory, as well as solution methods for some special cases.\\

For a discussion on the characterization of shape invariant potentials, see \cite{bhaduri}, \cite{quesne} and \cite{sandhya}.

\subsection{The Supersymmetric WKB Approximation}\label{subsec:swkb}
The \textbf{Wentzel-Kramers-Brillouin} (WKB) approximation is an approximate analytical method for solving ODEs like the time-independent Schrödinger equation. It was first formulated\footnote{A similar method, introduced in 1837 by \href{https://mathshistory.st-andrews.ac.uk/Biographies/Liouville/}{Joseph Liouville} (1809--1882) and \href{https://mathshistory.st-andrews.ac.uk/Biographies/Green/}{George Green} (1793--1841), had already been used in hydrodynamics. For this reason, it is known in some older texts as the Liouville-Green (LG) method.} in 1926 by \href{http://www.nasonline.org/publications/biographical-memoirs/memoir-pdfs/wentzel-gregor.pdf}{Gregor Wentzel} (1898--1978), \href{https://mathshistory.st-andrews.ac.uk/Biographies/Kramers/}{Hendrik Anthony Kramers} (1894--1952) and \href{http://www.nasonline.org/publications/biographical-memoirs/memoir-pdfs/brillouin-lon-n.pdf}{Léon Nicolas Brillouin} (1889--1969).\\

The WKB approximation is a method of \emph{asymptotic approximation}, and it relies on the coefficient of the highest derivative in the ODE being a small parameter (small enough that we can sensibly produce expansions in powers of it). See \cite[\S1.11.4, Pages 56--57]{polyanin} and \cite[\S3.6.2, Pages 173--174]{polyanin} for the general class of such methods. For our purposes, the parameter will be $\eta=\frac{\hbar}{\sqrt{2m}}$, which is always sufficiently small\footnote{Even with a crude estimate of $m=0.07$ eV for the electron neutrino (the lightest known massive particle), we get $\eta\approx2.11\times10^{-16}$.}. We will first outline the standard WKB approximation before invoking SUSY QM.\\

For more details on the WKB approximation, see \cite[Chapter 14]{muller-kirsten}.\\

In the standard WKB approximation, we first rewrite the time-independent Schrödinger equation \eqref{sch:1} as follows:
\begin{equation}\label{wkb:1}
    -\frac{\hbar^2}{2m}\psi''(x)=(E-V(x))\psi(x)
\end{equation}
We then rewrite the wavefunction $\psi(x)$ in the form:
\begin{equation}\label{wkb:2}
    \psi(x)=e^{\phi(x)} \ ,\qquad \phi'(x)=A(x)+iB(x)
\end{equation}
Where $A(x)$ and $B(x)$ are real-valued. Thus $A(x)$ encodes the amplitude of $\psi$, while $B(x)$ encodes the phase of $\psi$. More specifically:
\begin{align*}
    \textsf{Amplitude}=\exp\left(\int_a^x A(y)\dy\right) &&
    \textsf{Phase}=\int_a^x B(y)\dy
    \tag{up to normalization}
\end{align*}
With this, one would substitute \eqref{wkb:2} into \eqref{wkb:1} and use perturbation to approximate solutions for $A(x)$ and $B(x)$, in powers of $\hbar$ (or equivalently for our purposes, powers of $\eta$):
\begin{align*}
    A(x)=\sumn[-1] \hbar^n A_n(x) && B(x)=\sumn[-1] \hbar^n B_n(x)
\end{align*}
The leading terms have $n=-1$ as after substituting \eqref{wkb:2} into \eqref{wkb:1}, the leading order is $\frac{1}{\hbar}$.\\

We now turn to the supersymmetric WKB approximation. This time, we are aided by the fact that we have already defined a superpotential $W(x)$. Suppose $E_n$ is the energy of the $n$\textsuperscript{th} excited state of $H^{(1)}$.

\begin{center}
\begin{tikzpicture}
\draw[latex-latex] (-2.5,0)--(5,0) node[right]{$x$};
\fill[smooth,Cerulean,fill opacity=0.1,domain=-1.5:4,samples=200] (-1.5,0)--plot(\x,{0.1*(\x)^4-0.55*(\x)^3+0.45*(\x)^2+1.05*(\x)+0.7})--(4,0)--cycle; 
\draw[dashed,RawSienna!50!black] (-2.5,2.5)--(5,2.5) node[right]{$E$};
\draw[thick,smooth,RawSienna,domain=-1.618901:4.158273,samples=200] plot(\x,{0.1*(\x)^4-0.55*(\x)^3+0.45*(\x)^2+1.05*(\x)+0.7}) node[right]{$V^{(1)}$}; 
\foreach \n/\c in {1/red,2/Orange,3/Dandelion,4/Green,5/blue,6/violet}
\draw[thick,smooth,\c!40,domain=-1.5:4,samples=200] plot(\x,{0.3*sin(360/11*\n*(\x+1.5))+2.5});
\fill (-1.5,2.5) circle (1.5pt);
\fill (4,2.5) circle (1.5pt);
\draw (-1.5,2.5)--(-1.5,0) node[below=2pt,xshift=-4pt]{$a$};
\draw (4,2.5)--(4,0) node[below=0pt,xshift=4pt]{$b$};
\draw[decorate,decoration={brace,mirror,amplitude=6pt,raise=3pt}] (-1.5,0)--(4,0) node[midway,below=8pt,font=\small\sffamily] {Classically allowed region};
\node[font=\small\sffamily] at (1.25,3.05) {Approximations to eigenstates};
\end{tikzpicture}\\
\sffamily Visualization of the WKB approximation
\end{center}

\begin{theorem}[Supersymmetric WKB Approximation]
Suppose $E_n$ is the estimate of the energy of the $n$\textsuperscript{th} excited state of $H^{(1)}$. Define the \textit{classical turning points} $a$ and $b$ by $W(a)^2=W(b)^2=E_n$. Then the supersymmetric WKB approximation, to first order in $\eta$, is given by:
\begin{equation}\label{swkb}
    \int_a^b \sqrt{E_n-W(x)^2}\dx=n\pi\eta
\end{equation}
\end{theorem}
\begin{proof}
We start with the Bohr-Sommerfeld quantization condition:
\begin{equation}\label{bohrsommerfeld}
    \int_a^b p(x)\dx=n\pi\hbar
\end{equation}
Where $p(x)$ is the momentum of the particle. In the Schrödinger equation, this is given by the momentum operator $\widehat{p}=-\hbar^2\dv[2]{x}$. Comparing this with \eqref{wkb:1}, we get:
\begin{equation}
    p(x)=\sqrt{2m(E_n-V^{(1)}(x))}
\end{equation}
Thus we can rewrite \eqref{bohrsommerfeld} as follows:
\begin{equation}
    \int_a^b p(x)\dx=\int_a^b \sqrt{2m(E_n-V^{(1)}(x))}\dx=n\pi\hbar
\end{equation}
Using \eqref{ricsuperpot} to rewrite $V^{(1)}(x)$ as $-\eta W'(x)+W(x)^2$, we get:
\begin{align*}
    \int_a^b \sqrt{2m(E_n-W(x)^2+\eta W'(x))}\dx&=n\pi\hbar \\
    \int_a^b \sqrt{E_n-W(x)^2+\eta W'(x)}\dx&=n\pi\eta \tag{since $\eta=\frac{\hbar}{\sqrt{2m}}$}
\end{align*}
The integrand, to leading order in $\eta$, is $\sqrt{E_n-W(x)^2}$ (the square root does not affect this as $E_n-V^{(1)}(x)\ge0$ in the classically allowed region). Thus the supersymmetric WKB approximation, to first order in $\eta$, is given by:
\begin{equation*}
    \int_a^b \sqrt{E_n-W(x)^2}\dx=n\pi\eta \qedhere
\end{equation*}
\end{proof}

The supersymmetric WKB approximation has several advantages over the standard WKB approximation. For one, it works better at lower energy levels. The standard WKB approximation loses accuracy at lower energy levels as the correction (higher-order) terms become more significant. The supersymmetric WKB approximation gets around this problem by setting the ground state energy to zero, thereby giving us better estimates of the lower levels of the spectrum.\\

Another key advantage of the supersymmetric WKB approximation is that for shape invariant potentials, it is exact: The first-order approximation above gives all the exact energy levels! Well, that's what we think. For all known shape invariant potentials, this holds true (see \cite[Chapter 12]{mallow}), and we will show this explicitly for the harmonic oscillator in \Cref{harmonicwkb}. Whether it holds for \emph{all} shape invariant potentials remains an open problem.\\

For more details about the supersymmetric WKB approximation, see \cite{cooper} and \cite{mallow}.

\section{Supersymmetric Quantum Systems}\label[chapter]{sec:systems}
In this chapter, we will explore some important examples of quantum systems that exhibit supersymmetry. Some of these examples will have desirable properties, such as shape invariance which we discussed in \Cref{subsec:shapeinvariance}.

\subsection{The Inverse Square Potential}\label{subsec:invsqpot}
We will first look at the \emph{inverse square potential}, one of the potentials for which we can solve for the superpotential and the ground state analytically. While this potential does not have any normalizable eigenstates, it is a good example to demonstrate the non-uniqueness of the superpotential, and consequently the supersymmetric partner.\\

Suppose $V^{(1)}(x)=\frac{k(k+1)}{x^2}$ for some constant $k$ (the coefficient $k(k+1)$ is for convenience). Then \eqref{ricsuperpot} becomes:
\begin{equation}\label{invsqpot}
    -W'(x)+W(x)^2=\frac{k(k+1)}{x^2}
\end{equation}
We first look for a particular solution of the form $W(x)=\frac{a}{x}$ (where $a$ is a constant). Substituting this into \eqref{invsqpot}, we get:
\begin{equation*}
    \frac{a}{x^2}+\frac{a^2}{x^2}=\frac{k(k+1)}{x^2} \qquad\Longrightarrow\qquad a^2+a-k(k+1)=0 \quad\therefore\quad a=k \text{ or } a=-(k+1)
\end{equation*}
We now have two particular solutions of \eqref{invsqpot}, namely:
\begin{align*}
    W_+(x)=\frac{k}{x} && W_-(x)=-\frac{(k+1)}{x}
\end{align*}
These yield, respectively:
\begin{align}
    V_+^{(2)}(x)
    &=W_+'(x)+W_+(x)^2=-\frac{k}{x^2}+\frac{k^2}{x^2}=\frac{k(k-1)}{x^2} \\
    V_-^{(2)}(x)
    &=W_-'(x)+W_-(x)^2=\frac{(k+1)}{x^2}+\frac{(k+1)^2}{x^2}=\frac{(k+1)(k+2)}{x^2}
\end{align}
In either case, the partner potential is another inverse square potential, with a different proportionality constant. Note that $V_+^{(2)}$ is a `step down' from $V^{(1)}$ (as it is equivalent to replacing $k$ with $k-1$), while $V_-^{(2)}$ is a `step up' (as it is equivalent to replacing $k$ with $k+1$).

\begin{center}
\begin{tikzcd}[row sep=2mm,column sep=10mm]
\dfrac{0\cdot1}{x^2}
    \arrow[-stealth,bend left=30]{r}{-\tfrac{1}{x}}
    &
\dfrac{1\cdot2}{x^2}
    \arrow[-stealth,bend left=30]{l}{\tfrac{1}{x}}
    \arrow[-stealth,bend left=30]{r}{-\tfrac{2}{x}}
    &
\dfrac{2\cdot3}{x^2}
    \arrow[-stealth,bend left=30]{l}{\tfrac{2}{x}}
    \arrow[-stealth,bend left=30]{r}{-\tfrac{3}{x}}
    &
\dfrac{3\cdot4}{x^2}
    \arrow[-stealth,bend left=30]{l}{\tfrac{3}{x}}
    &[-10mm]
\cdots\cdots
\end{tikzcd}\\
Chain of partner potentials arising from the inverse square potential
\end{center}

\noindent By \Cref{riccatigs}, the general solution of \eqref{invsqpot} is given by:
\begin{equation*}
    W(x)
    =\frac{Ckx^{-k-1}-(k+1)x^{k}}{Cx^{-k}+x^{k+1}}
    =\frac{k}{x}-\frac{(2k+1)x^{2k}}{x^{2k+1}+C}
    =-\frac{(k+1)}{x}+\frac{C(2k+1)}{x(x^{2k+1}+C)}
\end{equation*}
Where $C$ is an arbitrary constant (the particular solutions $\frac{k}{x}$ and $-\frac{(k+1)}{x}$ occur when $C=\infty$ and $C=0$ respectively). This yields:
\begin{align*}
    V^{(2)}(x)=W'(x)+W(x)^2=\frac{k(k-1)}{x^2}+\frac{2(2k+1)x^{2k+1}(x^{2k+1}-2kC)}{x^2(x^{2k+1}+C)^2}
\end{align*}
We now use \eqref{psi0w} to solve for $\psi_0$:
\begin{align*}
    \psi_0(x)
    &=B\exp\left(-\int W(x) \dx\right)
    =B\exp\left(-\int \left(-\frac{(k+1)}{x}+\frac{C(2k+1)}{x(x^{2k+1}+C)}\right) \dx\right) \\
    &=B\exp\left(-\left(k\ln(x)-\ln(x^{2k+1}+C)\right)\right) \\
    &=B\exp\left(-k\ln(x)+\ln(x^{2k+1}+C)\right) \\
    &=B\left(x^{-k}\left(x^{2k+1}+C\right)\right) \\
    &=Bx^{k+1}+BCx^{-k}
\end{align*}
Here, $B$ is simply a normalization constant, while $C$ determines the `mixture' of the states $x^{k+1}$ and $x^{-k}$ (although this is not a mixed state in the usual sense). However, this is \emph{never} normalizable on $(0,\infty)$, and if both terms above are present, it is also not normalizable on $(0,1]$ or $[1,\infty)$. Even if we consider the system on some domain $[a,b]$ where $0<a<b<\infty$, we cannot get any normalizable eigenstates under self-adjoint boundary conditions (see \Cref{subsec:selfadjoint}).

\subsection{The \textrm{sech} Potential}\label{subsec:sech}
We now investigate the hyperbolic secant (sech) potential. This is interesting in its own right because, as we will see shortly, its supersymmetric partner is a constant potential (which corresponds to a free particle). Note that despite the name `sech potential', it is the \emph{ground state} that is a hyperbolic secant, not the potential itself.\\

We start with the ground state $\psi_0(x)=\sech(ax)$, where $a>0$ is a constant. Then we have:
\begin{equation*}
    W(x)
    =-\frac{\psi_0'(x)}{\psi_0(x)}
    =-\frac{-a\cdot\sech(ax)\tanh(ax)}{\sech(ax)}
    =a\cdot\tanh(ax)
\end{equation*}
We now compute the potential $V^{(1)}$ and the partner potential $V^{(2)}$:
\begin{align*}
    V^{(1)}(x)
    &=-W'(x)+W(x)^2
    =-a^2\sech(ax)^2+a^2\tanh(ax)^2
    =a^2\left(1-2\sech(ax)^2\right)
    \\
    V^{(2)}(x)
    &=W'(x)+W(x)^2
    =a^2\sech(ax)^2+a^2\tanh(ax)^2
    =a^2
\end{align*}

\begin{center}
\begin{tikzpicture}
\draw[latex-latex] (-4,0)--(4,0) node[right]{$x$};
\draw[latex-latex] (0,-1.4)--(0,1.4);
\draw[thick,smooth,red,domain=-4:4,samples=200] plot(\x,{1-2/(cosh(\x))^2});
\draw[thick,blue] (-4,1)--(4,1);
\node[red] at (2,0.5) {$V^{(1)}$};
\node[blue] at (3.6,1.3) {$V^{(2)}$};
\node[below left=-2pt] at (0,1) {$a^2$};
\end{tikzpicture}\\
\sffamily The sech potential and its supersymmetric partner
\end{center}

We see that the partner potential $V^{(2)}=a^2$ is a positive constant, so its states are simply plane waves:
\begin{equation*}
    \psi^{(2)}(x)=c_1e^{ikx}+c_2e^{-ikx} \qquad\qquad k=\sqrt{E-a^2}
\end{equation*}
Here we assume $E>a^2$, otherwise the eigenstates would be exponential and thus unbounded as $x\to\pm\infty$.\\

We can also get the full energy spectrum of $H^{(1)}$ immediately. Since $H^{(2)}$ has a continuous energy spectrum $(a^2,\infty)$ (every energy $E>a^2$ yields a plane wave solution, while $E<a^2$ do not), it follows from \Cref{susyladder} that $H^{(1)}$ must have the same energy spectrum, possibly including $E=0$. And by construction, $E=0$ \emph{must} be part of the spectrum, since it is the energy of the ground state $\psi_0$. This yields the full energy spectrum of the sech potential:
\begin{equation*}
    E\in\{0\}\cup(a^2,\infty)
\end{equation*}
This shows that the sech potential has \emph{one} bound state with $E=0$ (the ground state), and a continuum of scattering states with $E>a^2$.

\begin{center}
\begin{tikzpicture}[scale=1.2]
\begin{scope}[shift={(0,0)}]
\shade[top color=white,bottom color=red] (0,0.5) rectangle (1,2);
\draw[thick] (0,0)--(1,0);
\draw[dashed,thin] (0,0)--(-0.6,0) node[left]{$0$};
\draw[dashed,thin] (0,0.5)--(-0.6,0.5) node[left]{$a^2$};
\draw[-latex,thick,red!40] (0.5,2.1)--(0.5,2.4);
\node[left,red] at (0,2) {$V^{(1)}$};
\end{scope}
\begin{scope}[shift={(3,0)}]
\shade[top color=white,bottom color=blue] (0,0.5) rectangle (1,2);
\draw[dashed,thin] (1,0.5)--(1.6,0.5) node[right]{$a^2$};
\draw[-latex,thick,blue!40] (0.5,2.1)--(0.5,2.4);
\node[right,blue] at (1,2) {$V^{(2)}$};
\end{scope}
\end{tikzpicture}\\
\vspace{4pt}
\sffamily The energy levels of the \textrm{sech} potential and its supersymmetric partner
\end{center}

The Schrödinger equation for $H^{(1)}$ would be \emph{extremely} difficult to solve directly\footnote{I tried it, and believe me, you don't want to.}. Fortunately for us, we know that we can get the states of $H^{(1)}$ simply by applying $A\dg$ to the corresponding states of $H^{(2)}$. We will only do this for $e^{ikx}$ (the wave traveling from left to right), the case $e^{-ikx}$ follows similarly.
\begin{align*}
    \psi^{(1)}(x)
    =A\dg(e^{ikx})
    =-\dv{x}(e^{ikx})+W(x)e^{ikx}
    =e^{ikx}(-ik+a\cdot\tanh(ax))
\end{align*}
This can be viewed as a plane wave solution (the first term) with a `twist' by $a\cdot\tanh(ax)$ (the second term). More precisely, in the limit as $x\to\pm\infty$, we have:
\begin{align}\label{sechinout}
    \psi^{(1)}(x)\sim(-ik-a)e^{ikx} \text{  as } x\to-\infty
    &&
    \psi^{(1)}(x)\sim(-ik+a)e^{ikx} \text{  as } x\to+\infty
\end{align}
Thus the wavefunction resembles the plane wave solution on both ends, but with a different phase. The phase shift is given by:
\begin{equation*}
    \vp(a,k)=\arg\left(\frac{-ik+a}{-ik-a}\right)=\arctan\left(\frac{2ak}{k^2-a^2}\right)
\end{equation*}
Note that in \eqref{sechinout}, the limits in both directions only contain $e^{ikx}$ and not $e^{-ikx}$. Physically, this means that all of the incoming wave packet (whichever direction it comes from) is transmitted to the other side of the well, none is reflected.

\begin{center}
\begin{tikzpicture}
\def\a{1.7} 
\def\k{2} 
\draw[latex-latex] (-6.4,0)--(6.4,0) node[right]{$x$};
\draw[latex-latex] (0,-1.5)--(0,1.5) node[above]{$\Re(\psi)$};
\draw[thick,smooth,red,domain=-6:6,samples=200] plot(\x,{cos(180*\x)}); 
\draw[thick,smooth,blue,domain=-6:6,samples=200] plot(\x,{1/((\k)^2+(\a)^2)*((\k)^2-(\a)^2*tanh(\x))*cos(180*\x)-(\a*\k)/((\k)^2+(\a)^2)*(1+tanh(\x))*sin(180*\x)}); 
\node[right,red,font=\sffamily] at (2,1.9) {Free particle};
\node[right,blue,font=\sffamily] at (2,1.5) {Particle in a \textrm{sech} potential};
\end{tikzpicture}\\
\sffamily Wave packets coming from the left ($-\infty$)\\
We only show the real parts here, the imaginary parts behave similarly.
\end{center}

As can be seen above, the outgoing wave is identical to the incoming wave, with a phase shift of $\vp$. Note that the amplitude of the wave is smaller near $x=0$, indicating that the particle is less likely to be found inside the well (or in other words, it moves faster while inside the well, as one would expect).\\

Note that for fixed $k$, $\vp(a,k)$ is an increasing function of $a$. Qualitatively, the deeper the potential well, the larger the phase shift. On the other hand, for fixed $a$, $\vp(a,k)$ and a decreasing function of $k$. Qualitatively, the higher the energy of the wave packet, the smaller the phase shift.\\

The ground state of the sech potential is also important as a solution to the \emph{nonlinear Schrödinger equation}.

\begin{definition}
The (time-independent) \textbf{nonlinear Schrödinger equation} is given by:
\begin{equation*}
    -\frac{\hbar^2}{2m}\pdv[2]{\psi(x)}{x}+\kappa\abs{\psi(x)}^2\psi(x)=E\psi(x)
\end{equation*}
Where $\kappa$ is a real constant (which can be positive, negative or zero).
\end{definition}

This looks like the time-dependent Schrödinger equation, except that the potential $V(x)$ is replaced with $\abs{\psi(x)}^2$ (the probability density).\\

To relate the sech potential to the nonlinear Schrödinger equation, we first normalize $\psi_0$:
\begin{equation*}
    \int_{-\infty}^\infty \sech(ax)^2\dx=\frac{2}{a} \qquad\Longrightarrow\qquad \psi_0(x)=\sqrt{\frac{a}{2}}\sech(ax)
\end{equation*}
This allows us to rewrite $V^{(1)}$ as follows:
\begin{equation*}
    V^{(1)}(x)=a^2\left(1-2\sech(ax)^2\right)=a^2-4a\abs{\psi_0(x)}^2
\end{equation*}
Thus the ground state of the sech potential satisfies the nonlinear Schrödinger equation with $\kappa=-4a$ and $E=-a^2$.\\

The nonlinear Schrödinger equation describes the amplitude of solitary waves (solitons) in deep water. As such, the ground state of the sech potential can be thought of as a soliton solution of the nonlinear Schrödinger equation. See \cite{drazin} for details.

\subsection{The Particle in a Box}\label{subsec:particleinabox}
We now investigate the particle in a (one-dimensional) box, also known as the infinite square well. This potential is given by:
\begin{equation*}
    V(x)=\begin{cases} 0 & 0\le x\le L \\ \infty & \text{otherwise} \end{cases}
\end{equation*}
The Schrödinger equation can be solved directly to get:
\begin{align}\label{particleinaboxdirect}
    \psi_n(x)=\sqrt{\frac{2}{L}}\sin\left(\frac{(n+1)\pi}{L}x\right) && E_n=\frac{(n+1)^2\pi^2}{L^2}
\end{align}
We have $n+1$ above instead of $n$ as we want $n=0$ to denote the ground state. The ground state energy is $E_0=\frac{\pi^2}{L^2}$, so we will work with the shifted potential $V^{(1)}(x)=V(x)-E_0$. In other words:
\begin{align}\label{pbox:1}
    V^{(1)}(x)=\begin{cases} -\dfrac{\pi^2}{L^2} & 0\le x\le L \\ \infty & \text{otherwise} \end{cases}
    &&
    E_n^{(1)}=\frac{(n+1)^2\pi^2}{L^2}-\frac{\pi^2}{L^2}
    =\frac{n(n+2)\pi^2}{L^2}
\end{align}
The superpotential $W(x)$ is given by:
\begin{equation*}
    W(x)
    =-\frac{\psi_0'(x)}{\psi_0(x)}
    =-\frac{\sqrt{\frac{2}{L}}\frac{\pi}{L}\cos\left(\frac{\pi}{L}x\right)}{\sqrt{\frac{2}{L}}\sin\left(\frac{\pi}{L}x\right)}
    =-\frac{\pi}{L}\cot\left(\frac{\pi}{L}x\right)
\end{equation*}
And the partner potential $V^{(2)}(x)$ is given by:
\begin{equation}\label{boxpartner}
    V^{(2)}(x)
    =W'(x)+W(x)^2
    =\frac{\pi^2}{L^2}\left(1+2\cot\left(\frac{\pi}{L}x\right)^2\right)
\end{equation}
By \Cref{susyladder}, the $n$\textsuperscript{th} excited state of the partner is:
\begin{align*}
    \psi_n^{(2)}(x)
    &=\frac{1}{\sqrt{E_{n+1}^{(1)}}}A\psi_{n+1}^{(1)}(x)
    =\frac{L}{\sqrt{(n+1)(n+3)}\pi}\left({\psi_{n+1}^{(1)}}'(x)+W(x)\psi_{n+1}^{(1)}(x)\right) \\
    &=\sqrt{\frac{2}{(n+1)(n+3)L}}\left((n+2)\cos\left(\frac{(n+2)\pi}{L}x\right)-\cot\left(\frac{\pi}{L}x\right)\sin\left(\frac{(n+2)\pi}{L}x\right)\right)
\end{align*}

\begin{center}
\begin{tikzpicture}[scale=0.8]
\pgfmathsetmacro{\L}{3}

\begin{scope}[shift={(0,0)}] 
\pgfmathdeclarefunction{psi}{2}{\pgfmathparse{sqrt(2/\L)*sin((#2+1)*180/\L*#1)}}; 

\draw[latex-latex,very thick] (0,5)--(0,0)--(\L,0)--(\L,5);
\node at ({\L-0.5},5) {$V^{(1)}$};
\foreach \n/\c in {0/red,1/Dandelion,2/Green,3/blue,4/violet}
{\draw[thick,smooth,\c,domain=0.01:\L,samples=200] plot(\x,{0.5*psi(\x,\n)+\n}); 
\node[left,\c] at (0,\n) {$\psi_\n^{(1)}$};}
\end{scope}

\begin{scope}[shift={(8,0)}] 
\pgfmathdeclarefunction{psi}{2}{\pgfmathparse{sqrt(2/((#2+1)*(#2+3)*\L))*((#2+2)*cos((#2+2)*180/\L*#1)-cot(180/\L*#1)*sin((#2+2)*180/\L*#1)}}; 
\pgfmathdeclarefunction{V}{1}{\pgfmathparse{1+2*cot(180/\L*#1)^2}} 

\draw (0,5)--(0,0)--(\L,0)--(\L,5);
\node at (0.8,5) {$V^{(2)}$};
\draw[latex-latex,very thick,smooth,domain=0.266:{\L-0.266},samples=20] plot(\x,{0.2*V(\x)}); 
\foreach \n/\c in {0/red,1/Dandelion,2/Green,3/blue,4/violet}
{\draw[thick,smooth,\c,domain=0.01:\L,samples=200] plot(\x,{0.5*psi(\x,\n)+\n}); 
\node[right,\c] at (\L,\n) {$\psi_\n^{(2)}$};}
\end{scope}
\end{tikzpicture}\\
\sffamily The particle in a box and its supersymmetric partner
\end{center}

For each energy level $E_1^{(1)},E_2^{(1)},E_3^{(1)},...$, the partner eigenstate has a similar shape to the original one, with the same number of nodes and antinodes. The partner potential $V^{(2)}$ can be thought of as a `smoothing' of $V^{(1)}$, and its eigenstates resemble those of $V^{(1)}$ but `taper off' near the endpoints (their derivatives are zero at the endpoints).\\

You might wonder if we can get any other meaningful potentials using different choices of $W(x)$. Unfortunately, this method breaks down in the case of a particle in a box. To see this, note that the potential $V(x)$ is constant (at least in each region of the domain), and so the Schrödinger equation in each region is invariant under translation ($x\mapsto x+\alpha$). Thus the general solution for $W(x)$ is simply $W(x)=-\frac{\pi}{L}\cot\left(\frac{\pi}{L}(x+C)\right)$, where $C$ is an arbitrary constant (this is also the solution obtained using \Cref{riccatigs}), and so the partner potential in this case is unique up to translation.

\subsection{The Harmonic Oscillator}\label{subsec:harmonic}
We now investigate the harmonic oscillator. There are several ways to derive the supersymmetric treatment of the harmonic oscillator. One way is to start with the ground state:
\begin{equation}
    \psi_0(x)=e^{-ax^2}
\end{equation}
Where $a>0$ is a constant (specifically $a=\frac{m\omega}{2\hbar}$, where $\omega$ is the angular frequency of the harmonic oscillator). By \eqref{wpsi0}, the superpotential is given by:
\begin{equation*}
    W(x)=-\frac{\psi_0'(x)}{\psi_0(x)}=2ax
\end{equation*}
This yields the following potentials $V^{(1)}$ and $V^{(2)}$:
\begin{align}\label{harmoscv1v2}
    V^{(1)}(x)=-2a+4a^2x^2 && V^{(2)}(x)=2a+4a^2x^2
\end{align}
In a first course on quantum mechanics, the harmonic oscillator potential would be introduced as the parabola $V(x)=\frac{1}{2}m\omega^2x^2$ (or, in our notation, $V(x)=4a^2x^2$), and the ground state energy would be $\frac{1}{2}\hbar\omega$ (in our notation, $2a$). This explains the downward shift of $V^{(1)}$ by $2a$, as the ground state energy of $V^{(1)}$ must be zero by construction.

\newpage 

What is interesting is that $V^{(2)}(x)$ is the same parabola, shifted \emph{upward} by $2a$ instead of downward.

\begin{center}
\begin{tikzpicture}
\draw[latex-latex] (-2.5,0)--(2.5,0) node[right]{$x$};
\draw[latex-latex] (0,-1)--(0,4);
\draw[dashed] (0,-3/4)--(-1/2,-3/4) node[left]{$-2a$};
\draw[thick,smooth,red,domain=-2.5:2.5,samples=200] plot(\x,{(\x)^2/2-3/4});
\node[red] at (2.6,1.6) {$V^{(1)}$};
\draw[dashed] (0,3/4)--(-1/2,3/4) node[left]{$2a$};
\draw[thick,smooth,blue,domain=-2.5:2.5,samples=200] plot(\x,{(\x)^2/2+3/4});
\node[blue] at (1.6,2.8) {$V^{(2)}$};
\end{tikzpicture}\\
\sffamily The harmonic oscillator and its supersymmetric partner
\end{center}

The ground state of $V^{(1)}$, as we have seen, is $\psi_0(x)=e^{-ax^2}$. Using \eqref{adgpsi0}, we have:
\begin{equation*}
    A\dg\psi_0(x)=2W(x)\psi_0(x)=2(2ax)(e^{-ax^2})=4axe^{-ax^2}
\end{equation*}
If we define $\psi_1(x)=xe^{-ax^2}$ and apply $H^{(1)}$ to it, we get:
\begin{align*}
    H^{(1)}\psi_1(x)
    =-\psi_1''(x)+V^{(1)}(x)\psi_1(x)
    =4axe^{-ax^2}
    =4a\psi_1(x)
\end{align*}
Thus $\psi_1$ is also the first excited state of $H^{(1)}$, with energy $4a$.\\

If we apply $A$ and $A\dg$ separately to $\psi_1$, we get:
\begin{align*}
    A\psi_1(x)
    =\psi_1'(x)+W(x)\psi_1(x)
    =e^{-ax^2}
    &&
    A\dg\psi_1(x)
    =-\psi_1'(x)+W(x)\psi_1(x)
    =(4ax^2-1)e^{-ax^2}
\end{align*}
Thus $A\psi_1=\psi_0$ (as expected), while $A\dg\psi_1$ is a new state. If we define $\psi_2(x)=(4ax^2-1)e^{-ax^2}$ and apply $H^{(1)}$ to it, we get:
\begin{align*}
    H^{(1)}\psi_2(x)
    =-\psi_2''(x)+V^{(1)}(x)\psi_2(x)
    =8a(4ax^2-1)e^{-ax^2}
\end{align*}
Thus $\psi_2$ is also the second excited state of $H^{(1)}$, with energy $8a$.\\

We can continue this process to get the entire spectrum of eigenstates of $H^{(1)}$ and $H^{(2)}$. This is summarized in the next theorem:

\begin{theorem}\label{harmonicsoln}
The $n$\textsuperscript{th} excited state of $H^{(1)}$ is given by:
\begin{equation*}
    \psi_n(x)\propto(A\dg)^n\psi_0(x)=(2a)^{\tfrac{n}{2}}H_n\left(\sqrt{2a}x\right)e^{-ax^2}
\end{equation*}
With energy $E_n=4an$. Where $H_n$ is the $n$\textsuperscript{th} \textbf{Hermite polynomial}\footnotemark:
\begin{equation*}
    H_n(z)=(-1)^ne^{z^2}\dv[n]{z}\left(e^{-z^2}\right)
\end{equation*}
\end{theorem}
\vspace{-10pt}
\footnotetext{These are the \emph{physicist's} Hermite polynomials, as opposed to the \emph{probabilist's} Hermite polynomials, which have the same definition with $e^{\pm z^2/2}$ instead of $e^{\pm z^2}$. The function \href{https://functions.wolfram.com/Polynomials/HermiteH/}{\texttt{HermiteH}} in {\fontfamily{ppl}\selectfont\itshape Mathematica} yields the physicist's Hermite polynomials.}
\begin{proof}
Suppose $\psi$ is any eigenstate of $H^{(1)}$. Then we have:
\begin{equation*}
    A\dg\psi(x)
    =-\psi'(x)+2ax\psi(x)
    =-e^{ax^2}\left(e^{-ax^2}\psi'(x)-2axe^{-ax^2}\psi(x)\right)
    =-e^{ax^2}\dv{x}(e^{-ax^2}\psi(x))
\end{equation*}
Applying $A\dg$ to $\psi_0$ $n$ times, we get:
\begin{align*}
    (A\dg)^n\psi_0(x)
    &=(-1)^ne^{ax^2}\dv[n]{x}(e^{-ax^2}\psi_0(x)) \\
    &=(-1)^ne^{ax^2}\dv[n]{x}(e^{-2ax^2}) \tag{since $\psi_0(x)=e^{-ax^2}$}
\end{align*}
Substituting $z=\sqrt{2a}x$ and simplifying, we get:
\begin{align*}
    (A\dg)^n\psi_0(x)
    &=(2a)^{\tfrac{n}{2}}(-1)^ne^{\tfrac{z^2}{2}}\dv[n]{z}(e^{-z^2}) \\
    &=(2a)^{\tfrac{n}{2}}\left((-1)^ne^{z^2}\dv[n]{z}(e^{-z^2})\right)e^{-\tfrac{z^2}{2}}
    =(2a)^{\tfrac{n}{2}}H_n(z)e^{-\tfrac{z^2}{2}} \\
    &=(2a)^{\tfrac{n}{2}}H_n\left(\sqrt{2a}x\right)e^{-ax^2}
\end{align*}
We are left to prove that $(A\dg)^n\psi_0$ is the $n$\textsuperscript{th} excited state of $H^{(1)}$. It suffices to prove that $\psi_{n+1}\propto A\dg\psi_n$ (the result then follows by induction).
\begin{equation*}
    H^{(1)}A\dg\psi_n
    =A\dg AA\dg\psi_n
    =A\dg(AA\dg)\psi_n
    =A\dg H^{(2)}\psi_n
\end{equation*}
By \eqref{harmoscv1v2}, we have $H^{(2)}=H^{(1)}+4a$. This yields:
\begin{equation*}
    A\dg H^{(2)}\psi_n
    =A\dg(H^{(1)}+4a)\psi_n
    =A\dg H^{(1)}\psi_n+4aA\dg\psi_n
    =E_n^{(1)}A\dg\psi_n+4aA\dg\psi_n
    =(E_n^{(1)}+4a)A\dg\psi_n
\end{equation*}
Thus $A\dg\psi_n$ is an eigenstate of $H^{(1)}$ with energy $E_n^{(1)}+4a$. Since $H^{(2)}=H^{(1)}+4a$, the difference between consecutive energy levels must be $4a$. Thus $A\dg\psi_n$ is the $(n+1)$\textsuperscript{st} excited state of $H^{(1)}$, i.e. $\psi_{n+1}\propto A\dg\psi_n$.\\

Finally, since the ground state energy is zero (by definition), the energy of the $n$\textsuperscript{th} excited state of $H^{(1)}$ is given by $E_n=4an$.
\end{proof}
\begin{remark}
The normalized eigenstates are given by $\psi_n(x)=\tfrac{1}{\sqrt{2^nn!}}\left(\tfrac{2a}{\pi}\right)^{1/4}H_n\left(\sqrt{2a}x\right)e^{-ax^2}$.
\end{remark}

The previous theorem shows that all the eigenstates of $H^{(1)}$ can be obtained by repeatedly applying $A\dg$ to the ground state. This is no surprise, as $A\dg$ is simply the creation (or raising) operator for the harmonic oscillator. What is interesting for us is that $V^{(2)}$ is \emph{exactly} the same as $V^{(1)}$, except for an upward shift by $4a$. This means that the harmonic oscillator is not only shape invariant \eqref{shapeinvariance}, but satisfies a stronger form of invariance where $f(a)=a$. We will now show that the harmonic oscillator is the \emph{only} potential with this property.

\begin{theorem}
The harmonic oscillator is the only shape invariant potential where $f(a)=a$.
\end{theorem}
\begin{proof}
Suppose $f(a)=a$. Then we have:
\begin{align*}
    -W'(x,a)+W(x,a)^2=V^{(1)}(x,a) &&
    W'(x,a)+W(x,a)^2=V^{(1)}(x,a)+R(a)
\end{align*}
Taking the sum and difference of these equations, we get:
\begin{align*}
    2W(x,a)^2=2V^{(1)}(x,a)+R(a) &&
    2W'(x,a)=R(a)
\end{align*}
Since $R(a)$ is independent of $x$, the second equation implies $W(x,a)=\frac{R(a)}{2}x$ (plus a constant, which we can set to zero by translating $x$). Substituting this into the first equation, we get:
\begin{equation*}
    \frac{R(a)^2}{2}x^2=2V^{(1)}(x,a)+R(a) \qquad\therefore\qquad V^{(1)}(x,a)=\frac{R(a)^2}{4}x^2-\frac{R(a)}{2}
\end{equation*}
This is precisely the harmonic oscillator, where $R(a)=4a$ (any other function $R(a)$ would simply correspond to a reparametrization of $a$).
\end{proof}

We conclude this section with a treatment of the harmonic oscillator using the supersymmetric WKB approximation from \Cref{subsec:swkb}. In particular, we will show that the first-order approximation gives the exact energy levels for the harmonic oscillator.

\begin{theorem}\label{harmonicwkb}
The supersymmetric WKB approximation (to first order) yields the exact energy levels of the harmonic oscillator.
\end{theorem}
\begin{proof}
We first solve for the classical turning points (which we will call $x_1$ and $x_2$, since $a$ has already been used):
\begin{equation*}
    W(x_1)^2=W(x_2)^2=E_n \qquad\Longrightarrow\qquad x_1=-\frac{\sqrt{E_n}}{2a} \ , \ x_2=\frac{\sqrt{E_n}}{2a}
\end{equation*}
With this, \eqref{swkb} becomes:
\begin{equation*}
    \int_{-\tfrac{\sqrt{E_n}}{2a}}^{\tfrac{\sqrt{E_n}}{2a}} \sqrt{E_n-(2ax)^2}\dx=n\pi
\end{equation*}
The integral on the left simplifies to:
\begin{equation*}
    \frac{\pi}{4a}E_n=n\pi
\end{equation*}
Thus $E_n=4an$, which is exactly what we had from \Cref{harmonicsoln}.
\end{proof}

\subsection{The Hydrogen Atom}\label{subsec:hydatom}
We will now investigate the hydrogen atom using supersymmetry. In non-relativistic quantum mechanics, the hydrogen atom is a special case of a particle moving in a central (spherically symmetric) potential. After separating variables in the Schrödinger equation, we get the following equation for the radial component of the wavefunction:
\begin{equation}\label{centralpot}
    =-\eta^2\psi''(r)+\frac{l(l+1)\eta^2}{r^2}\psi(r)+V(r)\psi(r)=E\psi(r)
\end{equation}
Where $l$ is the orbital quantum number, i.e. $l=0,1,2,...$ correspond to the $s,p,d,...$ orbitals respectively. The Coulomb potential for an electron orbiting a proton is given by:
\begin{equation}\label{coulomb}
    V(r)=-\frac{e^2}{4\pi\ep_0r}
\end{equation}

\bubble[24]{Orchid}{From now on, we will denote the constant $\frac{e^2}{4\pi\ep_0}$ by $\Lam$.} 

We will not set $\Lam$ or $\eta$ equal to $1$ as we want to make comparisons with physical constants along the way, to check that out solution agrees with that of standard QM.\\

We can set the effective potential $V^{(1)}(r)$ by substituting \eqref{coulomb} into \eqref{centralpot} and subtracting the ground state energy $E_0$ (to be determined):
\begin{equation}\label{hydatom}
    V^{(1)}(r)=-\frac{\Lam}{r}+\frac{l(l+1)\eta^2}{r^2}-E_0
\end{equation}
Substituting this into \eqref{ricsuperpot}, we get:
\begin{equation*}
    -\eta W'(r)+W(r)^2=-\frac{\Lam}{r}+\frac{l(l+1)\eta^2}{r^2}-E_0
\end{equation*}
To solve for the superpotential $W(r)$, we guess a solution of the form $W(r)=A+\frac{B}{r}$, where $A$ and $B$ are constants. This yields:
\begin{equation*}
    V^{(1)}(r)
    =-\eta W'(r)+W(r)^2
    =-\eta\left(-\frac{B}{r^2}\right)+A^2+\frac{2AB}{r}+\frac{B^2}{r^2}
    =A^2+\frac{2AB}{r}+\frac{B^2+\eta B}{r^2}
\end{equation*}
Matching coefficients and solving, we get:
\begin{align}\label{hydatomAB}
    A^2=-E_0 && 2AB=-\Lam && B^2+\eta B=l(l+1)\eta^2 &&\Longrightarrow&& \fbox{$A=\dfrac{\Lam}{2(l+1)\eta}$ \qquad $B=-(l+1)\eta$}
\end{align}
We will ignore the other solution ($A=-\frac{\Lam}{2l\eta}$, $B=l\eta$) as we want a meaningful partner potential \eqref{hydatompartner} even when $l=0$. We can now find the ground state energy:
\begin{equation}\label{hydatomgse}
    E_0=-A^2
    =-\frac{\Lam^2}{4(l+1)^2\eta^2}
\end{equation}
The constant of proportionality is $\frac{\Lam^2}{4\eta^2}=\frac{m_e e^4}{32\pi^2\hbar^2\ep_0^2}\approx 13.6$ eV. Thus, setting $l=0$ above, we get that the ground state energy of the electron in a hydrogen atom is the famous $-13.6$ eV (the \emph{Rydberg energy}).\\

\noindent As for the partner potential $V^{(2)}(r)$, we have:
\begin{align}\label{hydatompartner}
    V^{(2)}(r)
    &=\eta W'(r)+W(r)^2
    =\eta\left(-\frac{B}{r^2}\right)+A^2+\frac{2AB}{r}+\frac{B^2}{r^2}
    =A^2+\frac{2AB}{r}+\frac{B^2-\eta B}{r^2} \nonumber\\
    &=\frac{\Lam^2}{4(l+1)^2\eta^2}-\frac{\Lam}{r}+\frac{(l+1)(l+2)\eta^2}{r^2}
\end{align}
This is interesting in its own right: The constant and $\frac{1}{r}$ terms are unchanged, while the coefficient of $\frac{1}{r^2}$ has been raised (this is reminiscent of the inverse square potential from \Cref{subsec:invsqpot}). This shows that $V^{(1)}$ is a shape invariant potential. Additionally, its partner potential $V^{(2)}$ is also the effective potential of the hydrogen atom, but with its orbital quantum number $l$ increased by $1$ (so going from $V^{(1)}$ to $V^{(2)}$, we are stepping up from the $s$-shell to the $p$-shell, or the $p$-shell to the $d$-shell, etc).\\

We can now use the tools of shape invariant potentials we developed in \Cref{subsec:shapeinvariance} to solve the hydrogen atom. Motivated by \eqref{shapeinvariance}, we first recast $V^{(1)}$ and $V^{(2)}$ in the form:
\begin{equation}
    V^{(2)}(r,l)=V^{(1)}(r,f(l))+R(l)
\end{equation}
Where $f(l)$ and $R(l)$ are functions of $l$ only (which we need to solve for). Substituting \eqref{hydatomgse} into \eqref{hydatom} and comparing with \eqref{hydatompartner}, we see that:
\begin{align}
    V^{(1)}(r,l)=\frac{\Lam^2}{4(l+1)^2\eta^2}-\frac{\Lam}{r}+\textcolor{red!80!black}{\frac{l(l+1)\eta^2}{r^2}}
    &&
    V^{(2)}(r,l)=\frac{\Lam^2}{4(l+1)^2\eta^2}-\frac{\Lam}{r}+\textcolor{red!80!black}{\frac{(l+1)(l+2)\eta^2}{r^2}}
\end{align}
The \textcolor{red!80!black}{last term} of each equation suggests that $f(l)=l+1$. Substituting this into $V^{(1)}$ above, we get:
\begin{equation}
    V^{(1)}(r,l+1)=\frac{\Lam^2}{4(l+2)^2\eta^2}-\frac{\Lam}{r}+\frac{(l+1)(l+2)\eta^2}{r^2}
\end{equation}
This yields:
\begin{equation}
    R(l)
    =\frac{\Lam^2}{4\eta^2}\left(\frac{1}{(l+1)^2}-\frac{1}{(l+2)^2}\right)
    =\frac{\Lam^2}{4\eta^2}\frac{2l+3}{(l+1)^2(l+2)^2}
\end{equation}
We are now ready to solve for the entire energy spectrum of the hydrogen atom. By \Cref{sipenergy} and the fact that $f^k(l)=f(f(\cdots f(l)\cdots))=l+k$, the energy of the $n$\textsuperscript{th} excited state is given by:
\begin{align*}
    E_n^{(1)}
    =E_0^{(n+1)}
    &=E_0^{(1)}+\sum_{k=0}^{n-1} R(l+k) \\
    &=-\frac{\Lam^2}{4(l+1)^2\eta^2}+\frac{\Lam^2}{4\eta^2}\sum_{k=0}^{n-1} \left(\frac{1}{(l+k+1)^2}-\frac{1}{(l+k+2)^2}\right) \\
    &=-\frac{\Lam^2}{4(l+n+1)^2\eta^2} \tag{telescoping sum}
\end{align*}
\begin{warning}
The number $n$ above is the \emph{radial} quantum number, NOT the principal quantum number. Here we are using $n$ to index the excited states, regardless of the value of $l$. So under our convention, the principal quantum number is $l+n+1$, which explains the factor $(l+n+1)^2$ in the denominator of $E_n$.
\end{warning}

Using \eqref{psi0w}, the ground state of $H^{(1)}$ (which we will now denote by $\psi_{0,l}^{(1)}$) is given by:
\begin{align*}
    \psi_{0,l}^{(1)}(r)
    &=C\exp\left(-\frac{1}{\eta}\int W(r)\,dr\right) \\
    &=C\exp\left(-\frac{1}{\eta}\int \left(\frac{\Lam}{2(l+1)\eta}-\frac{(l+1)\eta}{r}\right)\,dr\right) \tag{using $A$ and $B$ from \eqref{hydatomAB}} \\
    &=Cr^{l+1}e^{-\tfrac{\Lam}{2(l+1)\eta^2}r}
\end{align*}
Where $C$ is a normalization constant. The constant in the exponential is $\frac{\Lam}{2\eta^2}=\frac{e^2m_e}{4\pi\ep_0\hbar^2}=\frac{1}{a_0}$, where $a_0\approx 5.29\times10^{-11}$ m is the Bohr radius. This agrees with the solution of the $1s_0$ electron in standard QM (with the normalization constant $C=2a_0^{-\frac{3}{2}}$). The transitions of electrons from higher-energy states to this state produce an emission spectrum in the ultraviolet range, known as the \href{https://www.chemeurope.com/en/encyclopedia/Lyman_series.html}{Lyman series}.\\

We will now solve for the first excited state $\psi_{1,l}^{(1)}(r)$. Setting $m=2$ in \Cref{sipenergy}, we have $\psi_{0,l}^{(2)}=\psi_{0,l+1}^{(1)}$, i.e. the ground state\footnote{You might be wondering whether this is really the \emph{ground} state of $H^{(2)}$. If there were a lower energy state, we could use shape invariance to `climb down the ladder' and get a lower energy state of $H^{(1)}$, but by definition, $\psi_0^{(1)}$ is the ground state of $H^{(1)}$.} of $H^{(2)}$ is obtained from the ground state of $H^{(1)}$ by replacing $l$ with $f(l)=l+1$. In other words:
\begin{equation*}
    \psi_{0,l}^{(2)}(r)=\psi_{0,l+1}^{(1)}(r)=\textcolor{red!80!black}{r^{l+2}e^{-\tfrac{\Lam}{2(l+2)\eta^2}r}}
\end{equation*}
We have ignored the normalization constant here, we will include it at the end. Setting $n=0$ in the last equation of \Cref{susyladder}, we have:
\begin{equation*}
    \psi_{1,l}^{(1)}(r)=A\dg\psi_{0,l}^{(2)}(r)
\end{equation*}
The operator $A\dg$ here is given by:
\begin{equation*}
    A\dg=-\eta\dv{r}+W(r)=\textcolor{blue!80!black}{-\eta\dv{r}+\frac{\Lam}{2(l+1)\eta}-\frac{(l+1)\eta}{r}}
\end{equation*}
Putting it all together, we get:
\begin{align*}
    \psi_{1,l}^{(1)}(r)
    &=A\dg\psi_{0,l}^{(2)}(r) \\
    &=\left(\textcolor{blue!80!black}{-\eta\dv{r}+\frac{\Lam}{2(l+1)\eta}-\frac{(l+1)\eta}{r}}\right)\left(\textcolor{red!80!black}{r^{l+2}e^{-\tfrac{\Lam}{2(l+2)\eta^2}r}}\right) \\
    &\vdots \\ 
    &=\left(\frac{(2l+3)\Lam}{2(l+1)(l+2)\eta}r^{l+2}-(2l+3)\eta r^{l+1}\right)e^{-\tfrac{\Lam}{2(l+2)\eta^2}r}
\end{align*}
To get the first excited state of the $s$-orbital, i.e. the $2s_0$ electron, we substitute $l=0$:
\begin{equation*}
    \psi_{1,0}^{(1)}(r)
    =\left(\frac{3\Lam}{4\eta}r^2-3\eta r\right)e^{-\tfrac{\Lam}{4\eta^2}r}
    =-3\eta\left(r-\frac{\Lam}{4\eta^2}r^2\right)e^{-\tfrac{\Lam}{4\eta^2}r}
\end{equation*}
Substituting the Bohr radius $a_0=\frac{2\eta^2}{\Lam}$ and including the normalization constant, we get:
\begin{equation*}
    \psi_{1,0}^{(1)}(r)=2(2a_0)^{-\tfrac{3}{2}}\left(r-\frac{r^2}{2a_0}\right)e^{-\tfrac{r}{2a_0}}
\end{equation*}
Which again agrees with the solution of the hydrogen atom in standard QM. The transitions of electrons from higher-energy states to this state produce an emission spectrum in the visible and ultraviolet ranges, known as the \href{https://www.chemeurope.com/en/encyclopedia/Balmer_series.html}{Balmer series}.\\

Solving for \emph{all} the eigenstates is extremely tedious, and requires the use of \href{https://functions.wolfram.com/Polynomials/LaguerreL3/}{associated Laguerre polynomials}. As such, we will only outline the process here. We first make use of shape invariance to get the $k$\textsuperscript{th} superpotential (which we will now denote by $W_k(r,l)$):
\begin{align*}
    W_1(r,l)&=W(r,l)=\frac{\Lam}{2(l+1)\eta}-\frac{(l+1)\eta}{r} \\
    \therefore\quad W_k(r,l)&=W_1(r,l+k)=\frac{\Lam}{2(l+k)\eta}-\frac{(l+k)\eta}{r}
\end{align*}
Substituting this into \eqref{superpotansatz}, we get:
\begin{equation*}
    A_k\dg=-\eta\dv{r}+W_k(r,l)=-\eta\dv{r}+\frac{\Lam}{2(l+k)\eta}-\frac{(l+k)\eta}{r}
\end{equation*}
Finally, to get the eigenstates, we can use \eqref{iterpartnerstate2} in the following form:
\begin{align*}
    \psi_n^{(1)}
    &=A_1\dg A_2\dg \cdots A_n\dg \psi_{0}^{(n+1)}
\end{align*}
Where we have left out the constant factor as we would have to compute the normalization constant in the end anyway. By \Cref{sipenergy}, we have $\psi_0^{(n+1)}(r,l)=\psi_0^{(1)}(r,f^n(l))=\psi_0^{(1)}(r,l+n)$ (here $\psi_{\alpha,l}^{(\beta)}(r)$ and $\psi_\alpha^{(\beta)}(r,l)$ mean the same thing). The emission spectra for transitions to $n=2,3,4,...$ are known as the \href{https://www.chemeurope.com/en/encyclopedia/Paschen_series.html}{Paschen series}, \href{https://www.chemeurope.com/en/encyclopedia/Brackett_series.html}{Brackett series}, \href{https://www.chemeurope.com/en/encyclopedia/Pfund_series.html}{Pfund series} and so on. These frequencies all lie in the infrared range and below.\\

We have finally solved for the radial component of the wavefunctions of the hydrogen atom. To get the angular component, we would have to solve for the spherical harmonics using \href{https://functions.wolfram.com/HypergeometricFunctions/LegendreP2General/}{associated Legendre polynomials}, as one would in standard QM, and then multiply these functions together (which is valid as the Schrödinger equation for the hydrogen atom is separable in spherical coordinates, due to its $SO(3)$ symmetry). However, finding the radial component is already a large portion of the work required, which we have been able to greatly simplify using SUSY QM.\\

There is also an alternative treatment of the hydrogen atom using its $SO(4)$ symmetry, see \cite{weinberg}.

\section{SUSY QM with Matrices and Operators}\label[chapter]{sec:matricesop}
In \Cref{sec:schrodinger}, we investigated the idea of decomposing a Hamiltonian $H$ into $A\dg A$. In \Cref{sec:systems}, we looked at several key examples of this factorization and how it can be used to solve quantum systems. Now, we will translate these ideas into the language of matrices and operators on Hilbert spaces and explore the analysis of SUSY QM. We will also explain, through Python simulations (see \Cref{sec:numerical}), how SUSY QM can be used to set the stage for the \hyperref[eth]{eigenstate thermalization hypothesis}.

\subsection{The Matrix Picture}
We will start by recasting the building blocks of SUSY QM in terms of matrices. This will allow us to perform explicit computations and get a feel of what to expect in the general case.\\

We began \Cref{sec:schrodinger} with the Schrödinger Hamiltonian $H=-\eta^2\dv[2]{x}+V(x)$. In the matrix picture, the potential $V(x)$ would be a diagonal matrix, while the derivative operator would be as follows:
\begin{align}\label{derivativematrix}
    \partial=\frac{1}{2}
    \begin{pNiceMatrix}
        0 & 1 & & \Block{2-2}<\Large>{0} & \\
        -1 & \Ddots & \Ddots & & \\
        & \Ddots & & & \\
        \Block{2-2}<\Large>{0} & & & & 1 \\
        & & & -1 & 0
    \end{pNiceMatrix}
\end{align}
This comes from the central finite difference approximation:
\begin{equation*}
    \psi'(x)\approx\frac{\psi(x+h)-\psi(x-h)}{2h} \qquad\Longrightarrow\qquad (\partial\psi)_n=\frac{\psi_{n+1}-\psi_{n-1}}{2}
\end{equation*}
We use the central finite difference as it ensures the operator $-i\partial$ is self-adjoint (i.e. a Hermitian matrix).
\begin{align}\label{hamiltonianmatrix}
    -\partial^2+V=
    \begin{pNiceMatrix}[columns-width=0mm]
        \frac{1}{4}+v_1 & 0 & -\frac{1}{4} & \Block{2-2}<\large>{0} & \\
        0 & \frac{1}{2}+v_2 & \Ddots & \Ddots & \\
        -\frac{1}{4} & \Ddots & \Ddots & & -\frac{1}{4} \\
        \Block{2-2}<\large>{0} & \Ddots & & \frac{1}{2}+v_{n-1} & 0 \\
        & & -\frac{1}{4} & 0 & \frac{1}{4}+v_n
    \end{pNiceMatrix}
\end{align}
Again, we have set $\eta=1$. While this matrix \emph{can} be factored into $A\dg A$, it is generally very difficult to determine the ground state energy (the smallest eigenvalue), and this would prevent us from building up the supersymmetric ladder. Even in the simplest case $V\equiv0$), the ground state energy is zero when $n$ is odd and positive when $n$ is even (but tends to zero as $n\to\infty$).

\subsection{Numerical Supersymmetric Quantum Mechanics}\label{subsec:numericalsusyqm}
In 1902, \href{https://mathshistory.st-andrews.ac.uk/Biographies/Gibbs/}{Josiah Willard Gibbs} (1839--1903) defined the notion of a canonical ensemble of systems \cite{gibbs}. This was in some sense a precursor to the modern concepts used to describe systems in statistical mechanics via measure spaces. 30 years later, \href{https://mathshistory.st-andrews.ac.uk/Biographies/Von_Neumann/}{John von Neumann} (1903--1957) published his mathematical formalization of quantum mechanics \cite{neumann}, which set foundation for the modern treatment via operators on Hilbert spaces.\\

The \emph{mathematical} interplay between measure theory and functional analysis has long been known, and its applications stretch deep into other areas of mathematics, such as Fourier analysis, statistical sampling and chaos theory. However, the \emph{physical} interplay between these fields is much more elusive, and was not fully appreciated until other formulations of quantum mechanics came about, such as the $C^*$-algebra formulation \cite{sekhon}.\\

The \textbf{eigenstate thermalization hypothesis} (ETH) is a conjecture first proposed in 1994 by Mark Srednicki \cite{srednicki}. It seeks to justify the use of equilibrium statistical mechanics to study many-body quantum systems. Here, we will focus on a quantum system of finite extent with no degeneracy in its eigenstates\footnote{That is to say, the Hamiltonian is a matrix of finite size with all eigenvalues distinct.}.

\begin{cthm}[Eigenstate Thermalization Hypothesis]\label{eth}
Suppose $A$ is an observable of a quantum system (of finite extent, with no degeneracy) and $A_{ij}$ is its matrix in a basis of eigenstates of the system. Suppose the entries $A_{ij}$ satisfy the following conditions:
\begin{enumerate}[ref={(\arabic*)}]
    \item\label{eth:1} The diagonal entries $A_{ii}$ vary continuously with the energies of the eigenstates, and the difference $A_{i+1,i+1}-A_{ii}$ between consecutive diagonal entries is small compared to the size of the system.
    \item\label{eth:2} The off-diagonal entries $A_{ij}$ ($i\ne j$) are small compared to the diagonal entries, and also small compared to the size of the system.
\end{enumerate}
Then the expected value of $A$ will evolve to its value predicted by a canonical ensemble.
\end{cthm}
 
Condition \ref{eth:2} signifies that $A$ is `almost' a diagonal matrix, and so its diagonal entries also serve as approximations of its eigenvalues. Condition \ref{eth:1} signifies that as we move along the diagonal, the entries vary only by small amounts. By `small', we mean they vanish at least as fast as $\frac{1}{\sqrt{n}}$, where $n$ is the size of the matrix. This is a consequence of the limiting behavior of the variation in the entries\footnote{Put simply, it is due to the factor $\frac{1}{\sqrt{n}}$ that appears in the central limit theorem: sample mean $\sim\text{N}\big(\mu,\frac{\sigma}{\sqrt{n}}\big)$.}.\\

For more information on the eigenstate thermalization hypothesis, see \cite{statmechvideo}, \cite{chen} and \cite{dalessio}.\\

We will now show how SUSY QM, specifically the construction of supersymmetric partners of random Hamiltonians, naturally leads to the framework of the ETH described above. To do this, we first need to devise a way of computing supersymmetric partners numerically.

\begin{tcolorbox}[colback=white,colframe=Orchid!60,title=Finding a supersymmetric partner numerically,fonttitle=\sffamily\bfseries,coltitle=black,top=1mm,bottom=1mm,left=1mm,right=1mm,sharp corners,before skip=10pt,after skip=10pt]
\begin{enumerate}
    \item Start with a real symmetric\footnotemark\ matrix $H_1$ (not necessarily of the form \eqref{hamiltonianmatrix}).
    \item Find its ground state energy (its lowest eigenvalue) $E_0$ and compute $H_{1a}=H_1-E_0I$. This matrix has zero ground state energy.
    \item Use Cholesky decomposition to express $H_{1a}$ as $H_{1a}=A\dg A$, where $A$ is an upper triangular matrix.
    \item Compute $H_{2a}=AA\dg$.
    \item Add the original ground state energy $E_0$ back to $H_{2a}$.
\end{enumerate}
The matrix $H_2=H_{2a}+E_0I$ is the supersymmetric partner of $H_1$.
\end{tcolorbox}
\footnotetext{While any Hermitian matrix will do, it suffices for our purposes to focus on real symmetric matrices. This is also computationally less expensive, and not subject to additional floating-point errors introduced by dividing complex numbers.}

Unfortunately, this method cannot be used to find partner potentials. This is because the way we constructed the partner potentials relies on the supersymmetric property: The ground state of $H^{(1)}$ is unmatched, while every other eigenstate of $H^{(1)}$ is matched with an eigenstate of $H^{(2)}$ with the same energy.\\

If we implement this method anyway, we simply get a translated copy of the original potential, which demonstrates broken supersymmetry (the jagged edges in the partner potential are due to floating-point errors, and those at the endpoints are due to the asymmetry of the matrix $\partial$ at the endpoints):

\begin{tcolorbox}[blanker,halign=center]\label{sim:broken}
\begin{multicols}{2}

\includegraphics[width=7cm]{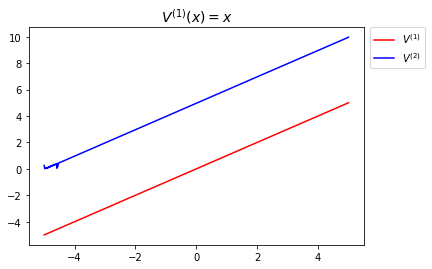}

\includegraphics[width=7cm]{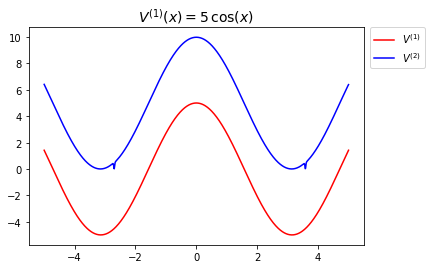}

\end{multicols}
\end{tcolorbox}

However, this method is still useful when dealing with \emph{random} Hamiltonians. Here, we forget about the Schrödinger equation and the potential, and consider \emph{any} Hermitian matrix to be a Hamiltonian\footnote{You might be wondering if every Hermitian matrix can be realized as the Hamiltonian of \emph{something}. This is indeed true, as we can prescribe any $n$ energy levels to an $n$-state system and express it in an arbitrary basis.}. For computational simplicity, however, we will stick to real symmetric matrices.

\newpage 

We will now generate a random real symmetric matrix and repeatedly apply the above algorithm to it. The entries of the matrix are drawn from a uniform distribution with mean $0$. The standard deviation is set so that, on average, 95\% of the entries\footnote{The entries of the original matrix, not the entries of the iterations (over which we have no control in general).} lie in the range $[-50,50]$, which is the range specified for the color bar.

\begin{tcolorbox}[blanker,halign=center]\label{sim:colorbar}
\begin{multicols}{3}

\includegraphics[width=5cm]{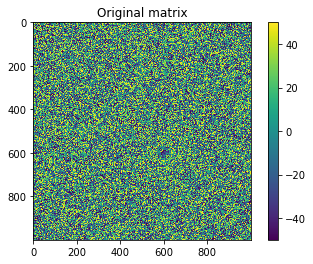}

\includegraphics[width=5cm]{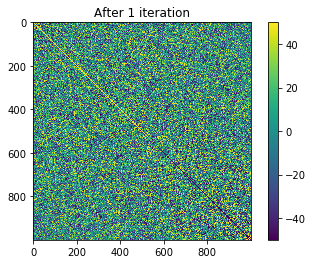}

\includegraphics[width=5cm]{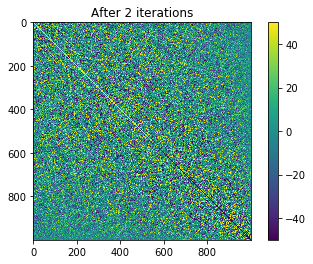}

\end{multicols}
\begin{multicols}{3}

\includegraphics[width=5cm]{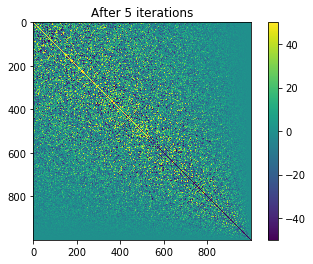}

\includegraphics[width=5cm]{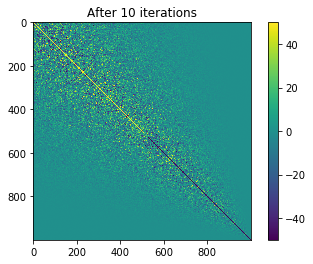}

\includegraphics[width=5cm]{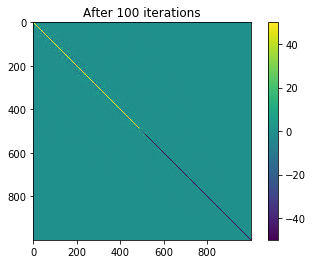}

\end{multicols}
\end{tcolorbox}

After one iteration, we can already see a distinct diagonal in the matrix. With more iterations, the diagonal becomes more pronounced as the `dust' around it clears up. We can make sense of the numbers on the diagonal as approximating the eigenvalues of the matrix\footnote{This is reminiscent of other eigenvalue-finding methods, such as the QR algorithm (which uses Gram-Schmidt orthogonalization) and the Jacobi algorithm (which uses Givens rotations).} (since the eigenvalues of a diagonal matrix are precisely its diagonal entries). We also observe the following:

\vspace*{10pt}

\begin{itemize}
    \item The upper left half of the diagonal consists of positive entries, while the lower right half consists of negative entries. This makes sense, as the entries of the original matrix were drawn from a distribution symmetric about $0$, and so we expect half the eigenvalues to be positive and half to be negative. We might also expect them to follow a semicircle distribution, according to Wigner's semicircle law \cite{wigner,wigner2}, but this only holds for the original (randomly generated) matrix, not its iterates.
    \item The dust clears up more slowly near the upper left corner than the lower right corner. In other words, the approximations for the positive eigenvalues take longer to converge than those for the negative eigenvalues. This makes sense, as the final step of each iteration is to add back the lowest eigenvalue (which is negative), so before this step, the eigenvalues near the upper left corner would be large and positive, while those near the lower right corner would be close to zero (for example, if they range from $30$ to $-30$ in the end, they would have ranged from $60$ to $0$ just before the final step, so the dust near $60$ would be larger than the dust near $0$).
\end{itemize}

\vspace*{10pt}

It is known that the largest eigenvalue of an $n\times n$ random real symmetric matrix is $O(\sqrt{n})$, see \cite{furedi} and \cite{meckes}. So we expect the difference between consecutive diagonal entries (which approximate the eigenvalues as we perform more iterations) to be $O\big(\frac{\sqrt{n}}{n}\big)=O\big(\frac{1}{\sqrt{n}}\big)$, which is suggestive of the assumptions of ETH.

\newpage 

We are now left to show that the iterates really do satisfy the assumptions of ETH. We first look at condition \ref{eth:1} for the diagonal entries. Below is a plot of the diagonals of the original matrix, as well as its 1\textsuperscript{st}, 2\textsuperscript{nd}, 3\textsuperscript{rd} and 100\textsuperscript{th} iterates. The intermediate ones are omitted to avoid clutter in the plot.

\begin{center}\label{sim:diagonaldecreasing}
    \includegraphics[width=9cm]{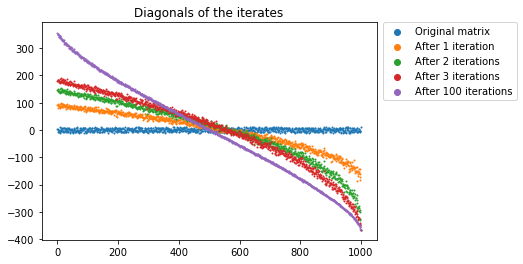}
\end{center}

As can be seen above, with more iterations, the diagonal entries line up in decreasing order from upper-left to lower-right. This allows us to estimate the average absolute difference $\abs{A_{i+1,i+1}-A_{ii}}$ with a much simpler quantity:
\begin{equation}\label{avgdiagdiff}
    \abs{A_{i+1,i+1}-A_{ii}}\approx\frac{A_{11}-A_{nn}}{n-1}
\end{equation}
Where $A_{11}$ is the upper-leftmost entry of $A$ and $A_{nn}$ is the lower-rightmost. We will make use of this fact in the next simulation.\\

We will now verify condition \ref{eth:1} of ETH by plotting the average difference between consecutive diagonal entries, given by \eqref{avgdiagdiff}, against the size of the matrix:

\begin{center}\label{sim:ethcondition1}
    \includegraphics[width=10.5cm]{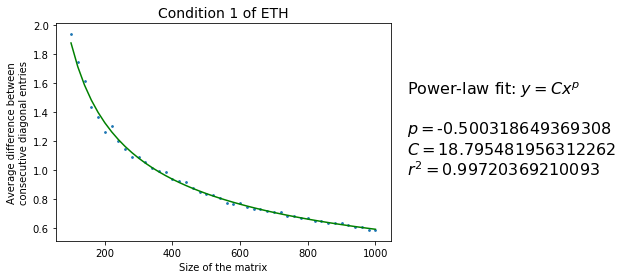}
\end{center}

As we can see, the average difference is proportional to $n^{-0.500}\approx\frac{1}{\sqrt{n}}$, with a very strong correlation ($r^2\approx0.997$). This verifies condition \ref{eth:1}.\\

We now turn to condition \ref{eth:2} for the off-diagonal entries. It is clear from the color bar plots that they are small compared to the diagonal entries (the `dust' clears up with more iterations). We will now show that they are also small compared to the size of the matrix, by plotting the average absolute value of all the off-diagonal entries\footnote{We use the average, rather than the maximum, as there are typically some outliers that take more iterations to fade away.} against the size:

\begin{center}\label{sim:ethcondition2}
    \includegraphics[width=10.5cm]{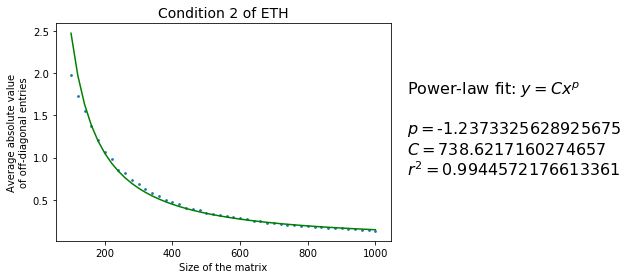}
\end{center}

As we can see, the average absolute off-diagonal entry is proportional to $n^{-1.237}\ll\frac{1}{\sqrt{n}}$, with a very strong correlation ($r^2\approx0.994$). This verifies condition \ref{eth:2}.\\

All in all, we have seen that by starting with a random real symmetric matrix and repeatedly taking supersymmetric partners, we end up with precisely the framework needed for ETH! While this method may not be useful in analyzing microcanonical ensembles directly, it is still remarkable that SUSY QM has led us to a conclusion in statistical mechanics.

\subsection{Self-adjoint Operators on Hilbert Spaces}\label{subsec:selfadjoint}
We will conclude this chapter with a brief discussion of boundary conditions and their effect on supersymmetric treatment of quantum systems. It was essential to all our work that the Hamiltonian was a self-adjoint operator. As we will see shortly, the precise definition of ``self-adjoint'' necessitates some care in defining our operators, particularly in choosing suitable boundary conditions for our systems.

\bubble{Orchid}{\textsf{Reminder:} We take inner products to be linear in the first argument, and conjugate linear in the second.}

\begin{definition}
Suppose $\Hs$ is a Hilbert space. An \textbf{operator} on $\Hs$ is a pair $(T,\Ds(T))$, where $\Ds(T)$ is a subspace of $\Hs$ and $T:\Ds(T)\to\Hs$ is a linear map.
\end{definition}
\begin{remark}
Some sources call this an \emph{unbounded operator}, to specify that it is not necessarily defined on all of $\Hs$. This is a red herring: Under this convention, an ``unbounded operator'' may also be bounded! As such, we will avoid this and reserve the term ``unbounded'' for operators that are not bounded.
\end{remark}

From now on, where unambiguous, we will simply denote an operator $(T,\Ds(T))$ by $T$.

\begin{definition}
An operator $T:\Ds(T)\to\Hs$ is \textbf{densely defined} if $\Ds(T)$ is dense in $\Hs$.
\end{definition}

\begin{example}
The derivative operator $D:C^1[a,b]\to L^2[a,b]$ is densely defined, since $C^1[a,b]$ is dense in $L^2[a,b]$.
\end{example}

The importance of densely defined operators lies in the construction of the adjoint operator, which we will do shortly. There are, however, other nice properties satisfied by densely defined operators, see \cite{ward} and \cite{weidmann}.

\begin{definition}
Suppose $\Hs$ is a Hilbert space and $T$ is an operator on $\Hs$. The \textbf{graph} of $T$ is given by $\Gs(T)=\{(x,Tx)\mid x\in\Ds(T)\}$.
\end{definition}

\begin{lemma}\label{graphequiv}
A set $G\sub\Hs\times\Hs$ is the graph of an operator on $\Hs$ if and only if it is a subspace of $\Hs\times\Hs$ and $(0,y)\notin G$ for any $y\in\Hs$, $y\ne0$.
\end{lemma}
\begin{proof}
($\Rightarrow$) Suppose $G$ is the graph of an operator $T$ on $\Hs$. Suppose $(x_1,y_1),(x_2,y_2)\in G$ and $\lam\in\C$. Then $Tx_1=y_1$ and $Tx_2=y_2$. Since $T$ is linear, we have $T(\lam x_1+x_2)=\lam y_1+y_2$, and so $(\lam x_1+x_2,\lam y_1+y_2)=\lam(x_1,y_1)+(x_2,y_2)\in G$. Thus $G$ is a subspace of $\Hs\times\Hs$. Now suppose $(0,y)\in G$ for some $y\in\Hs$, $y\ne0$. Then we have $T0=y\ne0$, a contradiction as $T$ is linear. Thus $(0,y)\notin G$ for any $y\in\Hs$, $y\ne0$.\\

($\Leftarrow$) Suppose $G$ is a subspace of $\Hs\times\Hs$ and $(0,y)\notin G$ for any $y\in\Hs$, $y\ne0$. Suppose $x\in\Hs$. We will show that there is at most one $y\in\Hs$ such that $(x,y)\in G$. Suppose $(x,y_1),(x,y_2)\in G$. Since $G$ is a subspace of $\Hs\times\Hs$, we have $(x,y_1)-(x,y_2)=(0,y_1-y_2)\in G$. By assumption, this implies $y_1-y_2=0$, i.e. $y_1=y_2$. Thus there is at most one $y\in\Hs$ such that $(x,y)\in G$.\\

Now define $D=\{x\in\Hs\mid (x,y)\in G \text{ for some } y\in\Hs\}$, and define $T:D\to\Hs$ by $Tx=y$. As we have just shown, $T$ is well-defined. We now show that $D$ is a subspace of $\Hs$. Suppose $x_1,x_2\in D$ and $\lam\in\C$. Then there exist $y_1,y_2\in\Hs$ such that $(x_1,y_1),(x_2,y_2)\in G$. This yields $\lam(x_1,y_1)+(x_2,y_2)=(\lam x_1+x_2,\lam y_1+y_2)\in G$, since $G$ is a subspace of $\Hs\times\Hs$. Thus $\lam x_1+x_2\in D$, and so $D$ is a subspace of $\Hs$. We now show that $T$ is linear. Suppose $x_1,x_2\in D$ and $\lam\in\C$. Then there exist $y_1,y_2\in\Hs$ such that $(x_1,y_1),(x_2,y_2)\in G$. This yields $\lam(x_1,y_1)+(x_2,y_2)=(\lam x_1+x_2,\lam y_1+y_2)\in G$, so $T(\lam x_1+x_2)=\lam y_1+y_2$. Thus $T$ is linear. By construction, $G$ is the graph of $T$.
\end{proof}

\begin{definition}
An operator $T$ on $\Hs$ is \textbf{closed} if its graph $\Gs(T)$ is closed in $\Hs\times\Hs$ (with respect to the product topology).
\end{definition}
\begin{remark}
Equivalently, $T$ is closed if for every sequence $(x_n)$ in $\Ds(T)$ such that $(x_n)$ converges to $x\in\Hs$ and $(Ax_n)$ converges to $y\in\Hs$, we have $x\in\Ds(T)$ and $Ax=y$.
\end{remark}

\begin{definition}
Suppose $\Hs$ is a Hilbert space and $S$ and $T$ are operators on $\Hs$. Then $T$ is an \textbf{extension} of $S$ (and $S$ is a \textbf{restriction} of $T$) if $\Ds(T)\sups\Ds(S)$ and $Tx=Sx$ for all $x\in\Ds(S)$.
\end{definition}
\begin{remark}
Equivalently, $T$ is an extension of $S$ if $\Gs(T)\sups\Gs(S)$.
\end{remark}

The closed graph theorem (see \cite[\S10.3.1, Theorem 1, Page 259]{kadets}) states that every closed operator $T:\Hs\to\Hs$ is bounded. As such, we cannot hope to define closed unbounded operators on the entire space $\Hs$, but rather only on a subspace. This is the reason for introducing the domain $\Ds(T)$ as part of the definition of an operator.

\begin{definition}
Suppose $T$ is a densely defined operator on $\Hs$. The \textbf{adjoint} of $T$ is the operator $T\dg:\Ds(T\dg)\to\Hs$ such that $\ip{Tx,y}=\ip{x,T\dg y}$ for all $x\in\Ds(T)$.\\
The domain $\Ds(T\dg)$ is the set of all $y\in\Hs$ such that the function $x\mapsto\ip{Tx,y}$ is continuous on $\Ds(T)$.
\end{definition}

More precisely, the adjoint of $T$ is constructed as follows:
\begin{tcolorbox}[colback=white,colframe=ForestGreen!60,title=Constructing the adjoint of a densely defined operator,fonttitle=\sffamily\bfseries,coltitle=black,top=1mm,bottom=1mm,left=1mm,right=1mm,sharp corners,before skip=10pt,after skip=10pt]
\begin{enumerate}
    \item Fix $y\in\Hs$.
    \item Define the linear functional $f_y:\Ds(T)\to\C$ by $f_y(x)=\ip{Tx,y}$.
    \item If $f_y$ is continuous on $\Ds(T)$ (i.e. if there is a constant $c>0$ such that $\abs{f_y(x)}\le c\norm{x}$ for all $x\in\Ds(T)$), declare $y\in E$. Otherwise, declare $y\notin E$.
    \item If $y\in E$, by the Hahn-Banach theorem, we can extend it to a continuous linear functional $F_y:\Hs\to\C$ on all of $\Hs$. Then, by the Riesz representation theorem, we can find some $z_y\in\Hs$ such that $F_y(x)=\ip{x,z_y}$ for all $x\in\Ds(T)$.
    \item For each $y\in E$, define $T\dg y=z_y$.
\end{enumerate}
Finally, we define $\Ds(T\dg)$ as the set $E$ of all $y\in\Hs$ such that $f_y$ is continuous on $\Ds(T)$, and we define $T\dg:\Ds(T\dg)\to\Hs$ by $T\dg y=z_y$.
\end{tcolorbox}
We need $T$ to be densely defined to ensure the vector $z_y$ is unique for each $y\in\Ds(T\dg)$.

\begin{proposition}\label{adjointclosed}
If $T$ is densely defined, then $T\dg$ is closed.
\end{proposition}
\begin{proof}
Suppose $(y_n)$ is a sequence in $\Ds(T\dg)$ such that $y_n\to y$ and $T\dg y_n\to z$. Then for all $x\in\Ds(T)$, we have $\ip{Tx,y}=\limn\ip{Tx,y_n}=\limn\ip{x,T\dg y_n}=\ip{x,z}$. Thus $y\in\Ds(T\dg)$ and $T\dg y=z$, and so $T\dg$ is closed.
\end{proof}

\begin{definition}
Suppose $T$ is an operator on $\Hs$.
\begin{itemize}
    \item $T$ is \textbf{self-adjoint} if $T\dg=T$.
    \item $T$ is \textbf{Hermitian}\footnotemark\ if $\ip{Tx,y}=\ip{x,Ty}$ for all $x,y\in\Ds(T)$.
\end{itemize}
\end{definition}
\vspace*{-10pt}
\footnotetext{Some sources, especially older analysis texts, call these \emph{symmetric} operators (even though they still play the role of Hermitian matrices, rather than symmetric matrices).}
\begin{remark}
Self-adjointness means two things: $\Ds(T\dg)=\Ds(T)$ \emph{and} $T\dg x=Tx$ for all $x\in\Ds(T)$.
\end{remark}

Every self-adjoint operator is Hermitian, but not every Hermitian operator is self-adjoint. This is because we could have $\ip{Tx,y}=\ip{x,Ty}$ for all $x,y\in\Ds(T)$ but not for all $x,y\in\Ds(T\dg)$. Every self-adjoint operator is also closed, which follows directly from \Cref{adjointclosed}. As such, given an operator that is not closed, it will help if we are able to extend it to a closed operator.

\begin{definition}
An operator $T$ on $\Hs$ is \textbf{closable} if it has a closed extension.
\end{definition}

\begin{proposition}
$T$ is closable if and only if $\overline{\Gs(T)}$ (the closure of $\Gs(T)$ in $\Hs\times\Hs$) is the graph of an operator on $\Hs$.
\end{proposition}
\begin{proof}
($\Leftarrow$) Suppose $\overline{\Gs(T)}$ is the graph of an operator $S:\Ds(S)\to\Hs$. Then $\Gs(S)=\overline{\Gs(T)}\sups\Gs(T)$, so $S$ is an extension of $T$. Also, $\Gs(S)=\overline{\Gs(T)}$ is closed in $\Hs\times\Hs$, and so $S$ is a closed operator. Thus $T$ is closable.\\

($\Rightarrow$) Suppose $T$ is closable. Then there is a closed operator $S:\Ds(S)\to\Hs$ such that $\Gs(S)\sups\Gs(T)$. Since $S$ is closed, $\Gs(S)$ is closed in $\Hs\times\Hs$, and since it contains $\Gs(T)$, it also contains $\overline{\Gs(T)}$. Since $\Gs(S)$ is the graph of $S$, by \Cref{graphequiv}, we have $(0,y)\notin\Gs(S)$ for all $y\ne0$. Thus $(0,y)\notin\overline{\Gs(T)}$ for all $y\ne0$. Moreover, since $\Gs(T)$ is a subspace of $\Hs\times\Hs$, so is $\overline{\Gs(T)}$. By \Cref{graphequiv}, $\overline{\Gs(T)}$ is the graph of an operator on $\Hs$.
\end{proof}

We usually do not have to worry about $T$ being closable or not, as most operators encountered in quantum mechanics are closable. See \cite[Example 7.1.4, Page 522]{simon} for an example of an operator that is not closable.

\begin{definition}
Suppose $T$ is a closable operator on $\Hs$. The \textbf{closure} of $T$ is given by $\Gs(\overline{T})=\overline{\Gs(T)}$.
\end{definition}
\begin{remark}
Since $T$ is closable, by the previous proposition, $\overline{\Gs(T)}$ is the graph of an operator on $\Hs$, so $\overline{T}$ is well-defined.
\end{remark}

\begin{definition}
An operator $T$ on $\Hs$ is \textbf{essentially self-adjoint} if its closure is self-adjoint.
\end{definition}
\begin{remark}
Equivalently, $T$ is essentially self-adjoint if it has \emph{exactly one} self-adjoint extension.
\end{remark}

We will now present an example of how this theory arises in the analysis of solving the Schrödinger equation.\\

Suppose $\Hs=L^2[a,b]$ (the space of square-integrable functions on $[a,b]$) and define the operator $T:\Ds(T)\to\Hs$ as follows:
\begin{equation*}
    (T\psi)(x)=-\psi''(x)+V(x)\psi(x)
\end{equation*}
This is simply the Hamiltonian operator from \Cref{sec:schrodinger}. Note that it still remains to define $\Ds(T)$. We need to be careful in doing so:

\vspace*{4pt}

\begin{itemize}
    \item If $\Ds(T)$ is \emph{too large}, say $L^2[a,b]$ or $L^\infty[a,b]$, the operator $T$ would not be well-defined (we cannot prescribe a meaning of $\psi''$ to an arbitrary function $\psi\in L^2[a,b]$).
    \item If $\Ds(T)$ is \emph{too small}, say $C^2[a,b]$ or $C^\infty[a,b]$, the operator $T$ would not be essentially self-adjoint, as these spaces are in a sense `too small' to capture the true behavior of $T$
\end{itemize}

\vspace*{4pt}

A suitable choice for $\Ds(T)$ is the Sobolev space $H^2[a,b]=W^{2,2}[a,b]$ (the space of all square-integrable functions on $[a,b]$ whose second derivative is also square-integrable).\\

Now that we have defined $T$, we can test if it is Hermitian by computing an inner product with arbitrary $\phi,\psi\in\Ds(T)$:
\begin{equation*}
    \ip{T\phi,\psi}
    =\int_a^b (-\phi''(x)+V(x)\phi(x))\overline{\psi(x)}\dx
\end{equation*}
Since $V(x)$ is real-valued, the second term is equal to $\int_a^b \phi(x)\overline{V(x)\psi(x)}\dx$. Integrating the first term by parts, we get:
\begin{align*}
    -\int_a^b \phi''(x)\overline{\psi(x)}\dx
    &=-\phi'(b)\overline{\psi(b)}+\phi'(a)\overline{\psi(a)}+\int_a^b \phi'(x)\overline{\psi'(x)}\dx \\
    &=-\phi'(b)\overline{\psi(b)}+\phi'(a)\overline{\psi(a)}+\phi(b)\overline{\psi'(b)}-\phi(a)\overline{\psi'(a)}-\int_a^b \phi(x)\overline{\psi''(x)}\dx
\end{align*}
This yields:
\begin{align*}
    \ip{T\phi,\psi}
    &=-\phi'(b)\overline{\psi(b)}+\phi'(a)\overline{\psi(a)}+\phi(b)\overline{\psi'(b)}-\phi(a)\overline{\psi'(a)}-\int_a^b \phi(x)\overline{\psi''(x)}\dx+\int_a^b \phi(x)\overline{V(x)\psi(x)}\dx \\
    &=-\phi'(b)\overline{\psi(b)}+\phi'(a)\overline{\psi(a)}+\phi(b)\overline{\psi'(b)}-\phi(a)\overline{\psi'(a)}+\ip{\phi,T\psi}
\end{align*}
For $T$ to be Hermitian, the boundary terms need to cancel for all $\phi,\psi\in\Ds(T)$. This can be done, for example, by imposing homogeneous Dirichlet and Neumann boundary conditions on $\phi$:
\begin{equation*}
    \phi(a)=\phi(b)=\phi'(a)=\phi'(b)=0
\end{equation*}
In which case all four boundary terms vanish and we do not need any boundary conditions on $\psi$. Or conversely, we could impose the same four conditions on $\psi$ and we would not need any boundary conditions on $\phi$. However, for $T$ to be \emph{self-adjoint} (as opposed to merely Hermitian), we additionally need $\Ds(T)=\Ds(T\dg)$, i.e. the boundary conditions must be symmetric in $\phi$ and $\psi$. There are two (essentially) standard choices for these:

\vspace*{4pt}

\begin{itemize}
    \item $\psi(a)=\psi(b)=0$ \hfill (Dirichlet boundary conditions)
    \item $\psi'(a)=\psi'(b)=0$ \hfill (Neumann boundary conditions)
\end{itemize}

\vspace*{4pt}

There are also other choices, such as Robin (mixed) boundary conditions under certain constraints, but these are rarely used in quantum mechanics due to their complexity. See \cite[Chapter 13]{ward}, \cite[\S2.5]{polyanin}, \cite[Chapter 7]{simon} and \cite[Chapter 8]{weidmann} for more details.

\newpage 

In \Cref{subsec:particleinabox}, when we solved the particle in a box, we used Dirichlet boundary conditions. This was necessary due to the nature of the problem: If we used Neumann boundary conditions instead, we would lose the vanishing of $\psi$ at the endpoints $x=0$ and $x=L$. While this is still physically valid, we would get the same solutions as in \eqref{particleinaboxdirect}, but with cosines instead of sines.\\

On the other hand, the supersymmetric partner of the particle in a box \eqref{boxpartner} behaves more nicely. Looking at its eigenstates, we see that both they \emph{and} their derivatives vanish at the endpoints, i.e. they satisfy both Dirichlet and Neumann boundary conditions simultaneously. This seems to contradict the self-adjoint boundary conditions we derived above. The explanation for this is that for the partner particle, the Hilbert space in question is no longer $L^2[a,b]$, but rather the space:
\begin{equation*}
    \Hs=\left\{f\in L^2[a,b]\biggm|\int_a^b \left(1+2\cot\left(\frac{\pi}{L}x\right)^2\right)\abs{f(x)}^2\dx<\infty\right\}
\end{equation*}
What changes here is that we insist our functions vanish sufficiently quickly at the endpoints so as to be square-integrable even against a function that blows up at those points.\\

For more information on self-adjoint operators on Hilbert spaces, see \cite{borthwick}, \cite{budde}, \cite{kadets}, \cite{pedersen}, \cite{schmudgen}, \cite{simon} and \cite{weidmann}.

\section{Conclusions}
So what have we done? In \Cref{sec:schrodinger}, we laid the foundations of SUSY QM. We rewrote the Schrödinger equation in terms of the superpotential and derived the time-independent quantum Hamilton-Jacobi equation. We then defined the supersymmetric partner of a quantum Hamiltonian and saw how its non-uniqueness led to several interesting behaviors, such as the lack of normalizable eigenstates and the periodicity of certain states. We showed in \Cref{eigendual} how to link the states of either Hamiltonian to the states of the other, and thereby how to construct the supersymmetric ladder.\\

We then constructed the supersymmetric Hamiltonian and saw how it generates a supersymmetry algebra as well as a condition for unbroken supersymmetry (that the ground state energy must be zero). Armed with this result, we extended the supersymmetric ladder to arbitrarily long chains of partner Hamiltonians and showed how the spectrum of one of them determines the spectra of all the others. Following which, we defined shape invariant potentials and showed in \Cref{sipenergy} how they impose additional structure on the supersymmetric ladder. Finally, we explored the supersymmetric WKB approximation and remarked the open problem of whether it yields exact results for all shape invariant potentials.\\

In \Cref{sec:systems}, we investigated some important examples of quantum systems and exploited the supersymmetric methods from \Cref{sec:schrodinger} to solve them. We saw how the sech potential leads to a wave packet that is entirely transmitted (no wave is reflected back) and how this corresponds to a soliton solution of the nonlinear Schrödinger equation. We saw how the particle in a box leads to a cotangent-squared partner potential, which unless one were \emph{really} clever (or lucky in guessing), they would have no hope of solving with standard QM methods.\\

We also explored the harmonic oscillator and saw that even though its supersymmetric partner is another harmonic oscillator (which \textit{a priori} is equally difficult to solve), the supersymmetric relation between the two is a silver bullet: It reduces the work of solving them to merely applying a single operator again and again (some physicists like to call this ``turning a handle''). We also looked at the nice behavior of the harmonic oscillator with regards to the supersymmetric WKB approximation. Finally, we exploited shape invariance to solve the hydrogen atom with relatively little work, which in a first course on quantum mechanics, would take at least three hours' worth of lectures.\\

In \Cref{sec:matricesop}, we presented a discrete (matrix) formulation of the Schrödinger operator and showed that as nice as it would be, we could not construct partner potentials due to inherently broken supersymmetry (which is made worse by the fact that we had to introduce a perturbation in the matrices to get the algorithm to work, see \Cref{sec:cholesky}). Not to be defeated, we applied the algorithm to random Hamiltonians and noticed an interesting phenomenon: Upon iteration, the partner Hamiltonians converge to a very specific form, which is exactly the form needed for the eigenstate thermalization hypothesis.\\

We then outlined some of the general theory of self-adjoint operators on a Hilbert space and saw how it necessitates boundary conditions in a delicate `balancing act'. Aided by the example of the particle in a box, we noted that this balance is highly sensitive to the operators in question as well as their domains.\\

In this project, we only explored SUSY QM for one-dimensional systems (okay, the hydrogen atom is three-dimensional, but we only made use of the radial dimension in our treatment of it). These ideas work in higher dimensions as well, but exactly solvable examples become few and far between. This is in part due to the lack of known shape invariant potentials in two or more dimensions. See \cite{mallow}, \cite{kuberski}, \cite{quesne} and \cite{sandhya} for more details.\\

The methods of SUSY QM can also be adapted to relativistic quantum mechanics, where it is used to efficiently construct solutions to the Dirac equation instead. This is done in \cite{cooper}, \cite{mallow}, \cite{gudmundsson} and \cite{kuberski}. It can also be applied to quantum electrodynamics (QED), see \cite{cooper} and \cite{mallow}. However, in the setting of QED, it no longer remains a much more efficient method and is usually abandoned in favor of other techniques, such as path integral methods.\\

Nonetheless, it is always nice to find different ways to achieve the same goal...

\vspace*{10mm}

\begin{tcolorbox}[enhanced,center,frame hidden,colback=white,
    borderline north={0.8pt}{0pt}{blue!40!black},borderline south={0.8pt}{0pt}{blue!40!black},
    top=0pt,bottom=0pt,left=0pt,right=0pt]
``You go far enough left, eventually you'll meet someone who has gone far enough right to get to the same place.''
\end{tcolorbox}

\begin{flushright} 
--- Thomas Shelby \cite{peakyblinders}
\end{flushright}

\appendix
\section{Cholesky Decomposition}\label[appendix]{sec:cholesky}
\textbf{Cholesky decomposition} (also known as \textit{Cholesky factorization}) is a method of factoring a positive semidefinite Hermitian matrix. It was first discovered by \href{https://mathshistory.st-andrews.ac.uk/Biographies/Cholesky/}{André-Louis Cholesky}\footnote{Cholesky was a major in the French army who died in battle in World War I. He was also a land surveyor, as were many of the pioneers of linear algebra, such as \href{https://en.wikipedia.org/wiki/Wilhelm_Jordan_(geodesist)}{Wilhelm Jordan} (1842--1899).} (1875--1918), and has applications in solving linear equation systems, numerical optimization problems and Monte Carlo simulations.

\begin{definition}
Suppose $A$ is a positive semidefinite Hermitian matrix. A \textbf{Cholesky decomposition} of $A$ is a decomposition of the form $A=LL\dg$, where $L$ is a lower triangular matrix with non-negative diagonal entries.
\end{definition}

In other words, given a positive semidefinite Hermitian matrix $A$, we would like to decompose it as follows:
\begin{equation*}
    A=\mat{a_{11},a_{12},\cdots,a_{1n}}{a_{21},a_{22},\cdots,a_{2n}}{\vdots,\vdots,\ddots,\vdots}{a_{n1},a_{n2},\cdots,a_{nn}}
    \qquad\longrightarrow\qquad A=LL\dg ,\qquad
    L=\mat{L_{11},0,\cdots,0}{L_{21},L_{22},\cdots,0}{\vdots,\vdots,\ddots,\vdots}{L_{n1},L_{n2},\cdots,L_{nn}}
\end{equation*}

The following formulas give us the required decomposition:

\begin{theorem}[Cholesky Decomposition]
The entries of $L$ are given by (for $1\le j<i\le n$):
\begin{align}\label{choleskyformula}
    L_{jj}=\sqrt{a_{jj}-\sum_{k=1}^{j-1} L_{jk}\overline{L_{jk}}} && L_{ij}=\frac{1}{L_{jj}}\left(a_{ij}-\sum_{k=1}^{j-1} L_{ik}\overline{L_{jk}}\right)
\end{align}
And $L_{ij}=0$ for $i<j$.
\end{theorem}
This can be derived by first solving the upper left entry $L_{11}\overline{L_{11}}=a_{11}$ for $L_{11}$ (which yields $L_{11}=\sqrt{a_{11}}$, since we require the diagonal entries of $L$ to be non-negative), and then working our way down and to the right. This can be done row-by-row (known as the \emph{Cholesky-Banachiewicz algorithm}) or column-by-column (known as the \emph{Cholesky-Crout algorithm}).\\

If $A$ is positive definite, the matrix $L$ is unique and its diagonal entries are strictly positive. See \cite[\S8.3, Theorem 3, Pages 387--388]{nicholson} for a proof.\\

If $A$ is merely positive \emph{semidefinite} (so all eigenvalues are non-negative, but some are zero), the Cholesky decomposition of $A$ is no longer unique. The formula \eqref{choleskyformula} also fails, as at least one of the diagonal entries $L_{jj}$ will be zero, leading to division by zero. This problem can be overcome by adding $\ep I$ (a small positive multiple of the identity matrix) to $A$, so that its smallest eigenvalue $0$ becomes slightly positive. This perturbation will of course lead to a discrepancy in the matrix $L$, but this can usually be ignored\footnote{The usual context where one would \emph{not} want to ignore this discrepancy is when finding the eigenvalues of $A$, but this issue will not arise in the first place. Even if $A$ is positive definite, its eigenvalues cannot be found simply by Cholesky decomposition (the diagonal entries of $L$ are, in general, not the square roots of the eigenvalues of $A$).}. If we subsequently wish to multiply $L$ and $L\dg$ in the reverse order, we can make up for this discrepancy by subtracting $\ep I$ from the result.\\

All in all, Cholesky decomposition provides a good way to numerically perform the supersymmetric decomposition $H^{(1)}=A\dg A$. Note that by convention, we choose the \emph{lower triangular} matrix $L$ to represent the Cholesky decomposition, instead of its (upper triangular) adjoint $L\dg$. Thus for our purposes, $L\dg$ plays the role of the operator $A$ in \eqref{superpotansatz}.

\section{The Spectral Theorems}\label[appendix]{sec:spectral}
In \Cref{sec:schrodinger}, we claimed that every Hamiltonian $H^{(1)}$, as long as its ground state energy is non-negative, can be factored into $H^{(1)}=A\dg A$ for some operator $A$. Here we will outline the rigorous result that makes this possible (the \hyperref[spectraltheorem]{spectral theorem for self-adjoint operators}), as well as its limitations for our purposes.\\

First, recall the spectral theorem for Hermitian matrices:
\begin{theorem}[Spectral Theorem for Hermitian Matrices --- Version 1]
Suppose $T$ is a Hermitian matrix. Then there is a diagonal matrix $D$ with real entries and a unitary matrix $U$ such that $T=UDU\dg$.
\end{theorem}

Essentially, this theorem says that every Hermitian matrix is unitarily similar to a real diagonal matrix. This is essentially why Hermitian matrices are so important in quantum mechanics\footnote{Their \emph{mathematical} importance had been known to Cauchy since the early 19\textsuperscript{th} century, long before quantum mechanics came about.}.\\

There are many versions of the above theorem. We now present one other version that will help motivate the \hyperref[spectraltheorem]{spectral theorem for self-adjoint operators}.

\begin{theorem}[Spectral Theorem for Hermitian Matrices --- Version 2]\label{spectralv1}
Suppose $T$ is an $n\times n$ Hermitian matrix and $\lam_1,\lam_2,...,\lam_n$ are its eigenvalues (including multiplicity). Then we have:
\begin{equation*}
    T=\sum_{k=1}^n \lam_k P_{\lam_k}
\end{equation*}
Where $P_{\lam_k}$ is the orthogonal projection matrix for the subspace $\ker(T-\lam_k I)$, i.e. it acts on a vector in $\C^n$ by keeping the component within $\ker(T-\lam_k I)$ and discards the component orthogonal to it.
\end{theorem}
\begin{remark}
This version of the spectral theorem is sometimes known as the \emph{resolution of the identity} version -- more on that later.
\end{remark}

Both versions of the spectral theorem above allow us to apply functions to a Hermitian matrix. For example, if $f:\C\to\C$ is a function, we can define the matrix $f(T)$ in either of the following ways:
\begin{align*}
    \textsf{Version 1: } T&=U
    \begin{pNiceMatrix}
        \lam_1 & & 0 \\
        & \Ddots & \\
        0 & & \lam_n
    \end{pNiceMatrix}
    U\dg
    &&\longrightarrow&&
    f(T)=U
    \begin{pNiceMatrix}
        f(\lam_1) & & 0 \\
        & \Ddots & \\
        0 & & f(\lam_n)
    \end{pNiceMatrix}
    U\dg \\
    \textsf{Version 2: } T&=\sum_{k=1}^n \lam_k P_{\lam_k}
    &&\longrightarrow&&
    f(T)=\sum_{k=1}^n f(\lam_k) P_{\lam_k}
\end{align*}
As long as the values $f(\lam_1),f(\lam_2),...,f(\lam_n)$ are well-defined, so is the matrix $f(T)$. For example, if $\lam_1,\lam_2,...,\lam_n\ne0$, we can set $f(z)=\frac{1}{z}$ to get a matrix $\frac{1}{T}$ (which is equal to the inverse matrix $T^{-1}$). This in turn allows us to essentially do calculus with matrices, a theory known as \textbf{matrix functional calculus}.\\

Extending this theorem to self-adjoint operators on a Hilbert space is a \emph{massive} leap (historically, it took over a hundred years). There are some intermediate steps along the way, such as the spectral theorems for compact self-adjoint operators, and for bounded self-adjoint operators. See \cite{borthwick}, \cite[Chapter 12]{ward}, \cite[Chapter 11]{kadets} and \cite[Chapter 5]{simon} for these (and more) versions of the spectral theorem.\\

The idea behind this extension is to turn the collection of projections in \Cref{spectralv1} into a \emph{measure}, and the sum into an \emph{integral}.

\begin{definition}
Suppose $\Omega$ is a non-empty set and $\As$ is a $\sigma$-algebra on $\Omega$. Also suppose $\Hs$ is a Hilbert space and $\Ps$ is the set of all orthogonal projection operators\footnotemark\ on $\Hs$. A \textbf{spectral measure} (or \textit{projection-valued measure}) on $(\Omega,\As)$ is a function $E:\As\to\Ps$ such that:
\begin{enumerate}[ref={(\arabic*)}]
    \item $E(\Omega)=I$
    \item\label{specmeasure:2} For any sequence $(A_n)_{n=1}^\infty$ of sets in $\As$, we have $E(\cuppn A_n)=\sumn E(A_n)$. \hfill (Countable additivity)
\end{enumerate}
Where the infinite sum in \ref{specmeasure:2} denotes strong convergence, i.e. convergence in the strong operator topology on $\Bs(\Hs)$.
\end{definition}
\vspace*{-10pt}
\footnotetext{An orthogonal projection operator on $\Hs$ is an operator $P:\Hs\to\Hs$ such that $P\dg=P$ (self-adjoint) and $P^2=P$ (idempotent). These conditions automatically imply that $P$ is bounded.}
\begin{remark}
We do not need to assume that $E(\emp)=0$, this already follows from \ref{specmeasure:2} by setting $A_n=\emp$ for all $n\in\N$.
\end{remark}

A spectral measure is analogous to a probability measure, except that the outputs are projections on a Hilbert space, rather than numbers between $0$ and $1$.\\

Now that we have defined spectral measures, we are ready to state the spectral theorem for self-adjoint operators:

\begin{theorem}[Spectral Theorem for Self-adjoint Operators]\label{spectraltheorem}
Suppose $T$ is a self-adjoint operator on a Hilbert space $\Hs$. Then there is exactly one spectral measure $E_A$ on $\Bs(\R)$ (the Borel $\sigma$-algebra on $\R$) such that:
\begin{equation*}
    T=\int_{\sigma(T)} \lam\,dE_T(\lam)
\end{equation*}
Where $\sigma(T)\sub\R$ is the spectrum of $T$, and the integral is a Bochner integral (an operator-valued Lebesgue integral).
\end{theorem}
See \cite[Theorem 5.7, Pages 89--91]{schmudgen} for a proof.

\begin{remark}
The spectral measure $E_T$ is sometimes known as the \emph{resolution of the identity} for $T$, as it `resolves' (splits up) the identity operator into a collection of orthogonal projections, on each of which $T$ behaves somewhat nicely. It plays a similar role to a partition of unity in topology.
\end{remark}

The spectral theorem for self-adjoint operators naturally leads to a functional calculus, known as the \textbf{Borel functional calculus}. For any real-valued Borel function $f:\R\to\R$, we can define $f(T)$ as follows:
\begin{equation*}
    f(T)=\int_{\sigma(T)} f(\lam)\,dE_T(\lam)
\end{equation*}
This is well-defined (by the existence and uniqueness of $E_T$), and allows us to generate various operator algebras from $T$. See \cite[\S5.3, Pages 91--99]{schmudgen} for more details.\\

If $T$ is a \emph{positive} self-adjoint operator, i.e. $\ip{Tx,x}\ge0$ for all $x\in\Hs$, then every $\lam\in\sigma(T)$ satisfies $\lam\ge0$. Thus we can set $f(\lam)=\sqrt{\lam}$ and define:
\begin{equation*}
    A=\int_{\sigma(T)} \sqrt{\lam}\,dE_T(\lam)
\end{equation*}
This is a positive self-adjoint operator that satisfies $A^2=T$ (and more importantly for our purposes, $A\dg A=T$, since $A\dg=A$).\\

Okay, so now we know that such an operator $A$ always \emph{exists}. Mathematically, this is an extremely profound and useful result. Physically, however, it is not very useful as we often want $A$ to have other properties than being self-adjoint. For instance, if $A$ were self-adjoint, then the partner Hamiltonian $H^{(2)}$ in \eqref{AAdg} would be identical to $H^{(1)}$, and so it would not be any more helpful to us. The good news is, $A$ is not unique\footnote{It is unique if we require it to be positive and self-adjoint, this is part of the statement of the spectral theorem.}, and many `natural' (in the context of quantum mechanics) choices of $A$ will give rise to these properties rather than positivity and self-adjointness.

\section{Numerical Simulations}\label[appendix]{sec:numerical}
Here we present the Python code used to run the simulations in \Cref{subsec:numericalsusyqm}, as well as the results of more simulations not included there.\\

\noindent {\sffamily\bfseries Preliminaries.} We first import the Python libraries \texttt{numpy} and \texttt{matplotlib.pyplot}. We also import the module \texttt{ortho\_group} from \texttt{scipy.stats}, which allows us to generate random real orthogonal matrices for some of our simulations\footnote{There is a similar module \texttt{unitary\_group} for generating random unitary matrices, but since we are focusing on real symmetric matrices, \texttt{ortho\_group} will suffice.}.

\begin{algotext}
\begin{minted}{python}
import numpy as np
import matplotlib.pyplot as plt
from scipy.stats import ortho_group
\end{minted}
\end{algotext}

We now define the derivative operator as the matrix in \eqref{derivativematrix}:

\begin{algotext}
\begin{minted}{python}
def D(n):
    M=np.array(np.zeros((n,n)))     #n×n matrix
    for i in range(n-1):
        M[i,i+1]=1/2                #Entries just above the main diagonal are 1/2
    for i in range(1,n):
        M[i,i-1]=-1/2               #Entries just below the main diagonal are 1/2
    return M
\end{minted}
\end{algotext}

The following algorithm takes a potential $V$ (expressed as a 1-dimensional array) and outputs the corresponding Hamiltonian $H^{(1)}$ with (almost) zero ground state energy. The purpose of \texttt{tol} is to keep the ground state energy slightly positive, which avoids problems with Cholesky decomposition. Its default value is set at $1\times 10^{-16}$.

\begin{algotext}
\begin{minted}{python}
def hamiltonian(V,tol=1e-16):
    n=np.shape(V)[0]                #Size of the array V
    H=-np.dot(D(n),D(n))+np.diag(V) #Initial Hamiltonian
    gse=min(np.linalg.eigvals(H))   #Ground state energy
    H1=H-(gse-tol)*np.eye(n)        #Subtracting the ground state energy
    return H1
\end{minted}
\end{algotext}

The following algorithm takes a Hamiltonian $H^{(1)}$ (whose ground state energy can be positive, negative or zero) and computes the partner Hamiltonian $H^{(2)}$ by Cholesky decomposition.

\begin{algotext}
\begin{minted}{python}
def susypartner(H1,tol=1e-16):
    n=np.shape(H1)[0]               #Size of the Hamiltonian H1
    gse=min(np.linalg.eigvals(H1))  #Ground state energy of H1
    H1a=H1-(gse-tol)*np.eye(n)      #Subtracting the ground state energy
    A1=np.linalg.cholesky(H1a)      #Cholesky decomposition
    H2a=np.dot(np.transpose(A1),A1) #Computing the partner Hamiltonian
    H2=H2a+(gse-tol)*np.eye(n)      #Adding back the ground state energy
    return H2
\end{minted}
\end{algotext}

\newpage 

The next two blocks of code produce the \hyperref[sim:broken]{potential plots} in \Cref{subsec:numericalsusyqm}, which demonstrate the failure of this numerical method to construct partner potentials due to broken supersymmetry.

\begin{algotext}[{\hypersetup{linkcolor=teal!20}\hyperref[sim:broken]{Broken SUSY Example 1: $V^{(1)}(x)=x$}}] 
\begin{minted}{python}
xs=np.linspace(-5,5,1000)
V1=xs                               #Original potential: V^(1)(x)=cos(x)
H1=hamiltonian(V1)                  #Computing the Hamiltonian H^(1) arising from V^(1)
H2=susypartner(H1)                  #Computing the supersymmetric partner of H^(1)
n=len(xs)
V2=np.diag(H2+np.dot(D(n),D(n)))    #Extracting the partner potential V^(2) from H^(2)

plt.plot(xs,V1,'r')                 #Plotting V^(1) against x
plt.plot(xs,V2,'b')                 #Plotting V^(2) against x
plt.title(r'$V^{(1)}(x)=x$',fontsize=14)
plt.legend([r'$V^{(1)}$',r'$V^{(2)}$'],bbox_to_anchor=(1.02,1),
        loc='upper left',borderaxespad=0)
plt.show()
\end{minted}
\end{algotext}

\begin{algotext}[{\hypersetup{linkcolor=teal!20}\hyperref[sim:broken]{Broken SUSY Example 2: $V^{(1)}(x)=5\cos(x)$}}] 
\begin{minted}{python}
xs=np.linspace(-5,5,1000)
V1=5*np.cos(xs)                     #Original potential: V^(1)(x)=5*cos(x)
H1=hamiltonian(V1)                  #Computing the Hamiltonian H^(1) arising from V^(1)
H2=susypartner(H1)                  #Computing the supersymmetric partner of H^(1)
n=len(xs)
V2=np.diag(H2+np.dot(D(n),D(n)))    #Extracting the partner potential V^(2) from H^(2)

plt.plot(xs,V1,'r')                 #Plotting V^(1) against x
plt.plot(xs,V2,'b')                 #Plotting V^(2) against x
plt.title(r'$V^{(1)}(x)=5\,\cos(x)$',fontsize=14)
plt.legend([r'$V^{(1)}$',r'$V^{(2)}$'],bbox_to_anchor=(1.02,1),
        loc='upper left',borderaxespad=0)
plt.show()
\end{minted}
\end{algotext}

\newpage 

We will now generate a random real symmetric matrix and repeatedly apply the algorithm \texttt{susypartner} to it. We will run four simulations, using two distinct methods to generate the random real symmetric matrix $A$:

\vspace*{10pt}

\begin{enumerate}
    \item We sample the entries of $A$ independently from a uniform distribution, then symmetrize the matrix (by redefining the entries above the diagonal to equal the corresponding entries below it).
    \item Same as above, but we sample from a normal distribution.
    \item We sample the diagonal entries of an auxiliary diagonal matrix $D$ independently from a uniform distribution. We also generate a random real orthogonal matrix $Q$ using the \texttt{scipy.stats} module \texttt{ortho\_group}. We then define the matrix $A=QDQ^\text{T}$ (which is guaranteed to be real and symmetric).
    \item Same as above, but we sample from a normal distribution.
\end{enumerate}

\vspace*{10pt}

In all four cases, the distributions will be symmetric (i.e. have mean $0$) and their variances will be set so that on average, 95\% of the entries of $A$ will lie in the range $[-50,50]$ that we set for the color bar.

\begin{algotext}[{\hypersetup{linkcolor=teal!20}\hyperref[sim:colorbar]{1. Entrywise sampling, Uniform distribution}}] 
\begin{minted}{python}
## Random 1000×1000 matrix with entries uniformly distributed in [-52.6316,52.6316]
A=np.random.uniform(low=-52.6316,high=52.6316,size=(1000,1000))
for i in range(np.shape(A)[0]):
    for j in range(np.shape(A)[1]):
        if i<j:
            A[i,j]=A[j,i]           #Symmetrizing the matrix

plt.imshow(A)                       #Plotting A itself
plt.title(f'Original matrix')
plt.colorbar()
plt.clim(-50,50)                    #Setting the color range to [-50,50]
plt.show()

for i in range(100):
    A=susypartner(A,1e-10)          #Tolerance increased to 1e-10
    plt.imshow(A)                   #Plotting successive supersymmetric partners of A
    if i==0:
        plt.title(f'After 1 iteration')
    else:
        plt.title(f'After {i+1} iterations')
    plt.colorbar()
    plt.clim(-50,50)                #Setting the color range to [-50,50]
    plt.show()
\end{minted}
\end{algotext}

The above code produces the \hyperref[sim:colorbar]{color bar plots} in \Cref{subsec:numericalsusyqm}.

\newpage 

\begin{algotext}[2. Entrywise sampling, Normal distribution]
\begin{minted}{python}
## Random 1000×1000 matrix with entries normally distributed
## with mean 0 and standard deviation 25.5107
A=np.random.normal(loc=0,scale=25.5107,size=(1000,1000))
for i in range(np.shape(A)[0]):
    for j in range(np.shape(A)[1]):
        if i<j:
            A[i,j]=A[j,i]           #Symmetrizing the matrix

plt.imshow(A)                       #Plotting A itself
plt.title(f'Original matrix')
plt.colorbar()
plt.clim(-50,50)                    #Setting the color range to [-50,50]
plt.show()

for i in range(100):
    A=susypartner(A,1e-10)          #Tolerance increased to 1e-10
    plt.imshow(A)                   #Plotting successive supersymmetric partners of A
    if i==0:
        plt.title(f'After 1 iteration')
    else:
        plt.title(f'After {i+1} iterations')
    plt.colorbar()
    plt.clim(-50,50)                #Setting the color range to [-50,50]
    plt.show()
\end{minted}
\end{algotext}

\begin{tcolorbox}[blanker,halign=center]
\begin{multicols}{3}

\includegraphics[width=5cm]{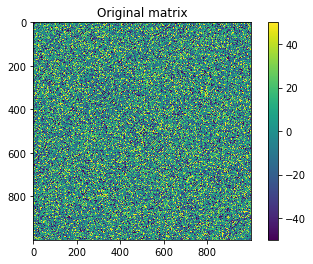}

\includegraphics[width=5cm]{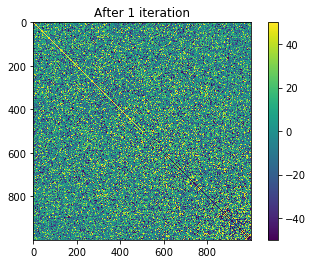}

\includegraphics[width=5cm]{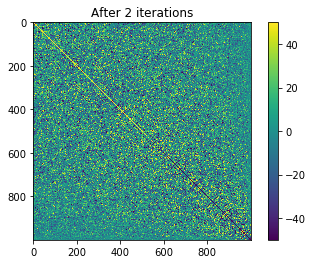}

\end{multicols}
\begin{multicols}{3}

\includegraphics[width=5cm]{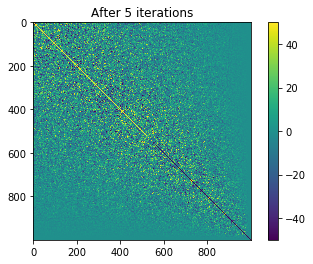}

\includegraphics[width=5cm]{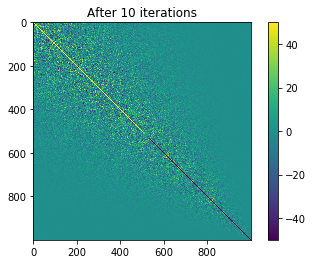}

\includegraphics[width=5cm]{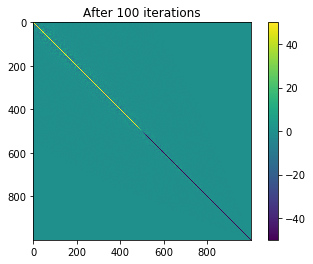}

\end{multicols}
\end{tcolorbox}

\newpage 

\begin{algotext}[3. Diagonal sampling, Uniform distribution]
\begin{minted}{python}
## Random 1000×1000 diagonal matrix with diagonal entries uniformly distributed
## in [-1675.32,1675.32]
D=np.diag(np.random.uniform(low=-1675.32,high=1675.32,size=1000))
Q=ortho_group.rvs(1000)             #Random 1000×1000 real orthogonal matrix
A=np.dot(np.dot(Q,D),np.transpose(Q))

plt.imshow(A)                       #Plotting A itself
plt.title(f'Original matrix')
plt.colorbar()
plt.clim(-50,50)                    #Setting the color range to [-50,50]
plt.show()

for i in range(100):
    A=susypartner(A,1e-10)          #Tolerance increased to 1e-10
    plt.imshow(A)                   #Plotting successive supersymmetric partners of A
    if i==0:
        plt.title(f'After 1 iteration')
    else:
        plt.title(f'After {i+1} iterations')
    plt.colorbar()
    plt.clim(-50,50)                #Setting the color range to [-50,50]
    plt.show()
\end{minted}
\end{algotext}

\begin{tcolorbox}[blanker,halign=center]
\begin{multicols}{3}

\includegraphics[width=5cm]{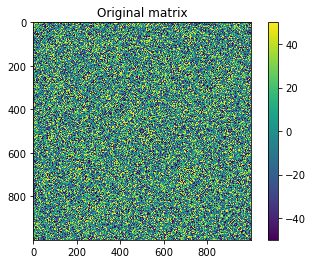}

\includegraphics[width=5cm]{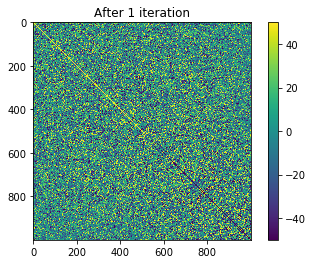}

\includegraphics[width=5cm]{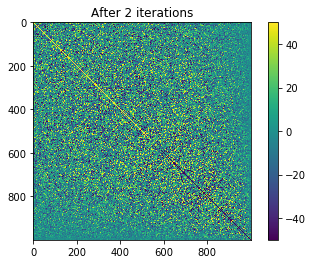}

\end{multicols}
\begin{multicols}{3}

\includegraphics[width=5cm]{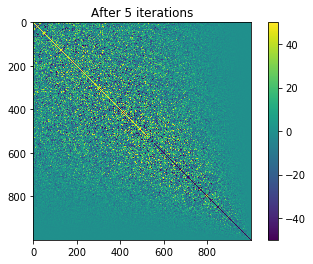}

\includegraphics[width=5cm]{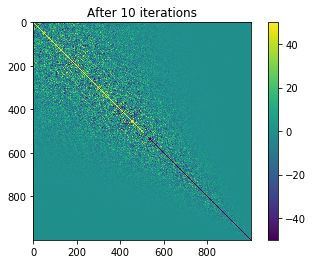}

\includegraphics[width=5cm]{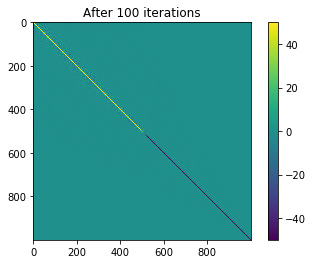}

\end{multicols}
\end{tcolorbox}

\newpage 

\begin{algotext}[4. Diagonal sampling, Normal distribution]
\begin{minted}{python}
## Random 1000×1000 diagonal matrix with diagonal entries normally distributed
## with mean 0 and standard deviation 812.030
D=np.diag(np.random.normal(loc=0,scale=812.030,size=1000))
Q=ortho_group.rvs(1000)             #Random 1000×1000 real orthogonal matrix
A=np.dot(np.dot(Q,D),np.transpose(Q))

plt.imshow(A)                       #Plotting A itself
plt.title(f'Original matrix')
plt.colorbar()
plt.clim(-50,50)                    #Setting the color range to [-50,50]
plt.show()

for i in range(100):
    A=susypartner(A,1e-10)          #Tolerance increased to 1e-10
    plt.imshow(A)                   #Plotting successive supersymmetric partners of A
    if i==0:
        plt.title(f'After 1 iteration')
    else:
        plt.title(f'After {i+1} iterations')
    plt.colorbar()
    plt.clim(-50,50)                #Setting the color range to [-50,50]
    plt.show()
\end{minted}
\end{algotext}

\begin{tcolorbox}[blanker,halign=center]
\begin{multicols}{3}

\includegraphics[width=5cm]{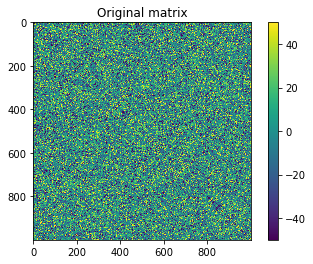}

\includegraphics[width=5cm]{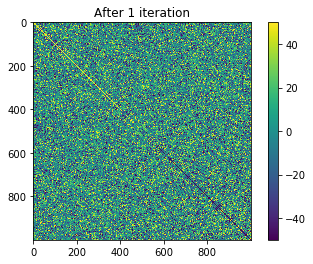}

\includegraphics[width=5cm]{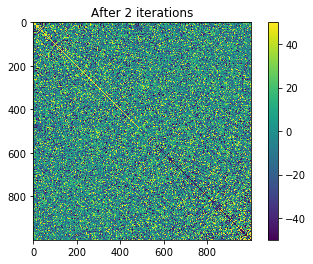}

\end{multicols}
\begin{multicols}{3}

\includegraphics[width=5cm]{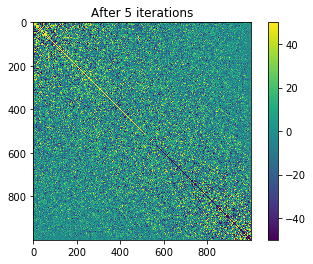}

\includegraphics[width=5cm]{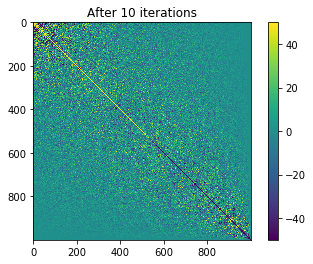}

\includegraphics[width=5cm]{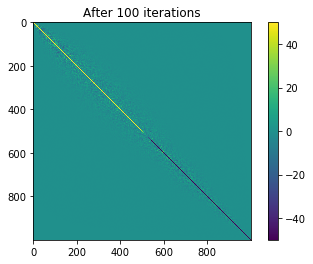}

\end{multicols}
\end{tcolorbox}

\newpage 

The following code produces the \hyperref[sim:diagonaldecreasing]{plot} of the diagonals of the iterates, showing that after many iterations, they line up in decreasing order from upper left to lower right. Here, the entries are still uniformly distributed, but in $[-10,10]$ instead (this does not matter, it merely amounts to a scaling of the \hyperref[sim:colorbar]{color bar plots}).

\begin{algotext}[{\hypersetup{linkcolor=teal!20}\hyperref[sim:diagonaldecreasing]{Diagonals of the iterates}}]
\begin{minted}{python}
## Random 1000×1000 matrix with entries uniformly distributed in [-10,10]
A=np.random.uniform(low=-10,high=10,size=(1000,1000))
for i in range(np.shape(A)[0]):
    for j in range(np.shape(A)[1]):
        if i<j:
            A[i,j]=A[j,i]           #Symmetrizing the matrix
plt.scatter(range(np.shape(A)[0]),np.diag(A),s=1,
        label=f'Original matrix')   #Plotting the diagonal of A

for i in range(100):
    A=susypartner(A,1e-10)
    if i==0:                        #Plotting the diagonal of the 1st iterate
        plt.scatter(range(np.shape(A)[0]),np.diag(A),s=1,
                label=f'After 1 iteration')
    elif i in [1,2,99]:         #Plotting the diagonals of the 2nd, 3rd and 100th iterates
        plt.scatter(range(np.shape(A)[0]),np.diag(A),s=1,
                label=f'After {i+1} iterations')
plt.title('Diagonals of the iterates')
plt.legend(markerscale=6,bbox_to_anchor=(1.02,1),
        loc='upper left',borderaxespad=0)
plt.show()
\end{minted}
\end{algotext}

The following code produces the \hyperref[sim:ethcondition1]{regression plot} used to verify condition \ref{eth:1} of ETH. Here, we have made a few changes:

\vspace*{10pt}

\begin{itemize}
    \item As with the previous block of code, we rescaled the support of the uniform distribution to $[-10,10]$.
    \item The difference $A_{i+1,i+1}-A_{i,i}$ between consecutive diagonal entries is estimated using \eqref{avgdiagdiff}.
    \item Since we need to examine the behavior of this differences as the size of the matrix increases, we need to generate random matrices of different sizes. Here, we have used $n=100,120,140,...,1000$.
    \item We only performed 5 iterations of the \texttt{susypartner} algorithm on each matrix (instead of 100). This is partly to reduce the computation time, but also because we know from the color bar plots that a few iterations already approximate the diagonal fairly well.
\end{itemize}

\begin{algotext}[{\hypersetup{linkcolor=teal!20}\hyperref[sim:ethcondition1]{Verifying Condition 1 of ETH}}]
\begin{minted}{python}
sizes=range(100,1020,20)            #List of matrix sizes to sample
differences=[]                      #List of differences
for n in sizes:
    #Random n×n matrix with entries uniformly distributed in [-10,10]
    A=np.random.uniform(low=-10,high=10,size=(n,n))
    for i in range(n):
        for j in range(n):
            if i<j:
                A[i,j]=A[j,i]       #Symmetrizing the matrix
    for k in range(5):              #Perform 5 iterations
        A=susypartner(A,1e-10)

    diagonal=np.diag(A)             #Extract the diagonal of the matrix
    diff=(A[0,0]-A[-1,-1])/(n-1)    #Average difference between consecutive entries
    differences.append(diff)        #Appending this difference to the list

plt.scatter(sizes,differences,s=3)  #Scatter plot of differences against matrix size
a,b=np.polyfit(np.log(sizes),np.log(differences),1) #Power-law regression
plt.plot(sizes,np.exp(b)*sizes**a,'g')              #Regression plot
plt.title('Condition 1 of ETH',fontsize=14)
plt.xlabel('Size of the matrix')
plt.ylabel('Average difference between\nconsecutive diagonal entries')
rsquared=(np.corrcoef(np.log(sizes),np.log(differences))[0,1])**2 #r^2
text='Power-law fit: $y=Cx^p$\n\n$p=${}\n$C=${}\n$r^2=${}'.format(
    a,np.exp(b),rsquared)
plt.text(1.05,0.3,text,fontsize=16,transform=plt.gca().transAxes)
plt.show()
\end{minted}
\end{algotext}

The following code produces the \hyperref[sim:ethcondition2]{regression plot} used to verify condition \ref{eth:2} of ETH. Here, we have made the same changes above, as well as the following:

\vspace*{10pt}

\begin{itemize}
    \item The off-diagonal entries are estimated by $\frac{2}{n(n-1)}\sum_{1\le j<i\le n} \abs{A_{ij}}$, the average (mean) of the absolute values of all entries below the diagonal. We only consider the entries \emph{below} the diagonal as the matrix is symmetric, so this will yield the same average as using all off-diagonal entries (but with shorter computation time).
    \item Unlike the simulation used to verify condition \ref{eth:1}, it is not sufficient here to perform 5 iterations of \texttt{susypartner}. This is because for larger matrices, more iterations are required for the off-diagonal entries to decay (in other words, the `dust' takes longer to settle). To this end, we perform $\frac{n}{20}$ iterations (which is always an integer $\ge5$ for $n=100,120,140,...,1000$). Again, we use $\frac{n}{20}$ instead of $n$ to reduce the computation time (however the time for the code below is still very long, so try at your own risk!).
\end{itemize}

\begin{algotext}[{\hypersetup{linkcolor=teal!20}\hyperref[sim:ethcondition1]{Verifying Condition 2 of ETH}}]
\begin{minted}{python}
sizes=range(100,1020,20)            #List of matrix sizes to sample
offdiags=[]                         #List of off-diagonals
for n in sizes:
    #Random n×n matrix with entries uniformly distributed in [-10,10]
    A=np.random.uniform(low=-10,high=10,size=(n,n))
    for i in range(n):
        for j in range(n):
            if i<j:
                A[i,j]=A[j,i]       #Symmetrizing the matrix
    for k in range(int(n/20)):      #Perform n/20 iterations
        A=susypartner(A,1e-10)

    below=np.absolute(A[np.tril_indices(n,-1)])     #Entries below the diagonal
    avgoffd=np.average(np.absolute(below))          #Average absolute value of entries
    offdiags.append(avgoffd)                        #Appending this average to the list

plt.scatter(sizes,offdiags,s=3)     #Scatter plot of averages against matrix size
a,b=np.polyfit(np.log(sizes),np.log(offdiags),1)    #Power-law regression
plt.plot(sizes,np.exp(b)*sizes**a,'g')              #Regression plot
plt.title('Condition 2 of ETH',fontsize=14)
plt.xlabel('Size of the matrix')
plt.ylabel('Average absolute value\nof off-diagonal entries')
rsquared=(np.corrcoef(np.log(sizes),np.log(offdiags))[0,1])**2 #r^2
text='Power-law fit: $y=Cx^p$\n\n$p=${}\n$C=${}\n$r^2=${}'.format(
    a,np.exp(b),rsquared)
plt.text(1.05,0.3,text,fontsize=16,transform=plt.gca().transAxes)
plt.show()
\end{minted}
\end{algotext}

The above blocks of code are implemented with the first method we used to generate random real symmetric matrices (entrywise sampling, uniform distribution). These have been tested with the other three methods, all yielding similar results.

\clearpage 
\phantomsection 
\addcontentsline{toc}{section}{References}
\printbibliography

\vfill

{
\definecolor{col}{HTML}{0AD1B0}
\begin{tcolorbox}[
    enhanced,width=6in,center,title=Photo Credits,
    frame hidden,colback=white,colbacktitle=white,
    fonttitle=\sffamily\bfseries,coltitle=black,
        borderline north={1.5pt}{0pt}{col},
        attach boxed title to top center={yshift=-0.25mm-\tcboxedtitleheight/2,yshifttext=2mm-\tcboxedtitleheight/2},
        boxed title style={boxrule=0.5mm,
        frame code={\path[tcb fill frame,col] ([xshift=-4mm]frame.west)--(frame.north west)--(frame.north east)--([xshift=4mm]frame.east)--(frame.south east)--(frame.south west)--cycle;},
        interior code={\path[tcb fill interior] ([xshift=-2mm]interior.west)--(interior.north west)--(interior.north east)--([xshift=2mm]interior.east)--(interior.south east)--(interior.south west)--cycle;}},
    top=\tcboxedtitleheight/2,bottom=0pt,left=0pt,right=0pt]
    \begin{itemize}[align=left,leftmargin=0.4in,font=\sffamily\selectfont]
        \item[{\hyperref[photo:fermi]{Fermi}:}] \url{https://www.smithsonianmag.com/history/enrico-fermi-really-father-nuclear-age-180967214/}
        \item[{\hyperref[photo:dirac]{Dirac}:}] \url{https://www.wsj.com/articles/the-pleasure-and-pain-of-scientific-predictions-11597935357}
        \item[{\hyperref[photo:bose]{Bose}:}] \url{https://dailyasianage.com/news/217177/satyendra-nath-bose}
        \item[{\hyperref[photo:einstein]{Einstein}:}] \url{https://www.biography.com/scientist/albert-einstein}
    \end{itemize}
\end{tcolorbox}
}

\end{document}